\documentclass[aps,prd,groupedaddress,nofootinbib]{revtex4-2} 
\usepackage{graphicx}
\usepackage{slashed}
\usepackage{braket}

\usepackage{bbold}
\usepackage[dvipsnames]{xcolor}
\usepackage{tikz}
\usetikzlibrary{decorations.pathmorphing}
\usetikzlibrary{shapes.geometric}
\usetikzlibrary{calc}
\usetikzlibrary{shapes.misc}
\usetikzlibrary{decorations.markings}
\tikzstyle{gluon}=[decorate, decoration={coil,aspect=0.8, amplitude=1.5pt,  segment length=3pt}]
\tikzstyle{lgluon}=[decorate, decoration={coil,aspect=-0.8, amplitude=1.5pt,  segment length=3pt}]

\usepackage{amsmath,amssymb}
\usepackage{verbatim}
\usepackage{hyperref} 
\hypersetup{
    colorlinks,
    citecolor=black,
    filecolor=black,
    linkcolor=black,
    urlcolor=black
}

\newcommand \nn {\nonumber}

\def\p{{\mathbf p}}
\def\P{{\mathbf P}}

\def\k{{\mathbf k}}
\def\x{{\mathbf x}}  
\def\y{{\mathbf y}}
\def\r{{\mathbf r}}
\def\z{{\mathbf z}}
\def\w{{\mathbf w}}
\def\v{{\mathbf v}}

\def\F{{\mathcal F}}
\def\C{{\mathcal C}}

\def\U{{\mathcal U}}
\def\T{{\mathcal T}}
\def\N{{\mathcal N}}

\def\ln{\text{ln}}
\def\Lam{\Lambda_{\text{QCD}}}

\def\Deik{{\mathcal{D}}}
\def\OpA{\hat{\mathbb{A}}^{(k)}}
\def\OpB{\hat{\mathbb{B}}^{(k)}}
\def\OpC{\hat{\mathbb{C}}^{(k)}}
\begin{document}

\title{Dijet production in DIS off a large nucleus at next-to-eikonal accuracy \\
in a Gaussian model within the CGC framework}

\author{Pedro Agostini$^{a,b}$, Tolga Altinoluk$^{a}$, N\'estor Armesto$^{b}$, Guillaume Beuf$\,^{a}$, Florian Cougoulic$\, ^{c}$ and Swaleha Mulani$^{a,d,e}$}

\affiliation{
 $^a$ Theoretical Physics Division, National Centre for Nuclear Research, Pasteura 7, Warsaw 02-093, Poland 
 \\
$^b$ Instituto Galego de F\'{\i}sica de Altas Enerx\'{\i}as IGFAE, Universidade de Santiago de Compostela, 15782 Santiago de Compostela, Galicia-Spain\\
$^c$ Institute of Theoretical Physics, Jagiellonian University, ul. \L ojasiewicza 11, 30-348, Krak\'ow, Poland \\
$^{d}$Department of Physics, University of Jyv\"askyl\"a, P.O. Box 35, 40014 University of Jyv\"askyl\"a, Finland\\
$^{e}$Helsinki Institute of Physics, P.O. Box 64, 00014 University of Helsinki, Finland} 

\date{\today}

\begin{abstract}
We develop a Gaussian model to evaluate the decorated dipole and quadrupole operators that arise beyond the eikonal approximation in the Color Glass Condensate framework. While the method is general and applicable to arbitrary beyond-eikonal Wilson line structures, we employ it for dijet production in deep inelastic scattering at next-to-eikonal accuracy. After validating the model at the eikonal level, we compute all next-to-eikonal  operator structures entering the dijet cross section. We show that some of them do not contribute to this observable, while others vanish identically.
Therefore, in the Gaussian model next-to-eikonal  corrections to dijet production in deep inelastic scattering originate solely from a given type of operators and from next-to-eikonal three-point correlators. The resulting expressions are provided in a form suitable for numerical implementation. 

\end{abstract}

\maketitle

\tableofcontents


\section{Introduction}
\label{sec:intro}

The Color Glass Condensate (CGC) (see~\cite{Gelis:2010nm, Kovchegov:2012mbw,Albacete:2014fwa, Blaizot:2016qgz} for reviews and references therein) is an effective field theory framework within Quantum Chromodynamics (QCD) that describes the high-energy regime, where the density of gluons inside a hadron or nucleus becomes so large that nonlinear interactions, leading to saturation effects, dominate. The CGC is mostly used to study processes such as deep inelastic scattering (DIS) at small Bjorken-$x$ and proton-nucleus ($pA$) collisions at the Relativistic Heavy Ion Collider (RHIC) and Large Hadron Collider (LHC). 

At sufficiently high energies, hadrons appear as dense clouds of gluons due to the gluon splitting with increasing energy: as the longitudinal momentum fraction ($x$) of the gluons decreases, each emitted gluon can further emit additional gluons and the density of the gluons inside the hadron increases rapidly. However, the self-interactions between the emitted gluons tame the rapid growth and cause a saturation effect. This saturation phenomena is defined via a momentum scale $Q_s$ which characterizes the transition from the dilute linear regime to the dense nonlinear regime. 

The CGC framework is formulated using classical Yang-Mills equations of motion, with the high-density gluon fields behaving like a classical field sourced by a static color charge distribution~\cite{McLerran:1993ni,McLerran:1993ka,McLerran:1994vd}. The evolution of these color charges with energy (or, equivalently, rapidity) is governed by nonlinear renormalization group equations, in particular the  Jalilian-Marian-Iancu-McLerran-Weigert-Leonidov-Kovner (JIMWLK) equation~\cite{Jalilian-Marian:1996mkd,Jalilian-Marian:1997qno,Jalilian-Marian:1997jhx,Jalilian-Marian:1997ubg,Kovner:2000pt,Weigert:2000gi,Iancu:2000hn,Iancu:2001ad,Ferreiro:2001qy} and its mean field approximation known as the Balitsky-Kovchegov (BK) equation~\cite{Balitsky:1995ub,Kovchegov:1999yj,Kovchegov:1999ua}.  

One of the key approximations adopted within the CGC framework is the eikonal approximation. It amounts to keeping only the leading terms in the high-energy limit, and discarding the ones that are power-suppressed by energy, in the computation of observables.  In the CGC framework, high energy is achieved by boosting the target along the $x^-$ direction with a boost parameter $\gamma_t$. 
In a high-energy dilute-dense scattering process, the eikonal approximation relies on three key assumptions: (i) due to Lorentz contraction, the highly boosted background field representing the target is localized in the longitudinal direction around $x^+=0$; 
(ii) the target’s background field is dominated by its leading component in $\gamma_t$, while subleading terms in $\gamma_t$ 
are disregarded; and (iii) Lorentz time dilation causes the background field to be independent of the light-cone coordinate $x^-$, effectively treating the target fields as static and neglecting their internal dynamics. Deviations from any of these assumptions introduce corrections to the eikonal approximation. To incorporate next-to-eikonal (NEik) corrections, one must account for terms of order $1/\gamma_t$. In addition to these three assumptions done in the eikonal approximation in the presence of the gluon background field of the target, another source of corrections to the eikonal approximation is to account for the quark background field of the target, via $t$-channel quark exchanges during the interaction of the projectile and the target.  While the eikonal approximation is a strong and reliable tool for studying scattering processes at the LHC with relatively high scattering energies, corrections to this approximation might be sizable, especially for phenomenological studies at RHIC and at the upcoming Electron-Ion Collider (EIC). 

Over the past decade, substantial progress has been made in deriving and applying subeikonal corrections within the CGC framework. NEik corrections to gluon and quark propagators have been computed in a gluon background field and they are used to computed various observables at NEik accuracy~\cite{Altinoluk:2014oxa,Altinoluk:2015gia,Altinoluk:2015xuy,Agostini:2019avp,Agostini:2019hkj,Altinoluk:2020oyd,Altinoluk:2021lvu,Agostini:2022ctk,Agostini:2022oge,Altinoluk:2022jkk,Agostini:2023cvc,Agostini:2024xqs,Altinoluk:2024zom,Altinoluk:2024dba}. The NEik corrections  that are stemming from $t$-channel quark exchanges between the projectile partons and the dense target have been studied in~\cite{Altinoluk:2023qfr,Altinoluk:2024tyx,Altinoluk:2025ang} to probe the quark content of the target.  The next-to-leading order (NLO) corrections to NEik deep inelastic scattering (DIS) structure function have been computed in~\cite{Altinoluk:2025ivn}. In~\cite{Kovchegov:2015pbl,Kovchegov:2016zex,Kovchegov:2016weo,Kovchegov:2017jxc,Kovchegov:2017lsr,Kovchegov:2018znm,Kovchegov:2018zeq,Kovchegov:2020kxg,Kovchegov:2020hgb,Adamiak:2021ppq,Kovchegov:2021lvz,Kovchegov:2021iyc,Cougoulic:2022gbk,Kovchegov:2022kyy,Borden:2023ugd,Kovchegov:2024aus,Borden:2024bxa}, not only the quark and gluon helicity evolutions but also observables such a single and/or double spin asymmetries are computed at NEik accuracy. Helicity dependent extensions of the CGC have been studied in  \cite{Cougoulic:2019aja,Cougoulic:2020tbc} at NEik order. On the other hand, the rapidity evolution of gluon transverse momentum dependent distribution functions (TMDs) that interpolates between moderate and small-x beyond eikonal accuracy has been studied in~\cite{Balitsky:2015qba,Balitsky:2016dgz,Balitsky:2017flc}. A similar interpolation has been also investigated for inclusive DIS~\cite{Boussarie:2020fpb,Boussarie:2021wkn} and for exclusive Compton scattering~\cite{Boussarie:2023xun}. Quark and gluon propagators have been studied at NEik accuracy within the high energy operator product expansion (OPE) formalism in~\cite{Chirilli:2018kkw,Chirilli:2021lif}. In \cite{Li:2023tlw,Li:2024fdb,Li:2024xra}, subeikonal corrections in the CGC are analyzed  using an effective Hamiltonian approach. Additionally, they have been explored through a framework that accounts for longitudinal momentum exchange between the projectile and target in~\cite{Jalilian-Marian:2017ttv,Jalilian-Marian:2018iui,Jalilian-Marian:2019kaf}. Finally, effects of subeikonal corrections have also been examined in the context of orbital angular momentum in~\cite{Hatta:2016aoc,Kovchegov:2019rrz,Boussarie:2019icw,Kovchegov:2023yzd,Kovchegov:2024wjs}. 

In this paper, we focus on dijet production in DIS off a dense target at NEik accuracy. The NEik corrections to this process were first computed in~\cite{Altinoluk:2022jkk,Altinoluk:2024zom} and their numerical analysis was performed in the dilute target limit in~\cite{Agostini:2024xqs}. As discussed in detail in~\cite{Altinoluk:2022jkk,Altinoluk:2024zom}, at NEik order the DIS dijet production cross section is expressed in terms of "decorated" dipole and quadrupole operators. These operators are composed of Wilson lines with local field strength tensor insertions along their longitudinal extent.
More precisely, in the present study, we restrict ourselves to NEik corrections related to the transverse components of the gluon background field or to the effects beyond infinite Lorentz contraction of the target, as a first step. By contrast, we do not yet consider the NEik corrections beyond the static approximation or the ones related to the quark background field. At eikonal order, the numerical analysis of DIS dijet production cross section, which is expressed in terms of standard dipole and quadrupole operators, is usually performed by adopting the McLerran-Venugopalan (MV) model~\cite{McLerran:1993ka,McLerran:1994vd} or the Golec-Biernat-Wustoff (GBW)~\cite{Golec-Biernat:1998zce,Golec-Biernat:1999qor} model to describe these operators. However, the standard implementation of these models cannot describe the decorated dipole and decorated quadrupole operators, and one should go beyond them to study DIS dijet production at NEik accuracy numerically. Motivated by the MV and GBW models and working along the lines in~\cite{Dumitru:2018vpr,Cougoulic:2020tbc,Agostini:2024xqs}, in this work we introduce a Gaussian model to describe the decorated dipole and quadrupole operators.

The manuscript is organized as follows. In Section~\ref{sec:DIS_dijet_prod} we briefly discuss DIS dijet production at NEik accuracy and provide the analytical expressions of the eikonal and NEik contributions to the production cross section. In Section~\ref{sec:Eik_operators_within_GM} we derive the expressions for the dipole and quadrupole operators at Eikonal order in a Gaussian model. In Section~\ref{sec:NEik_operators_within_GM} we generalize the derivation of the Gaussian model to the decorated dipole and decorated quadrupole operators that appear in the observables computed beyond eikonal accuracy. In Section~\ref{sec:X_sec_in_Gaussian_model} we revisit the NEik DIS dijet production process and provide the cross section within the Gaussian model which can be conveniently used for numerical studies. Finally, in Section~\ref{sec:summary} we provide a concise summary of our study and provide an outlook. Moreover, in Appendix~\ref{appA}, we provide the derivation of the color algebra used in the Gaussian model, in Appendix~\ref{App:Longi_integrals} we perform the longitudinal integrals that are needed for the evaluation of NEik type-1 and type-2 operators, and Appendix~\ref{app:Compare_to_literature} is devoted to comparison of our result for the eikonal quadrupole computed in the Gaussian model to the known results in the literature.   


\section{DIS dijet production at next-to-eikonal accuracy}
\label{sec:DIS_dijet_prod}


At eikonal order, the production cross section of a quark jet with transverse momenta $\k_1$ and rapidity $\eta_1$ and an antiquark jet with transverse momenta $\k_2$ and rapidity $\eta_2$ in DIS via a photon with  polarization $\lambda$ reads\footnote{Boldface letters will denote transverse coordinates or momenta, while their moduli will be denoted by the subscript $\perp$. Besides, a shorthand notation is introduced for the integration over the transverse coordinates, $\int_{\x}\equiv\int d^2\x$. Similarly, the shorthand notation used for the transverse momentum integrals in the rest of manuscript is defined as $\int_{\p}\equiv\int \frac{d^2\p}{(2\pi)^2}$.} 
\begin{align}
\label{Eik_dijet_Xsec}
\frac{d\sigma^{\gamma^{*}_{\lambda}+ A \rightarrow q \bar q + X}}{d^{2}\k_{1}d^{2}\k_{2}d\eta_{1}d\eta_{2}} \bigg |_{\text{Eik}} = \int_{\v, \v', \w, \w'} e^{i\k_{1} \cdot (\mathbf{v'}-\mathbf{v}) + i \k_{2} \cdot (\w' - \w)} \C_{\lambda}(\w'-\v', \w - \v) \\ \nn
\times \big[Q(\w', \v', \v, \w) - d(\w' , \v') - d(\v , \w)  + 1\big].
\end{align}
The dipole $d(\mathbf{v} , \w)$ and quadrupole $Q(\w', \mathbf{v'}, \mathbf{v}, \w)$ operators encode multiple scattering of the partons on the dense target. They are defined as 
\begin{eqnarray}
\label{eik_dip}
d(\mathbf{v} , \w) &=& \frac{1}{N_{c}} \Braket{ \text{Tr} [\U( \mathbf{v}) \U^{\dagger}(\w)]}, \\
 Q(\w', \mathbf{v'}, \mathbf{v}, \w) &=& \frac{1}{N_{c}} \Braket{\text{Tr}[\U(\w')\U^{\dagger}(\mathbf{v}')\U(\mathbf{v})\U^{\dagger}(\w)]},
 \label{eik_quad}
\end{eqnarray}
with the Wilson line $\U(\v)$ in the fundamental representation is given as the path-ordered exponential of the gluon background field $A^-$ of the target as\footnote{\label{FN_2}In the eikonal approximation, the limits of the $z^+$ integration in the Wilson line are usually taken to be from $-\infty$ to $+\infty$.}  
\begin{align}
\label{Standard_Wilson_line}
\U_{[x^{+},y^+]}(\z) = \mathcal{P}_{+} \text{exp}\bigg\{ -ig \int^{x^+}_{y^{+}} dz^{+} A^{-}(z^{+}, \z ) \bigg\}.
\end{align}
%
%
The functions $\C_{\lambda}$ in Eq. \eqref{Eik_dijet_Xsec} is defined as  
\begin{align}
\label{CL}
   & \C_{L}(\r_{1},\r_{2}) = \sum_{f} \frac{8 N_{c}\alpha_{\text{em}}e_{f}^{2}Q^{2}}{(2\pi)^{6}} \delta_{z} \, z_{1}^{3} z_{2}^{3} \, K_{0}(\epsilon_{f}|\r_{1}|) K_{0}(\epsilon_{f}|\r_{2}|),\\
  &  \C_{T}(\r_{1},\r_{2}) = \sum_{f} \frac{2 N_{c}\alpha_{\text{em}}e_{f}^{2}}{(2\pi)^{6}} \delta_{z} \, z_{1} z_{2} \, \big[m_{f}^{2} + (z_{1}^{2}+z_{2}^{2})\partial_{\r_{1}^{j}} \partial_{\r_{2}^{j}} \big] \, K_{0}(\epsilon_{f}|\r_{1}|) K_{0}(\epsilon_{f}|\r_{2}|),
\label{CT}
\end{align}
for incoming longitudinal and transverse photons, respectively.  
Here, we introduced the shorthand notations $\delta_{z} = \delta(z_{1} + z_{2}-1)$ and $\epsilon_{f}^{2} = m_{f}^{2} + z_{1} z_{2} Q^{2}$, with $m_{f}$ being the mass and $e_{f}$ being the fractional charge of the quark with flavor $f$. The longitudinal momentum fractions of the photon carried by the quark (and the antiquark) are defined as $z_{1,2} = k_{1,2}^{+}/q^{+}$, where $q^+$ is the longitudinal momenta of the incoming photon and $k_1^+$ and $k_2^+$ are the longitudinal momenta of the produced quark and antiquark jets, respectively. Finally, $K_0(\cdots)$ is the modified Bessel function of the second type. 
%
%
%


The NEik corrections to the DIS dijet cross section have been computed in~\cite{Altinoluk:2022jkk} in the case of a purely gluonic background field. Restricting ourselves for simplicity to the NEik corrections preserving the static approximation, the results from Ref.~\cite{Altinoluk:2022jkk} can be written as 
\begin{align}
\label{eq: Neik cross section}
   & \frac{d\sigma^{\gamma^{*}_{\lambda}+ A \rightarrow q \bar q + X}}{d^{2}\k_{1}d^{2}\k_{2}d\eta_{1}d\eta_{2}} \bigg |_{\text{NEik}}  = \frac{1}{q^{+}} \text{Re} \int_{\v, \v', \w, \w'} e^{i\k_{1} \cdot (\mathbf{v'}-\mathbf{v}) + i \k_{2} \cdot (\w' - \w)} \C_{\lambda}(\w'-\v', \w - \v) \\ \nn
   & \times \Bigg\{ \frac{1}{z_{1}} \bigg[ \frac{\k_{2}^{j} - \k_{1}^{j}}{2} + \frac{i}{2} \partial_{\w^{j}}\bigg] \big[Q_{j}^{(1)}(\w', \v', \v_{*}, \w) - d_{j}^{(1)}(\v_{*} , \w) \big] -\frac{i}{z_{1}} \big[Q^{(2)}(\w', \v', \mathbf{v}_{*}, \w) - d^{(2)}(\v_{*} , \w)\big] \\ \nn
   & -\frac{1}{z_{2}} \bigg[ \frac{\k_{2}^{j} - \k_{1}^{j}}{2} - \frac{i}{2} \partial_{\v^{j}}\bigg] \big[Q_{j}^{(1)}(\v', \w', \w_{*}, \v)^{\dagger} - d_{j}^{(1)}(\w_{*} , \v)^{\dagger} \big] -\frac{i}{z_{2}} \big[Q^{(2)}(\v', \w', \mathbf{w}_{*}, \v)^{\dagger} - d^{(2)}(\w_{*} , \v)^{\dagger}\big] \Bigg\} \\ \nn
   & + \delta^{\lambda T} \frac{d\sigma^{\text{trans.}}}{d^{2}\k_{1}d^{2}\k_{2}d\eta_{1}d\eta_{2}},
\end{align}
where $\delta^{\lambda T}$ is 1 when $\lambda = T$ and 0 when $\lambda = L$. The part of the cross section that contributes only to the case when the incoming photon is  transversely polarized is given by
\begin{align}
\label{NEik_X_Section_T}
    \frac{d\sigma^{\text{trans.}}}{d^{2}\k_{1}d^{2}\k_{2}d\eta_{1}d\eta_{2}} &= \frac{1}{q^{+}} \text{Re} \int_{\v, \v', \w, \w'} e^{i\k_{1} \cdot (\mathbf{v'}-\mathbf{v}) + i \k_{2} \cdot (\w' - \w)} \mathcal{D}^{ij}_{T,1}(\w'-\v', \w - \v) \\ \nn
   & \times \Bigg[ \frac{1}{z_{1}} \big[Q_{ij}^{(3)}(\w', \v', \v_{*}, \w) - d_{ij}^{(3)}(\v_{*} , \w) \big] + \frac{1}{z_{2}} \big[Q_{ij}^{(3)}(\v', \w', \mathbf{w}_{*}, \v)^{\dagger} - d_{ij}^{(3)}(\w_{*} , \v)^{\dagger}\big] \Bigg] \\ \nn
   & - \frac{1}{q^{+}} \text{Im} \int_{\z, \v', \w'} e^{i\k_{1} \cdot (\mathbf{v'}-\z) + i \k_{2} \cdot (\w' - \z)} \mathcal{D}^{j}_{T,2}(\w' -\v') \\ \nn
   &\times \int_{-\frac{L^{+}}{2}}^{\frac{L^{+}}{2}} dz^{+} \Braket{\frac{1}{N_{c}} \text{Tr} \big[\U(\w') \U^{\dagger}(\v') -1\big] \big[ \U_{\big[ \frac{L^{+}}{2}, z^{+}\big]} (\z) \overleftrightarrow{D_{\z^{j}} }(z^{+}) \U^{\dagger}_{\big[ \frac{L^{+}}{2}, z^{+}\big]}(\z) - \frac{1}{2} \U(\z) \overleftrightarrow{\partial_{\z^{j}}} \U^{\dagger}(\z)
   \big] },
\end{align}
where functions $\mathcal{D}^{ij}_{T,1}$ and $\mathcal{D}^{j}_{T,2}$ are defined as 
\begin{align}
     & \mathcal{D}_{T,1}^{ij}(\r_{1},\r_{2}) = \sum_{f} \frac{2 N_{c}\alpha_{\text{em}}e_{f}^{2}}{(2\pi)^{6}} \delta_{z} z_{1} z_{2}(z_1 -z_2)\partial_{\r_{1}^{i}} \partial_{\r_{2}^{j}} K_{0}(\epsilon_{f}|\r_{1}|) K_{0}(\epsilon_{f}|\r_{2}|),\\
  &  \mathcal{D}_{T,2}^{j}(\r) = {-} \sum_{f} \frac{N_{c}\alpha_{\text{em}}e_{f}^{2}}{2(2\pi)^{5}} \delta_{z} [1 + (z_{2} - z_{1})^{2}] \partial_{\r^{j}}  K_{0}(\epsilon_{f}|\r|) \, .
  \label{DT2}
\end{align}
As discussed in detail in the literature (see~\cite{Altinoluk:2022jkk, Agostini:2024xqs}), apart from the standard dipole and quadrupole operators that appear in the DIS dijet cross section at eikonal order, given in  Eqs.~\eqref{eik_dip} and \eqref{eik_quad}, respectively, at NEik order the DIS dijet cross section also involves the so-called {\it decorated dipole} and {\it decorated quadrupole} operators. These operators are the generalization of the standard eikonal operators and read 
%
%
%
\begin{equation}
\label{eq: dj1 and Qj1}
d^{(1)}_{j}(\v_{*} , \w) = \frac{1}{N_{c}} \Braket{ \text{Tr} [\U_{j}^{(1)}( \v) \U^{\dagger}(\w)]}, \hspace{0.5cm} Q_{j}^{(1)}(\w', \v', \v_{*}, \w) = \frac{1}{N_{c}} \Braket{\text{Tr}[\U(\w')\U^{\dagger}(\v')\U_{j}^{(1)}(\mathbf{v})\U^{\dagger}(\w)]},
\end{equation}
\begin{equation}
\label{eq: d2 and Q2}
d^{(2)}(\v_{*} , \w) = \frac{1}{N_{c}} \Braket{ \text{Tr} [\U^{(2)}( \v) \U^{\dagger}(\w)]},\hspace{0.5cm} Q^{(2)}(\w', \v', \mathbf{v}_{*}, \w) = \frac{1}{N_{c}} \Braket{\text{Tr}[\U(\w')\U^{\dagger}(\v')\U^{(2)}(\v)\U^{\dagger}(\w)]},
\end{equation}
\begin{equation}
\label{eq: dij3 and Qij3}
d^{(3)}_{ij}(\v_{*} , \w) = \frac{1}{N_{c}} \Braket{ \text{Tr} [\U_{ij}^{(3)}( \mathbf{v}) \U^{\dagger}(\w)]}, \hspace{0.5cm} Q_{ij}^{(3)}(\w', \mathbf{v'}, \v_{*}, \w) = \frac{1}{N_{c}} \Braket{\text{Tr}[\U(\w')\U^{\dagger}(\v')\U_{ij}^{(3)}(\v)\U^{\dagger}(\w)]},
\end{equation}
where one of the Wilson lines inside the trace has non-eikonal insertions along its longitudinal extend. The star in the transverse coordinate arguments of the decorated dipole and quadrupole operators indicate the position of the corresponding insertion. The three types of such decorated Wilson lines are defined as 
\begin{align}
    %
 &\U_{j}^{(1)}(\z) = \int_{-\frac{L^+}{2}}^{\frac{L^+}{2}} dv^{+}  \U_{[\frac{L^+}{2},v^+]}(\z) \overleftrightarrow{D_{\z^{j}}}(v^+)  
 \U_{[v^+, -\frac{L^+}{2}]}(\z),\\
 \label{eq: def U2}
&\U^{(2)}(\z) = \int_{-\frac{L^+}{2}}^{\frac{L^+}{2}} dv^{+}  \U_{[\frac{L^+}{2},v^+]}(\z) \overleftarrow{D_{\z^{j}}} (v^+) \overrightarrow{D_{\z^{j}}}(v^+)
 \U_{[v^+, -\frac{L^+}{2}]}(\z),\\
 &\U_{ij}^{(3)}(\z) = \int_{-\frac{L^+}{2}}^{\frac{L^+}{2}} dv^{+}  \U_{[\frac{L^+}{2},v^+]}(\z) g F_{ij}(v^{+},\z)  
 \U_{[v^+, -\frac{L^+}{2}]}(\z), 
 \end{align}
with the standard Wilson line defined in Eq.~\eqref{Standard_Wilson_line}. Moreover, the Wilson lines defined without specifying the two endpoints are evaluated according to
\begin{equation}
    \U(\x) \equiv \U_{[\frac{L^+}{2},-\frac{L^+}{2}]}(\x),
\end{equation}
where $L^+$ is the length of the support for the background field under consideration. Due to the fact that the background field is a pure gauge outside this support, the replacement $L^+ \rightarrow \infty$ can safely be taken in the final results (see~\cite{Altinoluk:2024zom} for a detailed discussion). The background covariant derivatives and the field strength insertions in the decorated Wilson lines are defined with the following conventions:
 \begin{align}
 \label{eq: def covariant derivatives}
&\overrightarrow{D_{\z^{j}}} (z^{+}) = \overrightarrow{\partial_{\z^{j}}} + ig A_{j}(z^{+},\z)
= \overrightarrow{\partial_{\z^{j}}} - ig A^{j}(z^{+},\z), \hspace{0.5cm}
\overleftarrow{D_{\z^{j}}} (z^{+}) = \overleftarrow{\partial_{\z^{j}}} + ig A^{j}(z^{+},\z), \\
&\overleftrightarrow{D_{\z^{j}}}  = 
\overrightarrow{\partial_{\z^{j}}}  - \overleftarrow{\partial_{\z^{j}}} (z^{+}) - 2igA^{j}(z^+, \z),\\
& F_{ij}(v^{+},\z) = \partial_{\z^{i}} A_{j}(z^+, \z) - \partial_{\z^{j}}A_{i}(z^+, \z) + ig[A_{i}(z^+, \z), A_{j}(z^+, \z)].
\label{F_ij}
 \end{align}

As mentioned earlier, numerical studies of DIS dijet production at eikonal accuracy are commonly performed using the MV model~\cite{McLerran:1993ka,McLerran:1994vd} and the GBW model~\cite{Golec-Biernat:1998zce,Golec-Biernat:1999qor}  to describe the dipole and quadrupole operators. However, extending such analyses to NEik accuracy requires going beyond these models. In the next section, we introduce a Gaussian model for the eikonal dipole and quadrupole operators, which will then be generalized to the NEik operators in Sec.~\ref{sec:NEik_operators_within_GM}. The primary goal of this work is to provide explicit  expressions for these decorated operators in a form suitable for future numerical studies of observables at NEik accuracy. 

%
%

\section{Wilson line operators at eikonal order within a Gaussian model}
\label{sec:Eik_operators_within_GM}
In this section we introduce, following the ideas in~\cite{Gelis:2001da,Agostini:2024xqs}, the Gaussian model, closely inspired in the MV model, for the Wilson line operators at eikonal order which will  be used later to derive the expressions of NEik operators such as decorated dipoles and decorated quadrupoles given in Eqs.~\eqref{eq: dj1 and Qj1},~\eqref{eq: d2 and Q2} and~\eqref{eq: dij3 and Qij3}. In the rest of the manuscript, $\langle \hat{\mathcal{O}} \rangle$ will denote\footnote{The same notation is used for denoting the CGC averaging of operators in Sec.~\ref{sec:DIS_dijet_prod}. However, in order not to overload the notation, hereafter $\langle \cdots\rangle$ will be used only for denoting the Gaussian averaging.} the Gaussian average of an operator $\hat{\mathcal{O}}$. 


\subsection{Monopole operator}
%
%
We start our analysis with the derivation of a closed-form expression for the averaging of the monopole operator $\Braket{\U_{[x^+, y^+]}(\x)}$ within the Gaussian
model. This derivation was originally performed in~\cite{Gelis:2001da} where the details can be found. Here, we just present the main steps of the derivation. 

In order to compute the Gaussian average of this monopole operator, we first represent the Wilson line as\footnote{ The short hand notation introduced in Eq. \eqref{Wilson_Line_exp} for the longitudinal integration corresponds to $ \int_{z_{1}^{+}>....>z_{n}^{+}}\equiv \int dz_1^+ dz_2^+ \dots dz^+_{n-1}dz^+_{n} \, \theta(z_1^+-z_2^+)\dots \theta(z^+_{n-1}-z_n^+)$.}  
%
\begin{align}
\label{Wilson_Line_exp}
    \U_{[x^{+},y^+]}(\z) = \sum_{n=0}^{\infty} (-ig)^{n} \int_{z_{1}^{+}>....>z_{n}^{+}}  A^{-}(z_{1}^{+}, \z )....... A^{-}(z_{n}^{+}, \z ) ,
\end{align}
where $A^{-}(z^{+}, \z) = A^{-}_
{a} (z^{+}, \z)T^{a}$, with $T^{a}$ the generators of the fundamental representation of $SU (N_c )$. For large
nuclei, using the central limit theorem, the distribution of the color sources can be approximated by a Gaussian distribution~\cite{Gelis:2001da}.
In~\cite{Agostini:2024xqs} it was shown that, working along the lines in~\cite{Dumitru:2018vpr,Cougoulic:2020tbc} and using the Wick theorem, the average of multiple gauge fields can be expressed in terms of the following two-point functions:
%
%
\begin{align}\label{eq:2pts_ev}
    \Braket{A^{\mu}_{a}(x^+,\x) A_{b}^{\nu}(y^+,\y)} = \delta^{ab} \delta(x^{+}-y^{+}) \mu^2(x^+) G^{\mu\nu}(\x - \y),
\end{align}
where the relevant components of $G^{\mu\nu}$ (up to NNEik accuracy) read
\begin{align}
\label{def:g--}
    &G^{--}(\r) = \int_{\P} e^{i\P \cdot \r} \frac{1}{\P^{4}},\\%
    \label{def:gi-}
    &G^{i-}(\r) = \frac{1}{2P_{q}^{-}}\int_{\P} e^{i\P \cdot \r}\frac{\P^{i}}{\P^{4}},\\
    \label{eq: def Gij}
    &G^{ij}(\r) = \frac{1}{(2P^{-}_{q})^{2}} \int_{\P} e^{i\P \cdot \r} \frac{\P^{i} \P^{j} + \epsilon^{im} \epsilon^{jn} \P^{m}\P^{n}}{\P^{4}}, 
\end{align}
with $\r\equiv \x-\y$. As discussed in detail in~\cite{Agostini:2024xqs}, while $G^{--}(\r)$ can contribute to the two-point correlator at eikonal order, $G^{i-}(\r)$ and $G^{ij}(\r)$ start contributing at NEik accuracy.  In Eq. \eqref{eq:2pts_ev}, $\mu^2(x^+)$ represents the transverse color charge density of the nucleus. It is more convenient 
to define $\tilde{\mu}^2$ as the color density integrated over the longitudinal nuclear profile, and its limit for a homogeneous profile $\mu(v^+) = \mu$ as 
\begin{equation}\label{eq:mu_tilde_def}
    \tilde{\mu}^2 \equiv \int\limits_{-\frac{L^+}{2}}^{\frac{L^+}{2}} dv^+ \mu^2(v^+) \longrightarrow \mu^2 \int\limits_{-\frac{L^+}{2}}^{\frac{L^+}{2}} dv^+ = \mu^2 L^+ \, .
\end{equation}

\begin{figure}[!h]
    \centering
    \includegraphics[width=.3\textwidth]{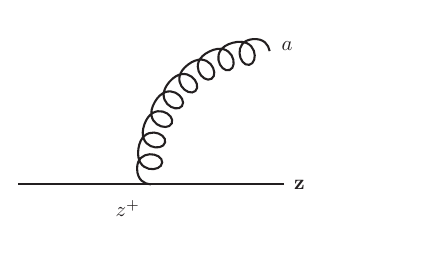}
    \caption{Diagrammatic representation of $T_a$ contracted with $A^{-}_{a}(z^+, \z)$.}
    \label{fig:T_a}
\end{figure}

\begin{figure}[!h]
    \centering
    \includegraphics[width=.2\textwidth]{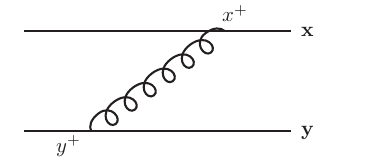}
    \caption{Diagrammatic representation of 2-point correlator.}
    \label{fig:two point correlator}
\end{figure}

The color structure of the monopole operator can be most conveniently obtained by introducing the following diagrammatic notation. 
First, we write the generator $T_a$ contracted with $A^{-}_{a} (z^{+}, \z)$ as shown in Fig.~\ref{fig:T_a}.
%
%
Next, we express the two-point correlator as a \textit{kinetic} term times a color structure, as given in Eq.~\eqref{eq:2pts_ev} and represented diagrammatically in Fig.~\ref{fig:two point correlator}. 
%
%
%
%

\begin{figure}[!h]
    \centering
    \includegraphics[width=.4\textwidth]{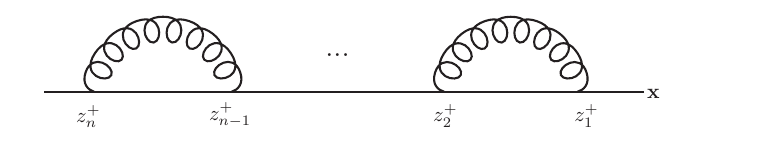}
    \caption{A \textit{tadpole contribution} that is resummed into the expectation value of the  operator $\langle \U_{[x^+,y^+]} \rangle$.}
    \label{fig:multiple sum monopole}
\end{figure}

Using this approach, since the transverse position of all the gauge fields in a monopole operator is the same, it can be cast into contributions as represented in Fig. \ref{fig:multiple sum monopole}. 
%
%

The fact that the two-point function is local in longitudinal coordinate $z^+$ dictates the Gaussian average to be local as well in $z^+$. Moreover, the domain of the $z^+$ integral is constrained to $z_1^+ > ... > z_n^+$. These two conditions ensure that the diagrams with nested or overlapping loops vanish due to the lack of support along the longitudinal direction. Therefore, for any integer $k$ the longitudinal coordinates obey $z_{2k}^{+} = z_{2k-1}^{+}$ and thus all the contributions to the average of the monopole operator are tadpoles. These contributions (shown in Fig. \ref{fig:multiple sum monopole}) to the average of the monopole operator can be solved straightforwardly by  using $T^a T_a = C_F \mathbb{1}_F$ and
%
%
%
%
%
\begin{align}
    \int_{y^{+}}^{z_{1}^{+}} dz_{2}^{+} \ \delta(z_{2}^{+}-z_{1}^{+}) = \frac{1}{2}\ ,
\end{align}
yielding for the Gaussian average of the monopole operator
\begin{align}
    \Braket{\U_{[x^+, y^+]}(\x)_{\alpha\beta}} &= \sum_{k=0}^{\infty} (-g^{2})^{k} \int_{z_{1}^{+}>.....>z_{2k}^{+}} \Braket{A^{-}(z_{1}^{+}, \x) A^{-}(z_{2}^{+}, \x)}...\Braket{A^{-}(z_{2k-1}^{+}, \x) A^{-}(z_{2k}^{+}, \x)}\nn\\
    &= \sum_{k=0}^{\infty}\bigg(\frac{-g^{2}}{2}\bigg)^{k} \frac{1}{k!} \, \mathbb{1}_{\alpha\beta}\, \big[ \mu^{2}(z^{+})\ C_{F} \ G^{--}(0)\big]^{k},
    \label{GA_monopole_with_mu}
\end{align}
with $1/k!$ accounting for the ordering of the longitudinal integrals. The expression for the Gaussian average of the monopole operator can be written in a more compact form by introducing the saturation momentum $Q_s$ (see for example~\cite{Agostini:2024xqs}) as
%
\begin{align}
\label{eq: def Qs}
    Q_{s}^{2} &\equiv \frac{g^{2}}{2\pi}C_{F}\int\limits_{-L^{+}/2}^{L^{+}/2} dz^{+} \mu^{2}(z^{+}) \simeq \frac{g^{2}}{2\pi} \, C_{F} \, \mu^2 \int\limits_{-L^{+}/2}^{L^{+}/2} dz^{+}\, ,
\end{align}
where the limit of homogeneous nuclear profile is assumed to write the second equality.   Using Eq.~\eqref{eq: def Qs} in Eq.~\eqref{GA_monopole_with_mu}, one arrives at the final closed-form expression for the average of the monopole operator in the Gaussian model:
%
\begin{align}
    \Braket{\U(\x)} = e^{-\pi Q_{s}^{2}G^{--}(0)}.
\end{align}
Although this section is devoted to the analysis of the Wilson line operators at eikonal order, we would like to comment on the Gaussian average of the monopole operator at NEik order before we proceed to our discussion for dipole operator. 
In the NEik case, the monopole vanishes since $G^{j-}(0) = 2i P_{q}^{-} \partial_{j}G^{--}(0) = 0$ and, as we will see later,  $G^{ij} (0)$,
that appears when we have two transverse field insertions, factorizes trivially.
%
%
\subsection{Dipole operator}
%
%
We can now continue our study with the dipole operator given in Eq.~\eqref{eik_dip}. The expression for the dipole operator within the Gaussian model and its detailed derivation can again be found in~\cite{Gelis:2001da}. As in the case of monopole operator presented in the previous subsection, here we show the main steps of the derivation as well as the final result for the expression at eikonal order.  

%

\begin{figure}[!h]
    \centering
    \includegraphics[width=.5\textwidth]{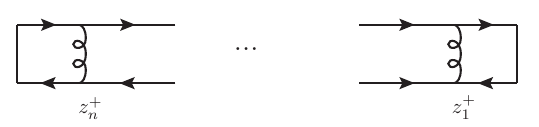}
    \caption{Diagram showing the interaction between quark or antiquark lines in the dipole.}
    \label{fig:nontadpole dipoles}
\end{figure}

\begin{figure}[!h]
    \centering
    \includegraphics[width=.5\textwidth]{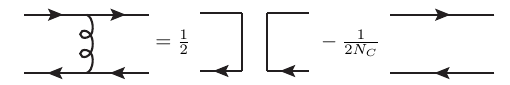}
    \caption{Diagrammatic representation of the Fierz identity.}
    \label{fig:Fierz identity}
\end{figure}

The derivation of the dipole operator closely follows the one performed for the monopole operator at eikonal order. Namely, we first expand the Wilson lines in terms of the gauge field. Then, we write the resulting correlators in terms of two-point functions given in Eq.~\eqref{eq:2pts_ev}. Under these conditions, the dipole operator can be factorized into a tadpole contribution and an interaction part as   
%
%
\begin{align}
\label{dip_fact}
    d(\x,\y) = \T_d\, \N_d,
\end{align}
with $\T_d$ the tadpole contribution to the dipole operator  is given by
%
\begin{align}
\label{dip_tad_pole}
    \T_d = \Braket{\U(\x)} \Braket{\U(\y)} = e^{-2\pi Q_{s}^{2}G^{--}(0)}, 
\end{align}
and $\N_d$ the non-tadpole contribution to the dipole operator that accounts for the interaction between both (quark or antiquark) lines. Diagrammatically, this part can be represented as  shown in Fig.~\ref{fig:nontadpole dipoles}.
Its explicit expression reads 
%
%
\begin{align}
    \N_d= \frac{1}{N_{c}} \text{Tr}
    \sum_{n=0}^{\infty} \int_{z_{1}^{+}>....>z_{n}^{+}} \prod_{i=1}^{n} [(T^a_\x \otimes T^a_\y)g^{2}\mu^{2}(z_{i}^{+})G^{--}(\x - \y)] .
\end{align}
%
The color structure can be simplified by using the $SU(N_c )$ Fierz identity
\begin{align}
    T^{a}_{\alpha_1 \alpha_2} T^{a}_{\beta_{2}\beta_{1}} = \frac{1}{2} \delta_{\alpha_{1}\beta_{1}} \delta_{\alpha_{2}\beta_{2}} - \frac{1}{2N_{c}} \delta_{\alpha_{1}\alpha_{2}}\delta_{\beta_{1}\beta_{2}}, 
\end{align}
\noindent
which can be represented in diagrammatic form according to Fig.~\ref{fig:Fierz identity}. 

\begin{figure}[!h]
    \centering
    \includegraphics[width=.5\textwidth]{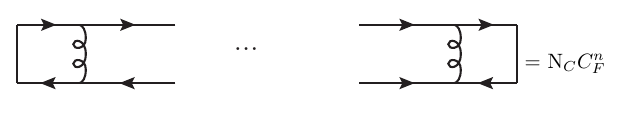}
    \caption{Birdtrack representation of $n$ gluons exchanged between a quark and an antiquark which contribute to the dipole $d(\x,\y)$.}
    \label{fig:C_F_dipoles}
\end{figure}

By iterating the Fierz identity to recast all intermediate gluons, we can write (using the birdtrack approach~\cite{Peigne:2024srm}) the contribution $\mathcal{N}_d$ to the dipoles as shown in Fig.~\ref{fig:C_F_dipoles}.
%

Finally, the non-tadpole ({\it interaction}) factor $\mathcal{N}_d$ in the dipole is given by
\begin{align}
\label{dip_non_tadpole}
    \N_d =  \sum_{n=0}^{\infty} \frac{1}{n!} \bigg[ C_{F} g^{2} \int \! dz^{+} \mu^{2}(z^{+})G^{--}(\x - \y) \bigg]^{n} = e^{2\pi Q_{s}^{2}G^{--}(\x -\y)}.
\end{align}
%
%
%

All in all, substituting the tadpole contribution given in Eq.~\eqref{dip_tad_pole} and the interaction part given in Eq.~\eqref{dip_non_tadpole} into the dipole function given in Eq.~\eqref{dip_fact}, the dipole operator within Gaussian model reads
\begin{align}
\label{def:dip_Gaussian_model}
     d(\x,\y) = \mathcal{T}_d\, \mathcal{N}_d = \text{exp}\bigg\{-2\pi Q_{s}^{2} [G^{--}(0) - G^{--}(\x-\y)]\bigg\}\, .
\end{align}
%
The relation to the usual expression of the dipole amplitude in the MV-model, by using the the definition of $G^{--}(\r)$ given in Eq. \eqref{def:g--}, follows from
\begin{align}\label{eq:eval_MV}
    G^{--}(0) - G^{--}(\x-\y) = \int \! \frac{d^{2}\P}{(2\pi)^{2}} \frac{1- e^{i\P \cdot \r}}{\P^{4}} = \frac{1}{2\pi} \int \! \frac{dP_{\perp}}{P^{3}_{\perp}} (1- J_{0}(P_{\perp}r_{\perp})) \approx \frac{1}{2\pi} \frac{r^{2}_{\perp}}{4} \ln \frac{1}{r_{\perp}\,\Lambda_{\text{QCD}}}
\end{align}
where in the last step we have used the small dipole approximation, $r_{\perp} P_{\perp} \ll 1$, and regulated the infrared divergences
by introducing $\Lambda_{\text{QCD}}$.
Thus, the dipole function becomes
%
\begin{align}
\label{eq: def dipole}
    d(\x,\y) = \text{exp}\bigg\{ - \frac{Q_{s}^{2}}{4} (\x - \y)^{2} \text{ln}\frac{1}{|\x -\y|\Lam} \bigg\}
\end{align}
as known in the MV model.

\subsection{Quadrupole operator}
\label{sec:Eik_quadrupole}

\begin{figure}
\begin{minipage}{.30\textwidth}
  \centering
  \includegraphics[width=1\linewidth]{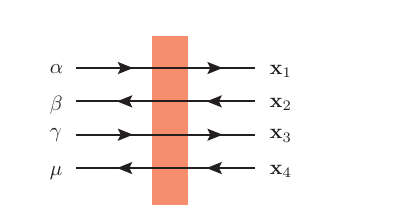}
  \caption{Diagrammatic representation of a shockwave between $y^+$ and $x^+$.}
    \label{fig:quadrupole}
\end{minipage}%
\hfill
\begin{minipage}{.30\textwidth}
  \centering
  \includegraphics[width=.8\linewidth]{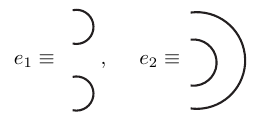}
  \caption{Diagrammatic representation of the basis elements $\langle e_1|$ and $\langle e_2|$.}
    \label{fig:e_1 and e_2}
\end{minipage}
\hfill
\begin{minipage}{.30\textwidth}
  \centering
  \includegraphics[width=.8\linewidth]{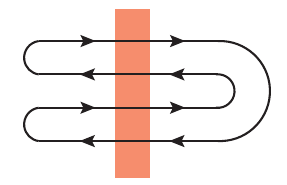}
  \caption{Diagrammatic representation of the quadrupole by closing color the trace through the shockwave.}
    \label{fig:closed quadrupole}
\end{minipage}
\end{figure}

The quadrupole operator is the next Wilson line structure that needs to be computed in the Gaussian model at eikonal order. Its definition is given in Eq.~\eqref{eik_quad}, where the longitudinal limits of the fundamental Wilson lines are taken from $-\infty$ to $+\infty$. Here, for the sake of the convenience of the derivation, we keep these limits explicit and also work with the operator $\mathcal{Q}_{[x^+,y^+]}$ which is composed of four fundamental Wilson lines with open color indices:
\begin{align}\label{eq:calQ_def}
   \mathcal{Q}_{[x^{+}, y^{+}]} =  \Braket{\U_{[x^{+}, y^{+}]}(\x_1)_{\alpha_1 \alpha_2} \U^{\dagger}_{[x^{+}, y^{+}]}(\x_2)_{\beta_1 \beta_2} \U_{[x^{+}, y^{+}]}(\x_3)_{\gamma_1 \gamma_2} \U^{\dagger}_{[x^{+}, y^{+}]}(\x_4)_{\mu_1 \mu_2} },
\end{align}   
which can be represented as in Fig.~\ref{fig:quadrupole} by using the birdtrack approach. In order to obtain various quadrupole operators from Eq.~\eqref{eq:calQ_def}, one should perform the color trace of the operator $\mathcal{Q}_{[x^+,y^+]}$. For the purpose of recovering any possible trace of this operator, we define a two-dimensional basis $\{|e_i\rangle\}$ in color space by 
\begin{align}
\label{def:basis_ei}
    \ket{e_1} \equiv \delta_{\alpha \beta} \delta_{\gamma \mu} \, 
     , \hspace{0.5cm} \ket{e_2}  \equiv \delta_{\alpha\mu}\delta_{\beta\gamma} \, ,
  \end{align}
which can be represented as in Fig.~\ref{fig:e_1 and e_2} by using the birdtrack approach. Upon adopting these notations, the quadrupole operator defined in Eq.~\eqref{eik_quad} is given by  
\begin{align}
    Q_{[x^{+}, y^{+}]}(\x_1, \x_2, \x_3, \x_4) = \frac{1}{N_c} \bra{e_2}  \mathcal{Q}_{[x^{+}, y^{+}]}\ket{e_1},
\end{align}
with $\w'=\x_1, \v'=\x_2,\v=\x_3$ and $\w=\x_4$ in Eq. \eqref{eik_quad}. Other possible operators, obtained from $\langle e_i | \mathcal{Q} | e_j \rangle $, are evaluated in Appendix~\ref{app:Compare_to_literature}.

\begin{figure}[!h]
    \centering
    \includegraphics[width=.3\textwidth]{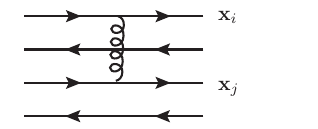}
    \caption{Diagrammatic representation of a two-gluon exchange at longitudinal point $x^{+}$ in the quadrupole between the transverse coordinates $\x_i$ and $\x_j$.}
    \label{fig:gluon exchange in quadrupole}
\end{figure}

As for the dipole operator, the interaction between two lines should be accounted for. For this purpose, we define 
all possible combinations where we can have two-gluon exchanges at a longitudinal point $x^+$ as follows:
\begin{align}
\label{eq: Sigma_eik}
\Sigma(x^{+}) = \mu^{2}(x^+) \sum_{i=1,j>i}^{4}W_{ij} \ ,
\end{align}
where $\mu^2(x^+)$ represents the transverse color charge density of the nucleus at longitudinal position $x^+$ and $W_{ij}$ is the kinetic factor given by
\begin{align}\label{eq:Wij_def}
    W_{ij} = (-1)^{i+j+1} g^{2}G^{--}(\x_{i}- \x_{j})\, . 
\end{align}
The factor $(-1)^{i+j+1}$ in the kinematic factor given in Eq.~\eqref{eq:Wij_def} is introduced to account for the change in sign when a quark (or antiquark) line is connected with another quark line. The diagrammatic representation of the interaction between two quark lines at transverse coordinates $\x_i$ and $\x_j$ via two-gluon exchanges at a longitudinal position $x^+$ is shown in Fig.~\ref{fig:gluon exchange in quadrupole}.

In order to iterate the previously defined interaction $\Sigma(v^+)$, we introduce the identity $\mathcal{I}$ of this color vector space as the linear map $V \otimes \bar{V} \otimes V \otimes \bar{V} \rightarrow V \otimes \bar{V} \otimes V \otimes \bar{V}$, where $V=\mathbb{C}^{N_c}$. In terms of Kronecker delta's, it reads
\begin{equation}
    \mathcal{I} = \delta_{\alpha_1 \alpha_2} \delta_{\beta_1 \beta_2} \delta_{\gamma_1\gamma_2} \delta_{\mu_1 \mu_2}\ ,
\end{equation}
where the labels follow the notation introduced in Fig.~\ref{fig:quadrupole}.
However, in the evaluation of $\mathcal{Q}_{[x^+,y^+]}$, only the gauge invariant subspace of this large vector space is of interest. Thus, only the singlet subspace of $\mathcal{I}\big|_{singlet}$ contributes and
it can be written using the basis $\{|e\rangle\}$ which (see Appendix~\ref{App:Color_Identity} for the details of the calculation) reads
%
%
\begin{align}
\label{eq: def c and identity}
\mathcal{I}\big|_{singlet} = 
    \mathbb{1} 
    = c_{ij}\ket{e_i}\bra{e_j}, \hspace{1cm} c= \frac{1}{N_{c}^{2}-1} \begin{pmatrix}
        1 & -\frac{1}{N_{c}}\\
        -\frac{1}{N_{c}} & 1 \\
    \end{pmatrix}\, .
\end{align}
Applying the identity $\mathbb{1}$ to both sides of Eq.~\eqref{eq: Sigma_eik}, we can write interaction term $\Sigma(x^{+})$ as a matrix in the $\ket{e_{i}}$ basis as 
\begin{align}
\label{eq: sigma in V eik}
    \Sigma(x^{+}) = \ket{e_i}c_{ij}\bra{e_j}\Sigma(x^{+})\ket{e_s}c_{st}\bra{e_t} \equiv \mu^{2}(x^{+}) \ket{e_i}V_{it}\bra{e_t},
\end{align}
where the matrix $V_{it} = c_{ij} \bra{e_j} \Sigma(x^+)\ket{e_t}$
can be computed by inserting the Fierz identity and only selecting the singlet component (see Appendix~\ref{App:Color_Interaction} for the details of the calculation). The result reads
%
\begin{align}\label{eq:int_V}
 V = \frac{1}{N_{c}^{2}-1} \begin{pmatrix}
     \frac{2\beta - \gamma + \alpha (N_{c}^{2}-2)}{2 N_{c}} & \frac{\alpha + \gamma - \beta (N_{c}^{2}+1)}{2 N_{c}^{2}} \\
     \frac{\alpha + \gamma - \beta(N_{c}^{2} + 1)}{2 N_{c}^{2}} & \frac{-\alpha + 2 \beta + \gamma (N_{c}^{2}-2)}{2 N_{c}}
 \end{pmatrix}   ,
\end{align}
with 
\begin{align}
\label{def:abg}
\alpha = W_{12} + W_{34}, \hspace{1cm} \beta = W_{13} + W_{24}, \hspace{1cm}  \gamma = W_{14} + W_{23}
\end{align}
and the kinetic factor $W_{ij}$ defined in Eq.~\eqref{eq:Wij_def}.
Up to now, we obtained all the required building blocks to compute the quadrupole operator systematically within the eikonal approximation. Similar to the case for the dipole operator, the quadrupole operator can also be factorized into tadpole and non-tadpole contributions as 
\begin{align}\label{eq:quad_T_N}
    Q_{[x^{+}, y^{+}]} = \T_Q\,  \N_Q\ ,
\end{align}
with $\T_Q$ being the tadpole contribution to the quadrupole operator which reads 
\begin{align}
\label{eq: def tadpole quadrupole}
    \T_Q = e^{-4\pi Q_{s}^{2}G^{--}(0)}.
\end{align}
On the other hand, the non-tadpole contribution to the quadrupole operator $\N_Q$ can be written as 
%
\begin{align}
\label{non_tadpole_quad_1}
    \N_Q =\frac{1}{N_c} \sum_{n=0}^{\infty}\bra{e_{2}}\int_{z_{1}^{+}>...>z_{n}^{+}} \Sigma(z^{+}_{1})...\Sigma(z_{n}^{+})\ket{e_{1}}.
\end{align}
Substituting the expression for the interaction matrix given in  Eq.~\eqref{eq: sigma in V eik} into non-tadpole contribution to the quadrupole operator given in Eq~ \eqref{non_tadpole_quad_1}, we get
\begin{align}
\label{non_tadpole_quad_2}
    \mathcal{N}_Q = 1 + \frac{1}{N_c} \langle e_2 |\left(\sum_{n=1}^\infty \int_{z_1^+> \cdots > z_n^+} \prod_{i=1}^n \mu^2(z_i^+)\ 
    |e_{k_i}\rangle V_{k_i\, \ell_i}\langle e_{\ell_i}| \right) |e_1 \rangle ,
\end{align}
with the matrix $V$ defined in Eq.~\eqref{eq:int_V}. Upon introducing the identity $\mathbb{1} = | e_{k_{n+1}} \rangle c_{k_{n+1},\ell_{n+1}} \langle e_{\ell_{n+1}}|$, with the matrix $c$ given in Eq.~\eqref{eq: def c and identity}, before the ket $|e_1\rangle$ in Eq.~\eqref{non_tadpole_quad_2}, one can write the non-tadpole contribution to the quadrupole as 
\begin{align}
\label{non_tadpole_quad_3}
    \mathcal{N}_Q &= 1 + \frac{1}{N_c} \langle e_2 | e_{k_1} \rangle \sum_{n=1}^\infty\left(\int_{z_1^+> \cdots > z_n^+} \prod_{i=1}^n \mu^2(z_i^+) V_{k_i\, \ell_i} \langle e_{\ell_i} | e_{k_{i+1}} \rangle \right) c_{k_{n+1},\ell_{n+1}} \langle e_{\ell_{n+1}}| e_1 \rangle \nn \\
    &= 1 + \frac{1}{N_c} \langle e_2 | e_{k_1} \rangle \sum_{n=1}^\infty\left(\int_{z_1^+> \cdots > z_n^+} \prod_{i=1}^n \mu^2(z_i^+) \F_{k_{i}\, k_{i+1}} \right) c_{k_{n+1},\ell_{n+1}} \langle e_{\ell_{n+1}}| e_1 \rangle,
\end{align}
where we have defined $\F_{ij} = V_{is}\braket{e_s | e_j}$ in the second equality. Using the explicit expression for the matrix $V$ given in Eq.~\eqref{eq:int_V}, the explicit form of the matrix $\F$ can be written as 
\begin{align}
    \F = \begin{pmatrix}
        \alpha C_{F} + \frac{1}{2N_{c}} (\beta - \gamma) & \frac{1}{2} (\alpha - \beta)\\
        \frac{1}{2} (\gamma - \beta)  & \gamma C_{F} + \frac{1}{2 N_{c}} (\beta - \alpha)
    \end{pmatrix},
\end{align}
with the factors $\alpha$, $\beta$ and $\gamma$ defined in Eq.~\eqref{def:abg} in terms of the kinetic factor $W_{ij}$ which is defined in Eq.~\eqref{eq:Wij_def}.  
Using the fact that 
\begin{align}\label{eq:left_right_quad_overlaps}
\langle e_2 | e_{k_1}\rangle = (N_c, N_c^2)_{k_1}, \hspace{0.5cm}\langle e_{\ell_{n+1}} | e_1 \rangle = (N_c^2,N_c)_{\ell_{n_1}}, \hspace{0.5cm}
c_{k_{n+1},\ell_{n+1}} \langle e_{\ell_{n+1}}| e_1 \rangle = (1,0)_{k_{n+1}} 
\end{align}
in Eq. \eqref{non_tadpole_quad_3}, the non-tadpole contribution to the quadrupole operator can be written as
%
\begin{align}
\label{NQ}
    \mathcal{N}_Q &= \frac{1}{N_c} 
    \begin{pmatrix}
        N_c & N_c^2
    \end{pmatrix}
    \left[
    {\cal P_{+}} \exp \int dz^+ \mu^2(z^+)\,  \F\, 
    \right]
    \begin{pmatrix}
        1 \\
        0
    \end{pmatrix},
\end{align}
where we have used 
\begin{equation}
    1 = \frac{1}{N_c}\langle e_2 | e_{k_1} \rangle c_{k_1\,\ell_1} \langle e_{\ell_1} | e_1 \rangle.
\end{equation}
The final expression of the non-tadpole contribution to the quadrupole operator can be written more conveniently after diagonalizing the eikonal interaction matrix $\F$ using
\begin{eqnarray}
\label{System_F}
    \F = R \cdot \text{diag}(\lambda_+,\lambda_-)  \cdot R^{-1}\ ,
\end{eqnarray}
where $R=[v_{+}, v_{-}]$ is the rotation matrix with $v_{\pm} = (a_{\pm}, b_{\pm})^{T}$ being the eigenvectors of the eikonal interaction matrix $\F$ with the corresponding eigenvalues $\lambda_{\pm}$. The eigensystem can be computed easily, leading to
\begin{align}
\label{eq: eigen system of F_1} 
    \lambda_{\pm} = \frac{1}{2 N_{c}} (\beta - \alpha - \gamma) + \frac{N_{c}}{4}(\alpha + \gamma \pm \sqrt{\Delta}),\\
    \label{eq: eigen system of F_2}
    a_{\pm} = \frac{(-(\alpha- \gamma) \pm \sqrt{\Delta}) N_{c}}{2(\beta-\gamma)}, \hspace{1cm} b_{\pm} = 1,\\
    \label{eq: eigen system of F_3}
    \Delta = (\alpha - \gamma)^{2} - \frac{4}{N_{c}^{2}}(\alpha - \beta)(\beta - \gamma),
\end{align}
with the factors $\alpha$, $\beta$ and $\gamma$ defined in Eq.~\eqref{def:abg}. 
Therefore, the non-tadpole contribution to the quadrupole operator can be written as 
\begin{align}
     \N_Q &= \frac{1}{N_c}\begin{pmatrix}
        N_{c} & N_{c}^{2}
    \end{pmatrix} \frac{1}{a_{+}-a_{-}} \begin{pmatrix}
        -a_{+} e^{\tilde{\mu}^{2}\lambda_{+} } + a_{-}e^{\tilde{\mu}^{2}\lambda_{-}} \nn \\
        e^{\tilde{\mu}^{2}\lambda_{+}} - e^{\tilde{\mu}^{2}\lambda_{-}}
    \end{pmatrix}\\
    \label{eq:quad_N_final}
    &= \frac{1}{N_c}\bigg\{\bigg[ \frac{\sqrt{\Delta} + \alpha - \gamma }{2 \sqrt{\Delta}} - \frac{\beta-\gamma}{\sqrt{\Delta}}\bigg] e^{\tilde{\mu}^{2}\lambda_{-}} - \bigg[ \frac{-\sqrt{\Delta} + \alpha - \gamma }{2 \sqrt{\Delta}} - \frac{\beta-\gamma}{\sqrt{\Delta}}\bigg] e^{\tilde{\mu}^{2}\lambda_{+}} \bigg\},
\end{align}
where $\tilde{\mu}^2$ is the color density integrated over the longitudinal nuclear profile defined in Eq.~\eqref{eq:mu_tilde_def}.

To summarize, the expression for the quadrupole operator within the Gaussian model at eikonal order is obtained according to Eq.~\eqref{eq:quad_T_N}, as the product of the tadpole contribution $\mathcal{T}_Q$ which is given in Eq.~\eqref{eq: def tadpole quadrupole} and the non-tadpole contribution $\mathcal{N}_Q$ which is  given in Eq.~\eqref{eq:quad_N_final}\footnote{The quadrupole operator in the Gaussian model was first computed in~\cite{Blaizot:2004wv}. Our result recovers the result from~\cite{Blaizot:2004wv} provided its non-tadpole contribution is included and is normalized with a factor $1/N_c$.}.

\section{Wilson line operators at next-to-eikonal order within a Gaussian model}
\label{sec:NEik_operators_within_GM}
We now focus on the derivation of the expressions for the non-eikonal operators within our Gaussian model.  As it can be seen from the NEik DIS dijet production cross section given in Eqs.~\eqref{eq: Neik cross section} and~\eqref{NEik_X_Section_T}, the operators that arise at NEik order involve one decorated Wilson line in addition to the standard eikonal Wilson lines.  Our aim, in this section, is to develop a generic method that can be used to derive expressions for any type of NEik operator that is composed of one decorated Wilson line and any number of eikonal Wilson lines, provided that these operators are projected into singlet states in order to preserve gauge invariance.  We will start with the illustration of the derivation for a specific case, namely for a NEik dipole operator that involves an $\langle A^jA^-\rangle$ interaction. Then, the derivation will be generalized in a way that it can be applied to any type of NEik operator. Finally, we will use this model to compute the expressions for the NEik dipole and quadrupole operators of the type 1, 2 and 3 that are given in Eqs.~\eqref{eq: dj1 and Qj1},~\eqref{eq: d2 and Q2}  and~\eqref{eq: dij3 and Qij3}, as well as the ones contained in the DIS dijet production cross section for transversely polarized photons given in Eq.~\eqref{NEik_X_Section_T}.

\subsection{Next-to-eikonal dipole involving the $\langle A^jA^-\rangle$ interaction}
Let us now introduce step by step how next-to-eikonal operators can be evaluated in the Gaussian approximation by considering first the case of the next-to-eikonal dipole $d^j(\x,\y)$, which is defined as 
%
\begin{align}
\label{NEik_dip_d^j}
    d^{j}(\x,\y) = \int_{-\frac{L^{+}}{2}}^{\frac{L^{+}}{2}} dv^{+} \Braket{\U_{[\frac{L^+}{2},v^+]}(\x) \left[-igA^j (v^{+}, \x) \right]
 \U_{[v^+, -\frac{L^+}{2}]}(\x) \U^{\dagger}_{[\frac{L^+}{2}, -\frac{L^+}{2}]}(\y) }.
\end{align}
This operator is the piece of the NEik dipole operators which  includes only the transverse field part of the decoration.   
The derivation of this operator follows that for the eikonal case, with the caveat that we have an additional interaction mediated by a factor $\Braket{A^j(v^+,\x) A^-(w^+,\y) }$.

Let us first examine the tadpole contribution to this operator $d^j(\x,\y)$. In addition to the eikonal tadpole contributions $\T_d$ computed in the preceding section and given in Eq.~\eqref{dip_tad_pole}, one must also account for the NEik tadpole contributions. These arise from the function $G^{j-}(\r)$ defined in Eq.~\eqref{def:gi-}. However, from this definition, it is straightforward to observe that its value at $\r=0$ -- corresponding to the tadpole contribution at NEik order -- is $G^{j-} (0) = 0$. Consequently, the NEik tadpole contribution to the operator $d^j(\x,\y)$ vanishes identically, leaving the eikonal $\T_d$ as the sole tadpole contribution to this NEik operator.  
Importantly, since a generic NEik operator consists of one decorated Wilson line together with an arbitrary number of eikonal Wilson lines, and since the above argument for the vanishing of the NEik tadpole term does not depend on the eikonal part of the operator, this conclusion extends to any NEik operator.

\begin{figure}[!h]
    \centering
    \includegraphics[width=.5\textwidth]{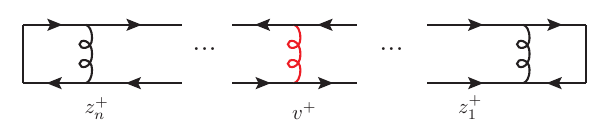}
    \caption{Diagrammatic representation of the contribution $\mathcal{N}^j_{NEik}$ to the next-to-eikonal dipole $d^j$. Here the non-eikonal interaction is shown with a red gluon line. The tadpole contribution is factored out into the previously defined $\T_d$ factor.}
    \label{fig:NEik dipoles}
\end{figure}

Additionally, the NEik non-tadpole contributions must be taken into account to evaluate $d^j(\x,\y)$ within the Gaussian approximation. These contributions, denoted by $\N^{j}_{\rm NEik}$, correspond to interactions between  quark and antiquark lines at NEik accuracy and are represented diagrammatically in Fig.~\ref{fig:NEik dipoles}. As evident from Eq.~\eqref{NEik_dip_d^j}, the operator $d^j(x,y)$ is genuinely NEik due to the insertion of a transverse background gluon field along the longitudinal extent of the Wilson line. Therefore, when expressed in terms of the two-point functions given in Eq.~\eqref{eq:2pts_ev}, the non-tadpole contribution to this operator requires inclusion of an additional interaction mediated by $\Braket{A^j(v^+,\x) A^-(w^+,\y) }$, alongside the standard eikonal interactions mediated by $\Braket{A^-(v^+,\x) A^-(w^+,\y) }$. 


From the derivation of $\N_d$, the eikonal non-tadpole contribution to the dipole operator, it follows that the NEik interaction $\Braket{A^j(v^+,\x) A^-(w^+,\y) }$ introduces an additional  color factor $C_F$ along with the function $G^{j-}(\x-\y)$, supplementing the eikonal non-tadpole contributions given in Eq. \eqref{dip_non_tadpole}. Thanks to the factorization of the kinetic terms, the NEik dipole operator $d^j(\x,\y)$ can thus be written as 
%
%
\begin{align}
\label{def:NEik_dip_d^j_inGA}
d^{j}(\x,\y) 
=  \left( g^2 C_F\int_{-L^+/2}^{L^+/2} dv^+ \mu^2(v^+)  G^{j-}(\x -\y) \right) d(\x,\y) 
= g^{2} C_{F} \tilde{\mu}^{2} G^{j-}(\x -\y) d(\x,\y),
\end{align}
with $d(\x,\y)$ being the eikonal dipole operator within the Gaussian approximation given in Eq.~\eqref{def:dip_Gaussian_model} and $\tilde{\mu}^2$ being the integrated color charge density defined in Eq.~\eqref{eq:mu_tilde_def}.
As in the case of the eikonal dipole, this result can be related to the MV model 
by noting that
\begin{align}
\label{eq: Gj- in MV}
    G^{j-}(\r) = \frac{1}{2P_{q}^{-}} \frac{1}{2\pi i} \partial_{\r^{j}} \frac{1}{2\pi} \int \! \frac{dP_{\perp}}{P^{3}_{\perp}} J_{0}(P_{\perp}r_{\perp})
    \simeq \frac{1}{2 P_{q}^{-}} \frac{i}{2\pi} \r^{j} \int\! \frac{dP_{\perp}}{P_{\perp}} = \frac{1}{2P_{q}^{-}} \frac{i}{2\pi}\r^{j} \ln \frac{1}{r_{\perp}\Lam}\ ,
\end{align}

\begin{align*}
    G^{j-}(\r) = \frac{-i}{2P^-_q}\partial_\r^j G^{--}(\r)
    \approx \left(\frac{i\,\r^j}{8\pi P^-_q}\right) \ln \frac{1}{r_{\perp}\,\Lambda_{\text{QCD}}},
\end{align*}

where we have used the small dipole approximation, $r_{\perp} P_{\perp} \ll 1$, and regulated the infrared divergences
by introducing $\Lambda_{\text{QCD}}$ as in the computation of eikonal dipole. Finally, by inserting Eq.~\eqref{eq: Gj- in MV} into Eq.~\eqref{def:NEik_dip_d^j_inGA} and using Eq.~\eqref{eq: def dipole} for the eikonal dipole within the MV model, we can write the result of the NEik dipole operator with an $ A^{j}$ insertion as 
%
\begin{align}
    d^{j}(\x,\y) = {\frac{1}{2}}\frac{iQ_{s}^{2}}{2P_{q}^{-}} (\x -\y)^{j} \ln \bigg( \frac{1}{|\x-\y|\Lam}\bigg) \text{exp}\bigg\{  - \frac{Q_{s}^{2}}{4} (\x - \y)^{2} \text{ln}\frac{1}{|\x -\y|\Lam}\bigg\}.
\end{align}
Before we finalize this section, we would like to emphasize that the generalization of the above result for the NEik operators with a $\partial_{j}A^{-}$ insertion or a double insertion is
straightforward, since the kinematic factors factorize and the color structure remains exactly the same as above. 

\subsection{Generalization to any next-to-eikonal operators}

Any of the NEik operator that we consider contains one decorated Wilson line accompanied by an arbitrary number of eikonal Wilson lines. As discussed earlier, these decorated Wilson lines incorporate insertions of background covariant derivatives like  $\overleftrightarrow{D}^j_\z$ or $\overleftarrow{D}^j_\z \overrightarrow{D}^j_\z$ or of the field strength tensor $F^{ij}(z^+,\z)$ between the semi-infinite Wilson lines evaluated at the same transverse coordinate $\z$. Since we would like to evaluate NEik operators that involve these decorated Wilson lines in the Gaussian approximation and these decorations involve the evaluation of derivatives, we introduce the following representation for convenience: 
%
%
%
%
\begin{equation}
    \partial_\x^j\, \U_{[b^+,a^+]}(\x) = -ig \int^{b^+}_{a^+} dc^+\, \U_{[b^+,c^+]}(\x) \left[\partial^j_\x A^-(c^+,\x)\right]\U_{[c^+,a^+]}(\x).
\end{equation}
Owing to these transverse partial derivatives and the transverse fields $A^j$ in the insertions 
it is straightforward to realize that the longitudinal structure of the NEik decorated Wilson lines contain either one or two longitudinal positions that are integrated over the longitudinal support of the corresponding Wilson line. This longitudinal coordinate dependence should be properly taken into consideration while performing the Gaussian average over the background field. 
As discussed in the preceding subsection, in addition to the eikonal interactions mediated by the two-point functions of the form $\langle A^-A^-\rangle$, NEik corrections require inclusion of interactions mediated by the two-point functions $ \langle (\partial^jA^-)A^-\rangle$ and  $\langle A^jA^-\rangle$ while keeping in mind that the insertion of either $\partial^jA^-$ or $A^j$ is constrained to the transverse location of the NEik Wilson line. 


Let us introduce an operator $\hat{\Omega}$ as the tensor product of an arbitrary number $n-1 \ge 0$ of fundamental Wilson lines and $n$ Wilson lines in the dual representation:
\begin{equation}
\label{def:Omega_Op}
    \hat{\Omega}_{[x^+,y^+]} = \bigotimes^{n-1}_{i=1} \U_{[x^+,y^+]}(\x_{i}) \ \bigotimes^{n}_{j=1} \U_{[x^+,y^+]}^\dagger(\y_j).
\end{equation}
The generic NEik operator that we would like to evaluate in the Gaussian approximation can then be written as 
\begin{equation}
\label{def:Gen_NEik_op}
    \hat{\mathcal{O}}^{(k)} = \U^{(k)}(\v) \otimes \hat{\Omega},
\end{equation}
with the superscript $k=1,2,3$ indicating the type of the decorated Wilson lines. In the most general case, the longitudinal structure of these decorated Wilson lines can have the following two forms: 
%
\begin{equation}
\label{Longitudinal_structure_Dec_WL}
    \U^{(k)} \propto 
    \begin{cases}
    \int dv^+\ \U_{\big[\frac{L^+}{2},v^+\big]} \, \OpA(v^+) \,  \U_{\big[v^+,-\frac{L^+}{2}\big]},  \\
    \int dv^+dw^+\ \theta(w^+-v^+)\,  \U_{\big[\frac{L^+}{2},w^+\big]}\,  \OpC(w^+) \, \U_{[w^+,v^+]} \, \OpB(v^+) \,  \U_{\big[v^+,-\frac{L^+}{2}\big]}  ,
    \end{cases}
\end{equation}
where the operators $\OpA$, $\OpB$, $\OpC$ result from the insertion of different decorations. Let us introduce $\{|v\rangle\}$ as a complete basis that spawn the subspace of all color singlet in $V^n\otimes \bar{V}^n$, and define the \textit{identity} $\mathcal{I}$ in this subspace according to
\begin{equation}
    \mathcal{I}\Big|_{singlet} = \sum_{nm} c_{nm}\left| v_n \right\rangle\left\langle v_m\right| = \mathbb{1}.
\end{equation}
Upon inserting the identity $\mathbb{1}$ before and after each internal longitudinal coordinate that are integrated over the corresponding longitudinal support (denoted as $v^+$ and $w^+$ in Eq.~\eqref{Longitudinal_structure_Dec_WL}) and using the definition of a generic NEik operator defined in Eq.~\eqref{def:Gen_NEik_op}, one can write the two types of generic NEik operators (with either one or two insertions along the longitudinal extend of the corresponding decorated Wilson line) as 
%
\begin{align}
    \hat{\mathcal{O}}^{(k)}_{1-point} &= \int dv^+   \left(\U_{[\frac{L^+}{2},v^+]}(\v) \otimes \hat{\Omega}_{\frac{L^+}{2},v^+]}\right) \left| v_n \right\rangle  c_{nm} \notag \\
    &\times \left\langle v_m\right| \OpA(v^+,\v)\sum_{\w=\{\x,\y\}} A^-(v^+,\w) \left| v_s \right\rangle c_{st} \left\langle v_t\right| \left(\U_{[v^+,-\frac{L^+}{2}]}(\v) \otimes \hat{\Omega}_{v^+,\frac{L^+}{2}]}\right)
\end{align}
and 
\begin{align}
    \hat{\mathcal{O}}^{(k)}_{2-point} &= \int dw^+dv^+ \theta(w^+-v^+)  \left(\U_{[\frac{L^+}{2},w^+]}(\v) \otimes \hat{\Omega}_{\frac{L^+}{2},w^+]}\right) \left| v_n \right\rangle c_{nm} \notag \\
    &\times \left\langle v_m\right| \OpC(v^+,\v)\sum_{\w=\{\x,\y\}} A^-(v^+,\w) \left| v_s \right\rangle c_{st}  
    \left\langle v_t\right| \left(\U_{[w^+,v^+]}(\v) \otimes \hat{\Omega}_{w^+,v^+]}\right) \left| v_f \right\rangle  c_{fg} \notag \\
    &\times \left\langle v_g\right| \OpB(v^+,\v)\sum_{\w'=\{\x,\y\}} A^-(v^+,\w') \left| v_p \right\rangle c_{pq}
    \left\langle v_q \right| \left(\U_{[v^+,-\frac{L^+}{2}]}(\v) \otimes \hat{\Omega}_{v^+,\frac{L^+}{2}]}\right)\,,  
\end{align}
where the summation over the indices labeling the basis is implicit. Moreover, the
summation over $\w$ and $\w'$ spans all the available transverse coordinates in the operator $\hat{\Omega}$. 
These operators are finally evaluated according to Eq.~\eqref{eq:2pts_ev} and projected onto the initial and final color state of interest, and are denoted as
\begin{equation}
    {\mathcal{O}}^{(k)}_{1-point} = \left\langle \hat{\mathcal{O}}^{(k)}_{1-point} \right\rangle, \qquad
    {\mathcal{O}}^{(k)}_{2-point} = \left\langle \hat{\mathcal{O}}^{(k)}_{2-point} \right\rangle.
\end{equation}
This approach provides a systematic evaluation of NEik operators within the Gaussian approximation, contingent oto the knowledge of the basis $\mathcal{I}$ and the action of $T^a \otimes T^a$ in that basis.

First, we note that at this level the longitudinal integrals are not straightforward to evaluate due to the non-diagonal nature of the NEik interaction, at longitudinal positions $v^+$ and/or $w^+$, in a generic basis $\{|v\rangle\}$. However, since the eikonal interactions are always mediated by two-point functions of the form $\langle A^-A^-\rangle$ given in Eq.~\eqref{eq:2pts_ev} with $\mu=\nu=-$,
the transverse coordinate dependence is encoded in the function $G^{--}$ defined in Eq.~\eqref{def:g--} and it remains unchanged along the longitudinal path. The nuclear longitudinal profile is contained in the function $\mu^2(x^+)$ that is factorized from the transverse dependence of the two-point function. Therefore, one can diagonlize the eikonal interactions and then perform the longitudinal integrals analytically. 
For this purpose, we introduce the operator $\mathcal{V}^{eik}(v^+)$, which resums all possible eikonal interactions between the partons in the system at a given longitudinal position $v^+$, as  
%
%
\begin{equation}
\label{Veik_def}
    {\mu^2(v^+)}\mathcal{V}^{eik}(v^+) =   \sum_{ij} (-1)^{i+j+1} \left(T^a_i \otimes  T^b_j\right) \left\langle A_a^-(v^+,\x_i) A_b^-(v^+,\x_j)\right\rangle\, . 
\end{equation}
We emphasize that tadpole contributions are automatically included in the definition of the eikonal interaction matrix $\mathcal{V}^{eik}$, since the case $i=j$ is not excluded in the summation appearing in its definition. 
In the $\{|v\rangle\}$ basis that we have introduced, this operator can be written as 
\begin{align}
\label{eik_int_v_basis}
    \mathcal{V}^{eik}(v^+)
    &= |v_a\rangle V^{eik}_{ae}(v^+) c^{-1}_{ef} c_{fg} \langle v_g | = |v_a\rangle F^{eik}_{af}(v^+) c_{fg} \langle v_g |,
\end{align}
with 
\begin{align}
 V^{eik}_{ae}(v^+) &=  c_{ab} \langle v_c|\mathcal{V}^{eik}(v^+) | v_d\rangle c_{de} 
 \, , \hspace{1,5cm} V^{eik}_{ae}(v^+)\, c^{-1}_{ef} = F^{eik}_{af}(v^+) \, , \hspace{1,5cm} c^{-1}_{ef}= \langle v_e|v_f\rangle \, . 
\end{align}
Thus, along a path with $n$ interactions at longitudinal positions $v^+_n > \cdots > v_2^+ > v_1^+$, we write
\begin{equation}
    \mathcal{V}^{eik}(v^+_m) \cdots \mathcal{V}^{eik}(v^+_1) = |v_a\rangle F^{eik}_{ab}(v^+_n) \{\cdots\}_{bc} F^{eik}_{cd}(v^+_1) c_{de} \langle v_d |.
\end{equation}
%
Let $R$ be the matrix that diagonalize the eikonal interaction matrix  $F^{eik}(v^+)$, so that 
\begin{equation}
    F^{eik}_{ij}(v^+) = (R \cdot f^{eik} \cdot R^{-1})_{ij}
\end{equation}
and we denote the eigenvalues by
\begin{equation}
    \left\{ \lambda_{1}, \lambda_{2}, \cdots \lambda_{m} \right\}.
\end{equation}
%
The locality of the $\langle A^-A^-\rangle$ two-point functions in longitudinal coordinate and the fact that the function $G^{\mu\nu}$ is independent of the longitudinal coordinate, ensure that the matrix $R$ that diagonalizes the eikonal interactions at a given longitudinal position also diagonalizes those interactions along the longitudinal path. This diagonal matrix can be written as       
%
\begin{equation}
    \Deik_{[x^+,y^+]} = \text{diag} \left[ \exp \left( \lambda_1
    \int_{y^+}^{x^+} dv^+ \mu^2(v^+)\right), \cdots, \exp \left( \lambda_m \int_{y^+}^{x^+} dv^+ \mu^2(v^+) \right)
    \right],
\end{equation}
which allows us to write the expectation value of the NEik decorated $1-point$ operator projected onto the color states $\langle v^{(+\infty)}| \cdots | v^{(-\infty)} \rangle$ as
\begin{align}
    \langle v^{(+\infty)} |{\mathcal{O}}^{(k)}_{1-point} | v^{(-\infty)} \rangle= \int dv^+   \langle v^{(+\infty)} | v_a\rangle &\underbrace{c_{ab} \langle v_b| \left(\U_{[\frac{L^+}{2},v^+]}(\v) \otimes \hat{\Omega}_{[\frac{L^+}{2},v^+]}\right) \left| v_n \right\rangle  c_{nj} \ c^{-1}_{jk}}_{(\exp {\mu^2} F^{eik})_{ak}} \notag \\
    \times &\underbrace{c_{km} \left\langle v_m\right| \OpA(v^+,\v)\sum_{\w=\{\x,\y\}} A^-(v^+,\w) \left| v_s \right\rangle c_{sq}\ c^{-1}_{qr} }_{{\mu^2(v^+)}(V^{NEik,\OpA} c^{-1})_{kr}={\mu^2(v^+)}(F^{NEik,\OpA})_{kr}}\notag \\
    \times &\underbrace{c_{rt} \left\langle v_t\right| \left(\U_{[v^+,-\frac{L^+}{2}]}(\v) \otimes \hat{\Omega}_{[v^+,-\frac{L^+}{2}]}\right)|v_c\rangle c_{cd}  c^{-1}_{dh} }_{(\exp {\mu^2} F^{eik})_{rh}} c_{hu}\langle v_u| v^{(-\infty)}\rangle.
    \label{eq:generic}
\end{align}
The NEik decorated $2-point$ operator can be written in an analogous way. Finally, 
in the rotated basis we write 
\begin{align}\label{eq:1p_Neik_op}
    \langle v^{(+\infty)} |{\mathcal{O}}^{(k)}_{1-point} | v^{(-\infty)} \rangle 
    = \int &dv^+   \langle v^{(+\infty)} | v_a\rangle c_{hu}\langle v_u| v^{(-\infty)}\rangle {\mu^2(v^+)} \left(R  \Deik_{[\frac{L}{2},v^+]}  R^{-1} F^{NEik,\OpA}_{v^+}  R  \Deik_{[v^+,-\frac{L^+}{2}]}  R^{-1} \right)_{ah}, \\
    \langle v^{(+\infty)} |{\mathcal{O}}^{(k)}_{2-point} | v^{(-\infty)} \rangle 
    = \int &dv^+ dw^+ \theta(w^+-v^+)  
    \ \langle v^{(+\infty)} | v_a\rangle  c_{hu}\langle v_u|v^{(-\infty)}\rangle {\mu^2(w^+)}{\mu^2(v^+)}\notag \\
    &\times \left(R \Deik_{[\frac{L}{2},w^+]} R^{-1} F^{NEik,\OpC}_{w^+}  R \Deik_{[w^+,v^+]}  R^{-1} F^{NEik,\OpB}_{v^+} R \Deik_{[v^+,-\frac{L^+}{2}]} R^{-1} \right)_{ah},
    \label{eq:2p_Neik_op}
\end{align}
where we have introduced the matrix $F^{NEik}$, defined in Eq.~\eqref{eq:generic}, that encodes the NEik interactions. This matrix is the NEik generalization of the eikonal interaction matrix given in Eq. \eqref{eik_int_v_basis}. Its explicit form and its relation to the NEik interaction matrix $V^{NEik}$ for the interaction type ${\mathbb{A}}$ reads 
\begin{align}
\label{F_NEik_A}
\mu^2(v^+)\left( F^{NEik,\, \OpA}_{v^+}\right)_{ac} &= c_{ab} \left\langle \langle v_b| \OpA(v^+,\v) \sum_{\w=\{\x,\y\}} A^-(v^+,\w)|v_c\rangle\right\rangle,  \\
{\mu^2(v^+)} \left(V^{NEik,\OpA}_{v^+}\right)_{ad} &= \mu^2(v^+)\left( F^{NEik,\, \OpA}_{v^+}\right)_{ac} c_{cd}\ .
\end{align}
The NEik interaction matrices $F^{NEik,\, \OpB}_{v^+}$ and $F^{NEik,\, \OpC}_{v^+}$ are defined analogous to Eq.~\eqref{F_NEik_A} with the interaction type $\OpA$ being replaced by type $\OpB$ and $\OpC$, respectively. Assuming a homogeneous longitudinal profile, the NEik matrices $\big(R^{-1} \cdot F^{NEik,\OpA}_{v^+} \cdot R\big)$ are either independent of $v^+$ or linear in $v^+$. Thus, the longitudinal integrals can be performed analytically according to the integrals listed in Appendix~\ref{App:Longi_integrals}. 



%
\subsection{Next-to-eikonal operators of type 1}
%
%
%
We can now use the scheme developed in the preceding subsection to compute the NEik operators of type 1 which have the  generic form
%
\begin{align}
\label{eq:NEik_op_type_1}
     \hat{\mathcal{O}}^{(1)}_j(\v_{*}) =  \int_{-\frac{L^+}{2}}^{\frac{L^+}{2}} dv^{+}\,  \U_{[\frac{L^+}{2},v^+]}(\v) \big\{\overrightarrow{\partial_{\v^{j}}}  - \overleftarrow{\partial_{\v^{j}}}  - 2igA^{j}(v^+, \v)\big\}
  \U_{[v^+, -\frac{L^+}{2}]}(\v) \otimes \hat{\Omega}\,.
\end{align}
This expression can be simplified by utilizing the fact that    
%
\begin{align}
\label{eq: rightderivative_v_integration}
    \int_{v^{+}} \U_{[\frac{L^+}{2},v^+]}(\v) \overrightarrow{\partial_{\v^{j}}} \U_{[v^+, -\frac{L^+}{2}]}(\v) &= \int_{v^{+},z_{1}^{+}} \Theta(v^{+}-z_{1}^{+}) \U_{[\frac{L^+}{2},v^+]}(\v) \U_{[v^+, z_{1}^{+}]}(\v) \Big({\partial_{\v^{j}}} \tilde{A}^{-}(z_{1}^{+},\v)\Big)
     \U_{[z_{1}^+, -\frac{L^+}{2}]}(\v) \\\nn
     &= \int_{z_{1}^{+}} \bigg(\frac{L^{+}}{2} -z_{1}^{+}\bigg) \U_{[\frac{L^+}{2},z_{1}^+]}(\v) \Big({\partial_{\v^{j}}} \tilde{A}^{-}(z_{1}^{+},\v)\Big)
     \U_{[z_{1}^+, -\frac{L^+}{2}]}(\v) ,
\end{align}
where the transverse derivative acts only on the gauge field\footnote{For brevity, we have introduced the following notation:
\begin{align}
    \int_{z^{+}} \equiv \int_{-\frac{L^{+}}{2}}^{\frac{L^{+}}{2}} dz^{+}, \hspace{1cm} \tilde{A}^{\mu} \equiv -igA^{\mu} \nn.
\end{align}
}.
%
Similarly, we can write
\begin{align}
\label{eq: leftderivative_v_integration}
    \int_{v^{+}} \U_{[\frac{L^+}{2},v^+]}(\v) \overleftarrow{\partial_{\v^{j}}} \U_{[v^+, -\frac{L^+}{2}]}(\v) 
     &= \int_{z_{1}^{+}} \bigg(z_{1}^{+} + \frac{L^{+}}{2} \bigg) \U_{[\frac{L^+}{2},z_{1}^+]}(\v) \Big({\partial_{\v^{j}}} \tilde{A}^{-}(z_{1}^{+},\v)\Big)
     \U_{[z_{1}^+, -\frac{L^+}{2}]}(\v) .
\end{align}
Substituting Eqs.~\eqref{eq: rightderivative_v_integration} and~\eqref{eq: leftderivative_v_integration} into Eq.~\eqref{eq:NEik_op_type_1}, one can write the generic form of the NEik operators of type 1 as  
%
\begin{align}
\label{eq:NEik_op_type_1_A}
     \hat{\mathcal{O}}^{(1)}_j(\v_{*}) = \int_{\frac{L^+}{2}}^{\frac{L^+}{2}} dv^{+}\  \U_{[\frac{L^+}{2},v^+]}(\v) \big\{2\tilde{A}^{j}(v^{+},\v) - 2v^{+} \partial_{\v^{j}} \tilde{A}^{-}(v^+, \v)\big\}
 \U_{[v^+, -\frac{L^+}{2}]}(\v) \otimes \hat{\Omega} \, . 
\end{align}
\paragraph{\underline{NEik type-1 dipole operator}.} The first NEik operator of type 1 that we need to evaluate is the decorated dipole of type 1, which can be obtained by considering ${\hat{\Omega} \rightarrow \U^\dagger(\w)}$ in the generic form given in Eq.\eqref{eq:NEik_op_type_1_A}, which then reads 
\begin{equation}
    d^{(1)}_j(\v_{*} , \w) = \frac{1}{N_c} \left\langle \text{Tr}\,  \hat{\mathcal{O}}^{(1)}_j\big|_{\hat{\Omega} \rightarrow \U^\dagger(\w)} \right\rangle.
\end{equation}
The tensor $\mathcal{I}$ that projects the system to color singlet is straightforward to evaluate in the case of a dipole, since there is a single way to project quark-antiquark system to a singlet. In the birdtrack approach, it is represented diagrammatically as 
\begin{equation}
    \mathcal{I} = \frac{1}{N_c}
    \vcenter{\hbox{
    \begin{tikzpicture}[scale=0.5]
        \draw[thick] (-1,.5) -- (-.5,.5) to[out=0,in=0] (-.5,-.5) -- (-1,-.5);
        \draw[thick,->] (-1,.5) -- (-.5,.5);
        \draw[thick] (1,.5) -- (.5,.5) to[out=180,in=180] (.5,-.5) -- (1,-.5);
        \draw[thick,-<] (1,.5) -- (.5,.5);
    \end{tikzpicture}
    }}.
\end{equation}
Therefore, the type 1 decorated dipole operator can be written as 
\begin{align}
    d^{(1)}_j(\v_{*} , \w) &= \int dv^+ \frac{1}{N_c} \left\langle \text{Tr}\, \U_{[\frac{L^+}{2},v^+]}(\v)\U^\dagger_{[\frac{L^+}{2},v^+]}(\w)\right\rangle \notag \\
    &\times 
    \frac{1}{N_c} \left\langle  \text{Tr}\, \big\{2\tilde{A}^{j}(v^{+},\v) - 2v^{+} \partial_{\v^{j}} \tilde{A}^{-}(v^+, \v)\big\} (-1) \tilde{A}^-(v^+,\w) \right\rangle
    \times
    \frac{1}{N_c} \left\langle \text{Tr}\, \U_{[v^+,-\frac{L^+}{2}]}(\v)\U^\dagger_{[v^+,-\frac{L^+}{2}]}(\w)\right\rangle.
\end{align}
Using the fact that two-point functions are local in longitudinal direction, neglecting the vanishing NEik tadpole contributions, and using our model for the field correlators given in Eq.~\eqref{eq:2pts_ev}, it is straightforward to write the NEik dipole operator of type 1 as 
\begin{align}
    d^{(1)}_{j}(\v_{*} , \w)  = 2g^{2}\, C_{F} \, d(\v,\w) \,  \int_{v^{+}} \mu^{2}(v^{+}) \bigg\{ G^{j-}(\v -\w) - v^{+}\partial_{\v^{j}} G^{--}(\v -\w)\bigg\},
\end{align}
where the $d(\v,\w)$ is the eikonal dipole operator within the Gaussian approximation defined in Eq.~\eqref{def:dip_Gaussian_model} with the functions $G^{--}$ and $G^{j-}$ respectively defined in Eqs.~\eqref{def:g--} and~\eqref{def:gi-}. 
Assuming that the transverse color charge density of the target $\mu^2(v^+)$ is constant, i.e., in the limit of the homogeneous profile, one can trivially perform the integration over the longitudinal extend of the target. Upon integration the term linear in $v^+$ vanishes, and one obtains
%
\begin{align}
\label{eq: d1_j final}
    d^{(1)}_{j}(\v_{*} , \w)  &= 2g^{2} \tilde{\mu}^{2} C_{F} d_{[\frac{L^{+}}{2}, -\frac{L^{+}}{2}]}(\v,\w) G^{j-}(\v -\w),
\end{align}
with $\tilde{\mu}^2$ is the color density integrated over the longitudinal nuclear profile defined in Eq.~\eqref{eq:mu_tilde_def}. As in the eikonal case, considering the small dipole limit and using the saturation momentum $Q_s$ defined in Eq.~\eqref{eq: def Qs}, the decorated dipole of type 1 can be written within the Gaussian approximation as  
\begin{align}
\label{type_1_dip_fin}
   d^{(1)}_{j}(\v_{*} , \w)
    &= {\frac{1}{2}} \frac{iQ_{s}^{2}}{P_{q}^{-}} (\v -\w)^{j} \,  \ln\bigg( \frac{1}{|\v - \w| \Lam}\bigg) \, \text{exp}\bigg\{- \frac{Q_{s}^{2}}{4} (\v - \w)^{2} \text{ln}\frac{1}{|\v -\w|\Lam}\bigg\} \, . 
\end{align}
%

\paragraph{\underline{NEik type-1 quadrupole operator}.} The next operator we consider is the decorated quadrupole operator of type 1, which reads 
\begin{equation}
    Q^{(1)}_j(\w',\v',\v_{*},\w) = \frac{1}{N_c} \left\langle \text{Tr}\,  \hat{\mathcal{O}}^{(1)}_j\big|_{\hat{\Omega} \rightarrow \U^\dagger(\w)\U(\w')\U^\dagger(\v')} \right\rangle.
\end{equation}
Using the identity defined in Eq.~\eqref{eq: def c and identity} (or, more explicitly, Eq. ~\eqref{eq:id_quadr}), we can write the quadrupole operator of type 1 in the basis $\{|e_1\rangle,|e_2\rangle\}$. 
Thanks to the locality of the two-point function in longitudinal direction within the Gaussian model, the decorated quadrupole operator of type 1 can be written in the $\{|e_1\rangle,|e_2\rangle\}$ basis as 
\begin{align}
    Q^{(1)}_j(\w',\v',\v_{*},\w) 
    = \frac{1}{N_c} &\int dv^+ \langle e_2 |\mathcal{Q}_{[\frac{L^+}{2},v^+]} \notag \\
    &    \left\langle  \big\{2\tilde{A}^{j}(v^{+},\v) - 2v^{+} \partial_{\v^{j}} \tilde{A}^{-}(v^+, \v)\big\} \otimes \sum_{\y = \{\w,\w',\v'\}} (\pm)_\y\tilde{A}^-(v^+,\y) \right\rangle 
    \mathcal{Q}_{[v^+,-\frac{L^+}{2}]}| e_1 \rangle ,
\end{align}
where the operator $\mathcal{Q}_{[x^+,y^+]}$ is defined in Eq. \eqref{eq:calQ_def} and $(\pm)_\y$ denotes the sign for a quark (+) or an antiquark ($-$) being at the transverse location $\y$.
Using the rotation matrix $R$ previously defined for the eikonal quadrupole in Eq.~\eqref{eq: eigen system of F_2} and the generic NEik decorated $1-point$ operator given in Eq.~\eqref{eq:1p_Neik_op}, we can write the type 1 quadrupole as 
%
\begin{equation}
\label{Op_Q1_fin}
    Q^{(1)}_j(\w',\v',\v_*,\w) = \frac{1}{N_c} \int dv^+ 
    \begin{pmatrix}
        N_c & N_c^2
    \end{pmatrix}
    \cdot
    \left( \mathcal{T}_{\mathcal{Q}} \cdot 
    R 
    S_{[\frac{L^+}{2}, v^+]}
    R^{-1}{\mu^2(v^+)} F^{NEik,j}_{v^+} R
    S_{[v^+,-\frac{L^+}{2}]}
    R^{-1}\right) \cdot
    \begin{pmatrix}
        1 \\ 0
    \end{pmatrix},
\end{equation}
where we used Eq.~\eqref{eq:left_right_quad_overlaps} to recast into the left and right vectors.
Since the NEik tadpole contributions vanish, the tadpole contribution to the quadrupole type 1 is governed by the eikonal $\mathcal{T}_\mathcal{Q}$ which is given in Eq.~\eqref{eq: def tadpole quadrupole},
and the non-tadpole eikonal interactions are implemented by 
%
\begin{equation}
\label{eikS_for_Q1}
    S_{[x^+,y^+]} = \text{diag} \left( \exp\left\{ \lambda_+\,\int_{x^+}^{y^+}dw^+ \mu^2(w^+)\right\}, \exp\left\{\lambda_-\,\ \int_{x^+}^{y^+}dw^+ \mu^2(w^+)\right\}\right),
\end{equation}
with the eigenvalues $\lambda_\pm$  provided in Eq.~\eqref{eq: eigen system of F_1}.
The NEik interaction introduces a single insertion of the matrix $F^{NEik,j}_{v^+}$ which is related to the interaction matrix $V^{NEik,j}_{v^+}$ via
\begin{equation}
\label{relation_F_V_NEik}
 {\mu^2(v^+)}\left(F^{NEik,j}_{v^+}\right)_{ij} 
    = {\mu^2(v^+)} \left(V^{NEik,j}_{v^+} c^{-1}\right)_{ij} \ ,
\end{equation}
and it is defined in Eq.~\eqref{eq:app_VNeik} with its components  given in terms of $\alpha, \beta$ and $\gamma$ defined as  
%
\begin{subequations}
\label{Q1_V_comp}
\begin{align}
    \alpha &= -{g^2}\left[ 2G^{j-}(\v_*,\w) - 2v^+\partial_{\v^j}G^{--}(\v_*,\w) \right], \\
    \beta  &= +{g^2}\left[ 2G^{j-}(\v_*,\w') - 2v^+\partial_{\v^j}G^{--}(\v_*,\w') \right], \\
    \gamma &= -{g^2}\left[ 2G^{j-}(\v_*,\v') - 2v^+\partial_{\v^j}G^{--}(\v_*,\v') \right],
\end{align}
\end{subequations}
where  function $G^{\mu\nu}$ follows from the expectation value of two fields in the Gaussian model according to Eq.~\eqref{eq:2pts_ev}.
For a homogeneous longitudinal profile, the integrations over $v^+$ can be easily performed using the integrals listed in Appendix \ref{App:Longi_integrals}. 



%
%
\subsection{Next-to-eikonal operators of type 2}
%

The NEik operators of type 2 can be written in a generic form as 
\begin{align}
\label{def:OP2_first}
    \hat{\mathcal{O}}^{(2)} (\v_{*}) = 
    \int dv^+ \left\{\U_{[\frac{L^+}{2},v^+]}(\v) 
    \bigg(\overleftarrow{\partial_{\v^{j}}} - \tilde{A}^{j}(v^{+},\v)\bigg) \bigg(\overrightarrow{\partial_{\v^{j}}} + \tilde{A}^{j}(v^{+},\v)\bigg)
    \U_{[v^+,-\frac{L^+}{2}]}(\v) 
    \right\} \otimes \hat{\Omega},
\end{align}
with the operator $\hat \Omega$ given in Eq.~\eqref{def:Omega_Op}. In Eq.~\eqref{def:OP2_first} the derivatives only act on the Wilson lines inside the curly bracket. It can be written as the sum of the following four terms: 
%
\begin{align}
\label{decomp_1}
    {\rm (i)} &= \int_{v^+} \U_{[\frac{L^+}{2},v^+]}(\v) \overleftarrow{\partial_{\v^{j}}}  \overrightarrow{\partial_{\v^{j}}}  \U_{[v^+, -\frac{L^+}{2}]}(\v) \nn
    \\
    &= \int_{w^+_2,w^+_1} \theta(w^+_2 - w^+_1)\, (w^+_2-w_1^+)\ 
    \U_{[\frac{L^+}{2},w_2^+]}
    \left[\partial^j \tilde{A}^-(w^+_2,\v)\right]
    \U_{[w^+_2,w^+_1]}
    \left[\partial^j \tilde{A}^-(w^+_1,\v)\right]
    \U_{[w^+_1,-\frac{L^+}{2}]} ,
    \\
%
\label{decomp_2}
   {\rm (ii)} &= \int_{v^+} \U_{[\frac{L^+}{2},v^+]}(\v) 
    \left[ -\tilde{A}^j(v^+,\v)\right] 
    \overrightarrow{\partial_{\v^{j}}}  
    \U_{[v^+, -\frac{L^+}{2}]}(\v) \nn
    \\
    &=  \int_{w^+_2,w^+_1} \theta(w^+_2 - w^+_1)\, 
    \U_{[\frac{L^+}{2},w_2^+]}
    \left[- \tilde{A}^j(w^+_2,\v)\right]
    \U_{[w^+_2,w^+_1]}
    \left[\partial_j \tilde{A}^-(w^+_1,\v)\right]
    \U_{[w^+_1,-\frac{L^+}{2}]}, \\
\label{decomp_3}
   {\rm  (iii)} &= \int_{v^+} \U_{[\frac{L^+}{2},v^+]}(\v) 
    \overleftarrow{\partial_{\v^{j}}}
    \left[ +\tilde{A}^j(v^+,\v)\right] 
    \U_{[v^+, -\frac{L^+}{2}]}(\v) \nn \\
    &=  \int_{w^+_2,w^+_1} \theta(w^+_2 - w^+_1)\, 
    \U_{[\frac{L^+}{2},w_2^+]}
    \left[\partial_j \tilde{A}^-(w^+_2,\v)\right]
    \U_{[w^+_2,w^+_1]}
    \left[\tilde{A}^j(w^+_1,\v)\right]    
    \U_{[w^+_1,-\frac{L^+}{2}]},
    \\
%
\label{decomp_4}
 {\rm   (iv)} &= -\int_{v^+}\U_{[\frac{L^+}{2},v^+]}(\v) \tilde{A}^{j}(v^{+}, \v)   \tilde{A}^{j}(v^{+}, \v)  \U_{[v^+, -\frac{L^+}{2}]}(\v).
\end{align}
It is more convenient to consider each of the above terms separately when evaluating NEik type 2 operators within the Gaussian approximation. In what follows, we evaluate the NEik dipole and quadrupole operators of type 2.  \\

\paragraph{\underline{NEik type-2 dipole operator}.} We first focus on the NEik dipole operator of type 2 which can be obtained by taking $\hat{\Omega} \rightarrow \U^\dagger(\w)$. It reads 
\begin{equation}
\label{def:d_2_gen}
    d^{(2)}(\v_{*} , \w) = \frac{1}{N_c} \left\langle \text{Tr}\,  \hat{\mathcal{O}}^{(2)}\big|_{\hat{\Omega} \rightarrow \U^\dagger(\w)} \right\rangle.
\end{equation}
We use the decomposition introduced in Eqs.~\eqref{decomp_1},~\eqref{decomp_2},~\eqref{decomp_3} and~\eqref{decomp_4} for the decorated Wilson line $\U^{(2)}(\v)$, write the NEik dipole of type 2 as 
\begin{align}
\label{d_2_decomp}
d^{(2)}(\v_{*} , \w)= d^{(2)}(\v_{*} , \w)\big|_{\rm (i)}+d^{(2)}(\v_{*} , \w)\big|_{\rm (ii)}
+d^{(2)}(\v_{*} , \w)\big|_{\rm (iii)}+d^{(2)}(\v_{*} , \w)\big|_{\rm (iv)}
\end{align}
and compute each term individually.

By using Eqs.~\eqref{decomp_1} and~\eqref{def:d_2_gen}, the first term in the decomposition of NEik dipole of type 2 given in Eq.~\eqref{d_2_decomp} reads
%
%
\begin{align}
d^{(2)}(\v_{*} , \w)\big|_{\rm (i)}
    &= \frac{1}{N_{c}}  \int_{v^{+}}  \Braket{ \text{Tr} \ [ \U_{[\frac{L^+}{2},v^+]}(\v) \overleftarrow{\partial_{\v^{j}}}  \overrightarrow{\partial_{\v^{j}}} 
 \U_{[v^+, -\frac{L^+}{2}]}(\v)\,\U^{\dagger}(\w)]} .
 \end{align}
Upon acting on the semi-infinite Wilson lines, the $v^+$ integral can be performed trivially, leading to 
 \begin{align}
d^{(2)}(\v_{*} , \w)\big|_{\rm (i)}
 &=  \frac{1}{N_{c}}  \int_{ z_{1}^{+},z_{2}^{+}} (z_{1}^{+}-z_{2}^{+}) \theta(z_{1}^{+}-z_{2}^{+}) \nn
 \\
& \hspace{0.5cm} \times 
\Braket{ \text{Tr} \ [ \U_{[\frac{L^+}{2},z_{1}^+]}(\v) \partial_{\v^{j}} \tilde{A}^{-}(z_{1}^{+},\v) \U_{[z_{1}^{+},z_{2}^{+}]}(\v) \partial_{\v^{j}} \tilde{A}^{-}(z_{2}^{+},\v) 
 \U_{[z_{2}^+, -\frac{L^+}{2}]}(\v)\,\U^{\dagger}(\w)]} .
 \end{align}
It is now straightforward to evaluate the Gaussian averaging by using the two-point correlator defined in Eq.~\eqref{eq:2pts_ev} and the functions $G^{--}$ defined in Eq.~\eqref{def:g--}. One simply gets 
%
 \begin{align}
 d^{(2)}(\v_{*} , \w)\big|_{\rm (i)}
 &= g^{4} \, C_{F}^{2} \,  \mu^{4} \, d(\v, \w)  
 \int_{ z_{1}^{+},z_{2}^{+}} (z_{1}^{+}-z_{2}^{+}) \,  \theta(z_{1}^{+}-z_{2}^{+}) \,  \big[\partial_{\v^{j}} G^{--}(\v - \w)\big]^{2},
 \end{align}
with  $d(\v, \w)$ the eikonal dipole defined in Eq.~\eqref{def:dip_Gaussian_model} and where we assumed a homogeneous longitudinal nuclear profile $\mu(v^+)=\mu$. From Eqs.~\eqref{def:g--} and~\eqref{def:gi-}, the relation between the functions $G^{--}(\r)$ and $G^{j-}(\r)$ is
%
%
\begin{align}
\label{relation_G--_Gj-}
\partial_{\v^{j}} G^{--}(\v - \w) =  2i\, P^{-}_{q} \, G^{j-}(\v - \w)\, .
\end{align}
Using this relation  and performing the integrations over $z^+_1$ and $z^+_2$, the first term in the decomposition of NEik
dipole of type 2 given in Eq.~\eqref{d_2_decomp} can be written as 
\begin{align}
 d^{(2)}(\v_{*} , \w)\big|_{\rm (i)} 
 &= - \frac{4}{6} \, (2\pi)^{2} \,  Q^{4}_{s}\, L^{+}\,  \big(P_{q}^{-}\big)^{2} \,  d(\v, \w) \,  \big[G^{j-}(\v - \w)\big]^{2},
\end{align}
where we have used Eq.~\eqref{eq: def Qs} for the definition of the saturation momenta $Q_s$. Finally, substituting Eq.~\eqref{eq: Gj- in MV} for the function $G^{j-}$ within the small dipole approximation, we can rewrite it as 
%

%
\begin{align}
\label{dip_2_i_GM}
    d^{(2)}(\v_{*} , \w)\big|_{\rm (i)}  = \frac{1}{4} \frac{1}{6} \, Q_{s}^{4} \, L^{+} \, (\v - \w)^{2} \, \ln^{2}\bigg( \frac{1}{|\v -\w|\Lam}\bigg) \, \text{exp}\bigg\{  - \frac{Q_{s}^{2}}{4} (\v - \w)^{2} \, \text{ln}\frac{1}{|\v -\w|\Lam}\bigg\}.
\end{align}

Let us now consider the second term in the decomposition of NEik dipole of type 2 given in Eq. \eqref{d_2_decomp} which reads 
\begin{align}
d^{(2)}(\v_{*} , \w)\big|_{\rm (ii)} &= \frac{1}{N_{c}} \int_{v^{+}} \Braket{\text{Tr} \big[ \U_{[\frac{L^+}{2},v^+]}(\v) \big( -\tilde{A}^{j}(v^{+}, \v)\big)\,  \overrightarrow{\partial}_{\v^{j}} \U_{[v^+, -\frac{L^+}{2}]}(\v) \,  \U^{\dagger}(\w) \big] }
\\
 &= \frac{1}{N_{c}} \int_{v^{+}, z_{1}^{+}} \theta(v^{+}-z_{1}^{+}) \left\langle \text{Tr} \Big[ \U_{[\frac{L^{+}}{2}, v^{+}]} (\v) \Big( - \tilde{A}^{j}(v^{+}, \v)\Big) \U_{[v^{+},z_{1}^{+}]}(\v)\right. \\
 &\hskip 5cm \left. \times \partial_{\v^j}\tilde{A}^{-}(z_{1}^{+}, \v) \U_{[z_{1}^{+},-\frac{L^{+}}{2}]}(\v) \U^{\dagger}(\w) 
    \Big]\right\rangle \nn.
\end{align}
By following the same steps and using the two-point correlator given in Eq.~\eqref{eq:2pts_ev}, one gets 
\begin{align}
d^{(2)}(\v_{*} , \w)\big|_{\rm (ii)} 
    &= -g^{4} \, C_{F}^{2} \, \mu^{4} \, 
    d(\v, \w) \, \big[ \partial_{\v^{j}}G^{--}(\v -\w) \big] \, G^{j-}(\v - \w) \int_{v^{+}, z_{1}^{+}} \theta(v^{+}-z_{1}^{+}).
\end{align}
Finally, performing the integrations over the longitudinal coordinates, using the relation between $G^{--}(\r)$ and $G^{j-}(\r)$ given in Eq.~\eqref{relation_G--_Gj-} and the definition of the saturation momenta in Eq.~\eqref{eq: def Qs}, one gets 
\begin{align}
d^{(2)}(\v_{*} , \w)\big|_{\rm (ii)} 
    &= -4i \,  \pi^{2} \, Q_{s}^{4} \,  P_{q}^{-} \, \big[G^{j-}(\v - \w)\big]^{2} \ {d(\v,\w)},
\end{align}
which reads 
\begin{align}
\label{dip_2_ii_GM}
d^{(2)}(\v_{*} , \w)\big|_{\rm (ii)}  = {\frac{1}{4}}  \frac{i}{4} \frac{Q_{s}^{4}}{P_{q}^{-}} \, 
(\v - \w)^{2} \,  \ln^{2}\bigg( \frac{1}{|\v -\w|\Lam}\bigg) \,  \text{exp}\bigg\{  - \frac{Q_{s}^{2}}{4} (\v - \w)^{2} \text{ln}\frac{1}{|\v -\w|\Lam}\bigg\} 
\end{align}
in the small dipole approximation. 

The third term in the decomposition in Eq.~\eqref{d_2_decomp} reads 
\begin{align}
\label{def:d2_ii}
d^{(2)}(\v_{*} , \w)\big|_{\rm (iii)} &= \frac{1}{N_{c}} \int_{v^{+}} \Braket{\text{Tr} \big[ \U_{[\frac{L^+}{2},v^+]}(\v)  \overleftarrow{\partial}_{\v^{j}}\tilde{A}^{j}(v^{+}, \v)  \U_{[v^+, -\frac{L^+}{2}]}(\v) \U^{\dagger}(\w) \big]  } \, .
\end{align}
By following the same steps as in the evaluation of $d^{(2)}(\v_{*} , \w)\big|_{\rm (ii)}$, one can easily obtain
\begin{align}
\label{def:d2_iii}
 d^{(2)}(\v_{*} , \w)\big|_{\rm (iii)}   = - {\frac{1}{4}} \frac{i}{4} \, \frac{Q_{s}^{4} }{P_{q}^{-}} \, (\v - \w)^{2} \, \ln^{2}\bigg( \frac{1}{|\v -\w|\Lam}\bigg) \, 
 \text{exp}\bigg\{  - \frac{Q_{s}^{2}}{4} (\v - \w)^{2} \text{ln}\frac{1}{|\v -\w|\Lam}\bigg\}. 
\end{align}
Comparing Eqs. \eqref{def:d2_ii} and \eqref{def:d2_iii}, one immediately realizes that 
\begin{align}
\label{vanishing_ii+iii}
d^{(2)}(\v_{*} , \w)\big|_{\rm (ii
)}+d^{(2)}(\v_{*} , \w)\big|_{\rm (iii)}=0 \, . 
\end{align}
%
%
%
%
%
%

Finally, using Eq.~\eqref{decomp_4}, the last term in the decomposition of the NEik dipole of type 2 can be written as  
\begin{align}
\label{dip_2_iv}
d^{(2)}(\v_{*} , \w)\big|_{\rm (iv)}
     &= -\frac{1}{N_{c}} \int_{v^{+}} \Braket{\text{Tr} \big[ \U_{[\frac{L^+}{2},v^+]}(\v) \tilde{A}^{j}(v^{+}, \v)   \tilde{A}^{j}(v^{+}, \v)  \U_{[v^+, -\frac{L^+}{2}]}(\v) \U^{\dagger}(\w) \big]  } .
\end{align}
When evaluating this operator within the Gaussian model presented in Eq~ \eqref{eq:2pts_ev}, the transverse gauge fields in Eq. \eqref{dip_2_iv} can only be contracted with themselves since they are localized at the same longitudinal position $v^+$, and thus one gets form Eq.~\eqref{eq:2pts_ev} an ill-defined factor $\delta(v^+\!-\!v^+)$. From dimensional analysis, we interpret this factor as $\delta(v^{+}- v^{+}) =  {\cal C}/L^{+}$, where ${\cal C}$ is a dimensionless constant of order one, which can be considered as a model parameter.
In such a way, one finds
%
\begin{align}
\label{dip_2_cont4}
d^{(2)}(\v_{*} , \w)\big|_{\rm (iv)} = g^{2} \, C_{F}\, \mu^{2} \, L^{+} \, \frac{{\cal C}}{ L^{+}} \, d(\v, \w) \, G^{jj}(0) = 2\pi \,  Q_{s}^{2} \, \frac{{\cal C}}{ L^{+}} \, d(\v, \w) \, G^{jj}(0)\, .
\end{align}
 Moreover, as before we also used the definition of the saturation momenta given in Eq.~\eqref{eq: def Qs}. With the definition of the function $G^{ij}(r)$ given in Eq.~\eqref{eq: def Gij}, one can readily write 
\begin{align}
    G^{jj}(\r) = {2} \frac{1}{(2P_{q}^{-})^{2}} \int_{\P} e^{i \P \cdot \r} \frac{1}{\P^{2}}
\end{align}
which, at $\r=0$, reads 
\begin{align}
 \label{Gjj_zero}
    G^{jj}(0) =\frac{1}{2\pi}\frac{1}{2(P_{q}^{-})^{2}} \ln\frac{\Lambda_{\text{UV}}}{\Lam},
\end{align}
where the cutoff $\Lambda_{\text{UV}}$ is introduced to regulate the ultraviolet divergence. Finally, using Eq.~\eqref{Gjj_zero}  for function $G^{jj}(0)$ and Eq.~\eqref{def:dip_Gaussian_model} for the eikonal dipole in Gaussian model and taking the small dipole limit, the fourth term in the decomposition of the NEik dipole of type 2 can be expressed as 
\begin{align}\label{eq:dipole_tupe_2_iv}
d^{(2)}(\v_{*} , \w)\big|_{\rm (iv)} = \frac{{\cal C}}{2 L^+}\, \frac{Q_{s}^{2}}{(P_{q}^{-})^{2}}\,  \ln\frac{\Lambda_{\text{UV}}}{\Lam} \,  \text{exp}\bigg\{  - \frac{Q_{s}^{2}}{4} \,  (\v - \w)^{2} \,   \text{ln}\frac{1}{|\v -\w|\Lam}\bigg\}. 
\end{align}

To sum up, substituting Eqs.~\eqref{dip_2_i_GM},~\eqref{dip_2_ii_GM},~\eqref{def:d2_iii} and~\eqref{eq:dipole_tupe_2_iv} into Eq. \eqref{d_2_decomp}, NEik dipole of type 2 within the Gaussian approximation reads  
%
\begin{align}
\label{d2_GM_fin}
    d^{(2)}(\v_{*}, \w) &= \bigg[  \frac{1}{24} \, Q_s^4 \, L^+\, (\v - \w)^{2} \, \ln^{2} \bigg( \frac{1}{|\v - \w| \Lam}\bigg) + \frac{{\cal C}}{2L^+}\, \frac{Q_{s}^{2}}{(P_{q}^{-})^{2}}\,  \ln\frac{\Lambda_{\text{UV}}}{\Lam}\bigg]  \nn \\
   & \hspace{8cm}
   \times  \text{exp}\bigg\{  - \frac{Q_{s}^{2}}{4} \,  (\v - \w)^{2} \, \text{ln}\frac{1}{|\v -\w|\Lam}\bigg\}. 
\end{align}

A comment is in order before ending the discussion for the NEik dipole of type 2 within the Gaussian model.  The observation in Eq.~\eqref{vanishing_ii+iii} can be extended beyond the evaluation of NEik dipole of type 2. Taking into account their explicit forms given in Eq.~\eqref{def:g--} and~\eqref{def:gi-}, the relation between the functions $G^{--}(\r)$ and $G^{j-}(\r)$ presented in Eq.~\eqref{relation_G--_Gj-} implies that each factor of  $\partial^j \tilde{A}^-$ can be substituted by $-2iP^-_q \tilde{A}^j$ within our model. Consequently, one readily sees that terms {\rm (ii)} and {\rm (iii)}, given in Eqs.~\eqref{decomp_2} and~\eqref{decomp_3}, respectively, cancel each other irrespective of the choice of the operator $\hat \Omega$. Therefore, only terms {\rm (i)} and {\rm (iv)}, given in Eqs.~\eqref{decomp_1} and~\eqref{decomp_4},  contribute to the evaluation of any NEik operator of type 2 in our Gaussian model. 
\\
\paragraph{\underline{NEik type-2 quadrupole operator}.} We now examine the NEik quadrupole operator of type 2, which can be obtained by taking ${\hat{\Omega} \rightarrow \U^\dagger(\w)\U(\w')\U^\dagger(\v')}$ in Eq.~\eqref{def:OP2_first}  and reads 
\begin{equation}
\label{def:NEik_quad_2_gen}
    Q^{(2)}(\v_{*},\w, \w', \v') = \frac{1}{N_c} \left\langle \text{Tr}\,  \hat{\mathcal{O}}^{(2)}\big|_{\hat{\Omega} \rightarrow \U^\dagger(\w)\U(\w')\U^\dagger(\v')} \right\rangle \, .
\end{equation}
One can use the decomposition introduced in Eqs.~\eqref{decomp_1},~\eqref{decomp_2},~\eqref{decomp_3} and~\eqref{decomp_4} to write the NEik quadrupole operator of type 2 as 
\begin{align}
\label{eq:quad_type_2_Decomp}
 Q^{(2)}(\v_{*},\w, \w', \v')&= Q^{(2)}(\v_{*},\w, \w', \v')\big|_{\rm (i)}+Q^{(2)}(\v_{*},\w, \w', \v')\big|_{\rm (ii)}\nn \\
 & \hspace{6cm}+Q^{(2)}(\v_{*},\w, \w', \v')\big|_{\rm (iii)}+Q^{(2)}(\v_{*},\w, \w', \v')\big|_{\rm (iv)}\ .
\end{align}

As discussed previously, the second and third terms cancel each other, i.e.,
\begin{align}
\label{quad_2+3}
   Q^{(2)}(\v_{*},\w, \w', \v')\big|_{\rm (ii)}+ Q^{(2)}(\v_{*},\w, \w', \v')\big|_{\rm (iii)}=0,
\end{align}
analogous to the case of the NEik dipole operator of type 2 as shown in Eq.~\eqref{vanishing_ii+iii}.

Now, using the identity defined in Eq.~\eqref{eq: def c and identity} (or more explicitly in Eq.~\eqref{eq:id_quadr}), and the eikonal quadrupole defined in Eq.~\eqref{eq:calQ_def}, we can write the quadrupole operator of type 2 in the basis $\{|e_1\rangle,|e_2\rangle\}$. When written in terms of the operator $\mathcal{Q}_{[x^+,y^+]}$ which is composed of four fundamental Wilson lines with opened color indices and defined in Eq.~\eqref{eq:calQ_def}, the first term in the decomposition can be written as 
\begin{align}
Q^{(2)}(\v_{*},\w, \w', \v')\big|_{\rm (i)} &= \frac{1}{N_c} \int_{w_2^+,w_1^+} (w^+_2-w^+_1)
\, \theta(w^+_2-w^+_1) \, \langle e_2 | \mathcal{Q}_{[\frac{L^+}    {2},w_2^+]} \notag \\
    & \times \Big\langle  \big\{ \partial_{\v^j}\tilde{A}^-(w^+_2,\v_*) \big\} \otimes \sum_{\y_2 = \{\w,\w',\v'\}} (\pm)_{\y_2} \tilde{A}^-(w^+_2,\y_2) \Big\rangle
    \mathcal{Q}_{[w_2^+,w_1^+]} \nn\\
    &\times \Big\langle  \big\{ \partial_{\v^j}\tilde{A}^-(w^+_1,\v_*) \big\} \otimes \sum_{\y_1 = \{\w,\w',\v'\}} (\pm)_{\y_1} \tilde{A}^-(w^+_1,\y_1) \Big\rangle
    \mathcal{Q}_{[w_1^+,-\frac{L^+}{2}]} |e_1\rangle,
\end{align}
%
where $(\pm)_\y$ denotes the sign for a quark or an antiquark being at the transverse location $\y$.
Using the rotation matrix $R$ for the eikonal quadrupole defined in Eq.~\eqref{eq: eigen system of F_2} and using the generic  $2-point$ function derived previously and given in Eq.~\eqref{eq:2p_Neik_op}, we can write the first term in decomposition of the NEik quadrupole of type 2 as  
%
\begin{align}
\label{eq:non-tad_Neik_quad_type_2}
   Q^{(2)}(\v_{*},\w, \w', \v')\big|_{\rm (i)} &=\frac{1}{N_c} \int_{w^+_2,w^+_1} 
   (w_2^+-w_1^+)\, \theta(w_2^+-w_1^+) \, {\mu^2(w_2^+)} \, {\mu^2(w_1^+)} \notag \\
    &\times
    \begin{pmatrix}
        N_c & N_c^2
    \end{pmatrix} \cdot \left( \mathcal{T}_\mathcal{Q} \cdot R S_{[\frac{L^+}{2},w_2^+]} R^{-1} F^{NEik,j}_{w_2^+} R S_{[w_2^+,w_1^+]} R^{-1} F^{NEik,j}_{w_1^+} R S_{[w_1^+,-\frac{L^+}{2}]} R\right) \cdot
    \begin{pmatrix}
        1 \\ 0
    \end{pmatrix},
\end{align}
where we used the properties of the basis vectors given in Eq.~\eqref{eq:left_right_quad_overlaps} to recast it into the left and right vectors.
The eikonal tadpole contribution $\mathcal{T}_{\mathcal{Q}}$ is given in Eq.~\eqref{eq: def tadpole quadrupole} and it is proportional to the identity. 
The eikonal non-tadpole interactions are implemented by
\begin{equation}
    S_{[x^+,y^+]} = \text{diag} \left( \exp\left\{ \lambda_+\,\int_{x^+}^{y^+}dw^+ \mu^2(w^+)\right\}, \exp\left\{\lambda_-\,\ \int_{x^+}^{y^+}dw^+ \mu^2(w^+)\right\}\right),
\end{equation}
with the eigenvalues $\lambda_\pm$ provided in Eq.~\eqref{eq: eigen system of F_1}.
The NEik interactions are encoded via double insertion of the NEik interaction matrix $F^{NEik}$ that satisfies
\begin{equation}
    \left(F^{NEik,j}_{w^+_{1/2}}\right)_{ij} 
    = \left(V^{NEik,j}_{w^+_{1/2}} c^{-1}\right)_{ij} \ ,
\end{equation}
where the matrix $V^{NEik}$ is defined in Eq.~\eqref{eq:app_VNeik}. Its components are given in terms of $\alpha$, $\beta$ and $\gamma$ that are defined as 
\begin{align}
    \alpha = - {g^2} \partial_{\v^j}G^{--}(\v_*,\w),\quad
    \beta  = + {g^2}\partial_{\v^j}G^{--}(\v_*,\w'), \quad
    \gamma = - {g^2}\partial_{\v^j}G^{--}(\v_*,\v').
\end{align}
As in the preceding cases, functions $G^{\mu\nu}$ are determined from the two-field  expectation value in the Gaussian model, Eq.~\eqref{eq:2pts_ev}.

The fourth term in the decomposition of the NEik quadrupole of type 2 is obtained by substituting the term {\rm (iv)}, given in Eq.~\eqref{decomp_4}, into Eq.~\eqref{def:NEik_quad_2_gen} and taking ${\hat{\Omega} \rightarrow \U^\dagger(\w)\U(\w')\U^\dagger(\v')}$. As explained in the evaluation of $d^{(2)}(\v_{*} , \w)\big|_{\rm (iv)}$, the transverse fields are at the same longitudinal position and contracted within themselves. Therefore, the color structure factorizes from the rest as shown in Eq.~\eqref{dip_2_cont4}.  Following the same steps, one can readily show that eikonal quadrupole factorizes in the fourth term of the NEik quadrupole of type 2 and it can be written as   
%
\begin{align}\label{eq:tad_Neik_quad_type_2}
Q^{(2)}(\v_{*},\w, \w', \v')\big|_{\rm (iv)}  =  \frac{{\cal C}}{2L^+}\, \frac{Q_{s}^{2}}{(P_{q}^{-})^{2}} \, \ln\frac{\Lambda_{\text{UV}}}{\Lam} \,  Q(\v,\w, \w',\v'),
\end{align}
with $Q(\v,\w, \w',\v')$ being the eikonal quadrupole that is defined in Eq.~\eqref{eq:quad_T_N} with the tadpole and non-tadpole contributions are given in Eqs.~\eqref{eq: def tadpole quadrupole} and~\eqref{eq:quad_N_final}, respectively.  

To sum up, NEik quadrupole operator of type 2 is given by Eq.~\eqref{eq:quad_type_2_Decomp} with Eq~ \eqref{eq:non-tad_Neik_quad_type_2} for the first term in the decomposition (which governs the NEik interactions), Eq.~\eqref{eq:tad_Neik_quad_type_2} for the fourth term in the decomposition (which corresponds to NEik tadpole contribution), and Eq.~\eqref{quad_2+3} for the sum of the second and the third contributions in the decomposition (which vanishes identically in the Gaussian model).  


\subsection{Next-to-eikonal operators of type 3}
\label{subsec:type_3}
The NEik operators of type 3 can be written in a generic form as  
\begin{align}
\label{def:OP3_first}
    \hat{\mathcal{O}}^{(3)}_{ij} (\v_{*}) = 
    \int dv^+ \left\{\U_{[\frac{L^+}{2},v^+]}(\v)\, g\, F_{ij}(v^+,\v)\,  
    \U_{[v^+,-\frac{L^+}{2}]}(\v) 
    \right\} \otimes \hat{\Omega},
\end{align}
with the operator $\hat{\Omega}$ is defined in Eq.~\eqref{def:Omega_Op}. Using the explicit expression of the field strength tensor given in Eq.~\eqref{F_ij}, the generic form of NEik operators of type 3 reads 
\begin{align}
\label{def:OP3_second}
    \hat{\mathcal{O}}^{(3)}_{ij} (\v_{*}) = 
    \int dv^+ \left\{\U_{[\frac{L^+}{2},v^+]}(\v)
    \bigg(  \partial_{\v^{i}} A_{j}(v^+, \v) - \partial_{\v^{j}}A_{i}(v^+, \v) + ig\big[A_{i}(v^+, \v), A_{j}(v^+, \v)\big]\bigg)
    \U_{[v^+,-\frac{L^+}{2}]}(\v) 
    \right\} \otimes \hat{\Omega}.
\end{align}
The first two terms inside the parenthesis on the right hand side of this equation correspond to the Abelian part of the decoration of type 3 which is antisymmetric under the exchange of $i\leftrightarrow j$. On the other hand, as discussed at the end of the evaluation of NEik dipole of type 2, the relation between the functions $G^{--}(\r)$ and $G^{j-}(\r)$ presented in Eq.~\eqref{relation_G--_Gj-} suggests that, in our Gaussian model, each factor of $2iP^-_q {A}^j$ can be approximated by $\partial_j {A}^-$. Using this approximation, one can write the Abelian part of the NEik operators of type 3 as 
\begin{align}
\label{def:OP3_Abelian}
    \hat{\mathcal{O}}^{(3)}_{ij} (\v_{*})\Big|_{\rm Abelian} = \frac{i}{2}\frac{1}{P_q^-}
    \int dv^+ \left\{\U_{[\frac{L^+}{2},v^+]}(\v)
    \bigg(  \partial_{\v^{i}} \partial_{\v^{j}} A^-(v^+, \v) - \partial_{\v^{j}} \partial_{\v^{i}}A^{-}(v^+, \v) 
    \bigg)
    \U_{[v^+,-\frac{L^+}{2}]}(\v) 
    \right\} \otimes \hat{\Omega},
\end{align}
which vanishes. 

On the other hand, the non-Abelian part of the NEik operators of type 3 reads 
\begin{align}
\label{def:OP3_non_Abelian}
    \hat{\mathcal{O}}^{(3)}_{ij} (\v_{*})\Big|_{\rm non-Abelian} = 
    \int dv^+ \left\{\U_{[\frac{L^+}{2},v^+]}(\v)
    \bigg(  
     ig\big[A_{i}(v^+, \v), A_{j}(v^+, \v)\big]\bigg)
    \U_{[v^+,-\frac{L^+}{2}]}(\v) 
    \right\} \otimes \hat{\Omega}.
\end{align}
The structure of this decoration closely resembles the {\rm (iv)} term in the decomposition of the type-2 NEik operators presented in Eq.~\eqref{decomp_4}. In evaluating the non-Abelian contribution of operator $\hat{\mathcal{O}}^{(3)}_{ij}$ within the Gaussian model defined in Eq.~\eqref{eq:2pts_ev}, the transverse gauge fields appearing in Eq.~\eqref{def:OP3_non_Abelian} may contract only with one another, as they are localized at the same longitudinal coordinate $v^+$. According to  Eq.~\eqref{eq:2pts_ev}, the corresponding  two-point correlator of the transverse fields is proportional to $\delta^{ij}$. The total non-Abelian contribution to the type-3 NEik operator is, however, antisymmetric under the exchange of $i\leftrightarrow j$, and therefore vanishes identically in the Gaussian model. Hence, we conclude that type-3 NEik operators do not contribute within this framework. 

\subsection{Contribution from the three-point correlation}


Apart from the three types of NEik operators discussed above, in the NEik cross section of DIS dijet production off transversely polarized photons presented in Eqs.~\eqref{eq: Neik cross section} and~\eqref{NEik_X_Section_T},  there are extra Wilson line operators of the form 
%
\begin{align}
\label{Op_3Point_def}
\mathcal{O}^{\rm 3- point}_{[\frac{L^+}{2},-\frac{L^+}{2}]}(\w',\v',\z)&=
\int_{-\frac{L^{+}}{2}}^{\frac{L^{+}}{2}} dz^{+}
\bigg\langle\frac{1}{N_{c}} \text{Tr} \bigg[\Big(\U(\w') \U^{\dagger}(\v') -1\Big) 
\nn \\
& \hspace{4cm}
\times 
\Big( \U_{\big[ \frac{L^{+}}{2}, z^{+}\big]} (\z) \overleftrightarrow{D_{\z^{j}} }(z^{+}) \U^{\dagger}_{\big[ \frac{L^{+}}{2}, z^{+}\big]}(\z) - \frac{1}{2} \U(\z) \overleftrightarrow{\partial_{\z^{j}}} \U^{\dagger}(\z)
   \Big)\bigg] \bigg\rangle   
\end{align}
which, for convenience, can be written as the sum of two terms:
\begin{align}
\label{Op_3point_Sum}
\mathcal{O}^{\rm 3- point}_{[\frac{L^+}{2},-\frac{L^+}{2}]}(\w',\v',\z) = 
\mathcal{O}^{\rm 3- point}_{[\frac{L^+}{2},-\frac{L^+}{2}]}(\w',\v',\z)\bigg|_{\overleftrightarrow{D}}+ 
\mathcal{O}^{\rm 3- point}_{[\frac{L^+}{2},-\frac{L^+}{2}]}(\w',\v',\z)\bigg|_{\overleftrightarrow{\partial}} \, . 
\end{align}
The first operator involves the left-right covariant derivative $\overleftrightarrow{D}_j$ and it reads 
\begin{align}
\label{op_3point_Cov}
 \mathcal{O}^{\rm 3- point}_{[\frac{L^+}{2},-\frac{L^+}{2}]}(\w',\v',\z)\bigg|_{\overleftrightarrow{D}}  
    &\equiv \int_{-\frac{L^{+}}{2}}^{\frac{L^{+}}{2}} dz^{+}
\frac{1}{N_{c}}\bigg\langle \text{Tr}\bigg[ \Big(\U(\w') \U^{\dagger}(\v') -1\Big) \Big( \U_{\big[ \frac{L^{+}}{2}, z^{+}\big]} (\z) \overleftrightarrow{D_{\z^{j}} }(z^{+}) \U^{\dagger}_{\big[ \frac{L^{+}}{2}, z^{+}\big]}(\z) \Big)\bigg] \bigg\rangle \nn \\
    &= \frac{1}{N_c}\int_{-\frac{L^+}{2}}^{\frac{L^+}{2}} dz^+\ \bigg\langle 
    \text{Tr}\bigg[ \Big(\U(\w')\U^\dagger(\v') - 1\Big)\,
     \nn \\
    & \hspace{2cm}
    \times 
    \U_{[\frac{L^+}{2},z^+]}(\z)
    \Big\{
    2\tilde{A}^j(z^+,\z)-2\Big(z^+ + \frac{L^+}{2}\Big)\partial_j \tilde{A}^-(z^+,\z)
    \Big\} \U_{[\frac{L^+}{2},z^+]}^\dagger(\z)
    \bigg]\bigg\rangle,
\end{align}
where we have explicitly computed the action of the covariant derivate on the semi-infinite Wilson lines and used the shorthand notation $\tilde{A}^\mu = - ig A^\mu$ as in the preceding sections. The second operator involves left-right partial derivative $\overleftrightarrow{\partial}_j$ and reads 
\begin{align}
\label{op_3point_par}
\mathcal{O}^{\rm 3- point}_{[\frac{L^+}{2},-\frac{L^+}{2}]}(\w',\v',\z)\bigg|_{\overleftrightarrow{\partial}}
    &\equiv \int_{-\frac{L^{+}}{2}}^{\frac{L^{+}}{2}} dz^{+}
\frac{1}{N_{c}} \bigg\langle \text{Tr} \bigg[\Big(\U(\w') \U^{\dagger}(\v') -1\Big) \Big( - \frac{1}{2} \U(\z) \overleftrightarrow{\partial_{\z^{j}}} \U^{\dagger}(\z)   \Big)\bigg]\bigg\rangle \nn 
\\
    &= \frac{1}{N_c}\int_{-\frac{L^+}{2}}^{\frac{L^+}{2}} dy^+\ \bigg\langle \text{Tr}\bigg[ \Big(\U(\w')\U^\dagger(\v') - 1\Big)\,
    \U_{[\frac{L^+}{2},y^+]}(\z) \Big[ 
    L^+ \partial_j \tilde{A}^-(y^+,\z)
    \Big] \U_{[\frac{L^+}{2},y^+]}^\dagger(\z)
    \bigg]\bigg\rangle.
\end{align}
In the first equality, the integral over $dz^+$ simply introduces a factor $L^+$. In the second equality, the longitudinal point corresponding to the action of the partial derivative is labeled by $y^+$.
Substituting Eqs.~\eqref{op_3point_Cov} and~\eqref{op_3point_par} into Eq.~\eqref{Op_3point_Sum}, one gets the NEik 3-point operator as 
\begin{align}
\label{Op_3_point_sum_simpf}
\mathcal{O}^{\rm 3- point}_{[\frac{L^+}{2},-\frac{L^+}{2}]}(\w',\v',\z)
   & = \frac{1}{N_c}\int_{-\frac{L^+}{2}}^{\frac{L^+}{2}} dz^+\ \bigg\langle \text{Tr}\bigg[ \Big(\U(\w')\U^\dagger(\v') - 1\Big)\nn \\
    & \hspace{3cm}
    \times \; 
    \U_{[\frac{L^+}{2},z^+]}(\z) \Big[ 
    2\tilde{A}^j(z^+,\z)-2z^+\partial_j \tilde{A}^-(z^+,\z)
    \Big] \U_{[\frac{L^+}{2},z^+]}^\dagger(\z)
    \bigg]\bigg\rangle \\
     & = \frac{1}{N_c}\int_{-\frac{L^+}{2}}^{\frac{L^+}{2}} dz^+\ \bigg\langle \text{Tr}\bigg[ \U(\w')\U^\dagger(\v')
    \U_{[\frac{L^+}{2},z^+]}(\z) \Big[ 
    2\tilde{A}^j(z^+,\z)-2z^+\partial_j \tilde{A}^-(z^+,\z)
    \Big] \U_{[\frac{L^+}{2},z^+]}^\dagger(\z)
    \bigg]\bigg\rangle
\nn \\
 & - \frac{1}{N_c}\int_{-\frac{L^+}{2}}^{\frac{L^+}{2}} dz^+\ \bigg\langle \text{Tr}\bigg[ 
    \U_{[\frac{L^+}{2},z^+]}(\z) \Big[ 
    2\tilde{A}^j(z^+,\z)-2z^+\partial_j \tilde{A}^-(z^+,\z)
    \Big] \U_{[\frac{L^+}{2},z^+]}^\dagger(\z)
    \bigg]\bigg\rangle .
    \label{loc_3-point}
\end{align}

Note that, in the Gaussian approximation, the eikonal field $\tilde{A}^-$ cannot interact with a singlet point-like system. Therefore, the insertion of the operator
\begin{eqnarray}
    \mathcal{O}(z^+,\z) = \left[ 
    2\tilde{A}^j(z^+,\z)-2z^+\partial^j \tilde{A}^-(z^+,\z)
    \right]
\end{eqnarray}
between the two semi-infinite Wilson lines can only be projected onto the adjoint subspace. The term in the second line of Eq.~\eqref{Op_3_point_sum_simpf} can be written as 
\begin{align}
\label{Color_Id_Oa}
    \U_{\big[\frac{L^+}{2},z^+\big]}(\z) \,  \mathcal{O}(z^+,\z) \, \U_{\big[\frac{L^+}{2},z^+\big]}^\dagger(\z) &= \U_{\big[\frac{L^+}{2},z^+\big]}(\z) \, T^a \,  \U_{\big[\frac{L^+}{2},z^+\big]}^\dagger(\z) \ \mathcal{O}^a(z^+,\z)
    = T^b \,  U^{ba}_{\big[\frac{L^+}{2},z^+\big]}(\z) \, \mathcal{O}^a(z^+,\z),
\end{align}
where we introduced the adjoint Wilson line $U(\z)$ in the second line.
In addition, the second term in the NEik 3-point operator given in Eq.~\eqref{loc_3-point} has a single transverse coordinate dependence and therefore can only give a tadpole contribution. However, this contribution vanishes since the adjoint generator $T_{adj}^a = -if^{abc}$ is traceless. Therefore, the only non-vanishing contribution to the NEik 3-point operator is presented in the first line of Eq.~\eqref{loc_3-point} which, in the birdtrack approach, can be represented as  
%
\begin{center}
\begin{tikzpicture}
    \draw[thick,->] (-2,1) -- (-.5,1);
    \draw[thick,-<] (-2,.5) -- (.5,.5);
    \draw[thick] (-2,1) -- (2,1) to[out=0,in=0] (2,.5) -- (-2,.5) to[in=180,out=180] (-2,1);
    \draw[gluon] (0,0) -- (2.15,0) to[out=0,in=0] (2.15,.75);
    \draw[fill=Blue!50!white] (0,0) circle (0.1);
    \node at (-3,0) {$\z$};
    \node at (-3,1) {$\w'$};
    \node at (-3,.5) {$\v'$};
    \draw[cyan,dashed] (0,1.5) -- (0,-.5) node[below] {$z^+$};
    \node[below] at (-2,-.5) {$-\infty$};
    \node[below] at (2,-.5) {$+\infty$};
\end{tikzpicture},
\end{center}
where the blue bullet represents the insertion of the operator $\mathcal{O}^a(z^+,\z)$ given in Eq.~\eqref{Color_Id_Oa}. Since there is only one single singlet in $V\otimes \bar{V}$ (resp. $V \otimes \bar{V} \otimes A$)\footnote{Let us recall that $V$ denotes the fundamental vector space, $\bar{V}$ the corresponding dual vector space, and $A$ the adjoint vector space.}, we can directly project onto this subspace before and after the longitudinal insertion position $z^+$ to write the NEik 3-point operator as
\begin{align}
\label{Op_3_point_fact}
\mathcal{O}^{\rm 3- point}_{[\frac{L^+}{2},-\frac{L^+}{2}]}(\w',\v',\z)   = & 
    \int dz^+\ 
    \frac{1}{N_c}
    \Big\langle 
    \text{Tr} \Big[ \U_{[z^+,-\frac{L^+}{2}]}(\w') \, \U_{[z^+,-\frac{L^+}{2}]}^\dagger(\v') \Big] \Big\rangle 
    \nn \\
    & \hspace{0.7cm}
    \times \frac{1}{N_c} 
    \Big\langle \frac{C_F\, N_c}{N_c^2-1}\, \mathcal{O}^a(z^+,\z) 
    \, 
    \big[ \tilde{A}^{a-}(z^+,\w')-\tilde{A}^{a-}(z^+,\v')\big]  \Big\rangle 
    \nn \\
    & \hspace{0.7cm}
    \times \frac{2}{N_c^2-1} 
    \Big\langle 
    \text{Tr} \Big[ T^b \U_{[\frac{L^+}{2},z^+]}(\w') T^a \U_{[\frac{L^+}{2},z^+]}^\dagger(\v') \Big] U^{ba}_{[\frac{L^+}{2},z^+]}(\z)    
    \Big\rangle.
\end{align}
Before proceeding further with the discussion of the NEik 3-point operator, we first comment  on the color factors appearing in Eq.~\eqref{Op_3_point_fact}. The factor of $1/N_c$ in the first line originates from the overall normalization of the operator, already included in the definition of the NEik 3-point operator in Eq.~\eqref{Op_3Point_def}. In the second line, the color factor $1/N_c$ arises from the $q\bar q$ singlet projection, while the factor $1/(N_c^2-1)$ compensates the $\delta^{aa}$ which comes from the expectation value $\langle \mathcal{O}^a \tilde{A}^a\rangle$. The remaining factor, $C_FN_c$, follows from the trace over two fundamental generators, $tr(T^aT^a)$. In the last line of Eq.~\eqref{Op_3_point_fact}, the factor $2/(N_c^2-1)$ is associated with the $q\bar q g$ singlet projection. All in all, the NEik 3-point operator can be written as 
\begin{align}
\label{3point_Fact_def}
\mathcal{O}^{\rm 3- point}_{[\frac{L^+}{2},-\frac{L^+}{2}]}(\w',\v',\z)   = & 
\int dz^+ 
\, C^{(3)}_{\big[\frac{L^+}{2},z^+\big]}(\w',\v';\z)
\, d_{\big[z^+,-\frac{L^+}{2}\big]}(\w',\v') \nn \\
   & \hspace{2cm} 
   \times \Big\langle \frac{C_F}{N_c^2-1}\, \mathcal{O}^a(z^+,\z)\,  \big[\tilde{A}^{a-}(z^+,\w')-\tilde{A}^{a-}(z^+,\v')\big]  \Big\rangle,
\end{align}
where $d_{\big[z^+,-\frac{L^+}{2}\big]}(\w',\v')$ is the eikonal dipole defined in the range from $-L^+/2$ to the  point $z^+$ which reads 
\begin{align}
\label{semi_dip}
d_{\big[z^+,-\frac{L^+}{2}\big]}(\w',\v') =
 \frac{1}{N_c}
    \Big\langle 
    \text{Tr} \Big[ \U_{[z^+,-\frac{L^+}{2}]}(\w') \, \U_{[z^+,-\frac{L^+}{2}]}^\dagger(\v') \Big] \Big\rangle \, ,
\end{align}
and $C^{(3)}_{\big[\frac{L^+}{2},z^+\big]}(\w',\v';\z)$ is the eikonal 3-point correlation function which is defined from the longituidnal point $z^+$ to $L^+/2$ and it reads 
\begin{align}
C^{(3)}_{\big[\frac{L^+}{2},z^+\big]}(\w',\v';\z) = 
\frac{2}{N_c^2-1} 
    \Big\langle 
    \text{Tr} \Big[ T^b \U_{[\frac{L^+}{2},z^+]}(\w') T^a \U_{[\frac{L^+}{2},z^+]}^\dagger(\v') \Big] U^{ba}_{[\frac{L^+}{2},z^+]}(\z)    
    \Big\rangle \, .
\end{align}
Finally, the last line of Eq.~\eqref{3point_Fact_def} corresponds to the NEik interaction contribution to the 3-point function. 

Let us first consider the eikonal 3-point correlation in the Gaussian approximation. The invariance condition for a system of three partons reads 
\begin{equation}
    T_{\w'} + T_{\v'} + T_\z = 0,
\end{equation}
where we label by $T_i$ the generator in the representation of the parton at the corresponding coordinate $i=\{\w',\v',\z\}$.
Thus, the eikonal interaction reads\footnote{As an illustration, one can write $2T_{\w'}\otimes T_{\v'} = (T_{\w'} + T_{\v'})^2-T_{\w'}^2-T_{\v'}^2 = T^2_\z -T_{\w'}^2-T_{\v'}^2 = N_c - 2 C_F =\frac{1}{N_c}$.}
\begin{align}
\label{txt}
    &(T_{\w'}^a\otimes T_{\v'}^a)  w_{\w'\v'} + (T_{\w'}^a\otimes T_\z^a) w_{\w'\z} + (T_{\v'}^a \otimes T_\z^a) w_{\v'\z} 
    \longrightarrow \mathbb{1}\, \left\{\frac{1}{2N_c}w_{\w'\v'} -\frac{N_c}{2}(w_{\w'\z} +w_{\v'\z}) \right\} \equiv \mathbb{1}\, \Gamma^{(q\bar{q}g)}_{\w'\v'\z}\ .
\end{align}
Here, $w_{ij}$ is closely related to the kinematic factor $W_{ij}$ defined in Eq.~\eqref{eq:Wij_def}. The two are equivalent up to an overall sign and tadpole contributions. In particular, $W_{ij}$ incorporates the sign factor and therefore alternates accordingly, whereas the kinematic factor $w_{ij}$ introduced above is defined without the sign factor. Moreover, in the kinematic factor $W_{ij}$ tadpole contributions are excluded and computed separately, while $w_{ij}$ includes both tadpole and non-tadpole contributions. Finally, in Eq.~\eqref{txt}, $\mathbb{1}$ denotes the identity of the corresponding space. All in all, the $q\bar qg$ interaction function $\Gamma^{(q\bar{q}g)}_{\w'\v'\z}$ is given by 
\begin{align}
\label{gamma_3P}
    \Gamma^{(q\bar{q}g)}_{\w'\v'\z} = \frac{1}{2N_c} [G^{--}(0) - G^{--}(\w'-\v')] - \frac{N_c}{2}[G^{--}(0) - G^{--}(\w'-\z) ] - \frac{N_c}{2}[G^{--}(0) - G^{--}(\v'-\z)],
\end{align}
with the function $G^{\mu\nu}$ defined in Eq.~\eqref{eq:2pts_ev} and its eikonal component given in Eq.~\eqref{def:g--}.   
In the small dipole approximation, the combination that appears in all the terms above is computed in Eq.~\eqref{eq:eval_MV}. These results are in agreement with the eikonal $q\bar qg$ interaction computed in~\cite{Marquet:2025jdr} in the context of resummation of soft gluons for inclusive heavy-quark pair production.     

Because the eikonal interaction in the 3-point correlation function is proportional to the identity in color space, as shown in Eq.~\eqref{txt}, its Gaussian average is color-trivial due to the locality of its longitudinal coordinate dependence. After incorporating the transverse color charge density $\mu^2(z^+)$ and the appropriate coupling constant factor, the three point correlator readily resums the eikonal interactions into     
\begin{eqnarray}
\label{C3_non-hom}
    C^{(3)}_{[a^+,b^+]}(\w',\v';\z) = \exp\left({\int^{a^+}_{b^+} du^+ {g^2}\, \mu^2(u^+) \, \Gamma^{(q\bar{q}g)}_{\w\v\z}} \right),
\end{eqnarray}
with a generic support of the background fields extending from $b^+$ to $a^+$. In the limit of a homogeneous longitudinal nuclear profile, $\mu(u^+)$ can be replaced by a constant  $\mu$, which renders the longitudinal integral trivial, yielding the following expression for the eikonal 3-point correlation function:   
\begin{eqnarray}
\label{C3-hom_GM}
C^{(3)}_{[a^+,b^+]}(\w',\v';\z) = \exp \Big[ (a^+-b^+)\, {g^2} \, \mu^2 \, \Gamma^{(q\bar{q}g)}_{\w'\v'\z}\Big]\, .
\end{eqnarray}
The eikonal dipole defined in Eq. \eqref{semi_dip} can be evaluated straightforwardly within the Gaussian approximation by noting that the $q\bar qg$ interaction function $\Gamma^{(q\bar{q}g)}_{\x\y\z}$, given in Eq.~\eqref{gamma_3P}, is related to the $q\bar q$ interaction function $\Gamma^{(q\bar{q})}_{\y\z}$ through 
\begin{align}
\lim_{\x \rightarrow \y} \, \Gamma^{(q\bar{q}g)}_{\x\y\z} = \frac{N_c}{C_F} \, \Gamma^{(q\bar{q})}_{\y\z} \, . 
\end{align}
Its explicit expression reads 
\begin{align}
\label{Gamma_2P}
\Gamma^{(q\bar{q})}_{\w'\v'} \equiv -C_F [G^{--}(0) - G^{--}(\w'-\v')] \, . 
\end{align}
The reasoning and methodology employed to resum the eikonal intercations for the eikonal 3-point correlation function carry over directly to the dipole operator in Eq.~\eqref{semi_dip}. Applying the same steps leads to the following expression:
\begin{align}
\label{d2_non-hom}
 d_{[a^+,b^+]}(\w',\v') = \exp \left( \int_{b^+}^{a^+} du^+ \, {g^2}\, \mu^2(u^+) \, \Gamma^{(q\bar{q})}_{\w'\v'}\right),
\end{align}
with the $q\bar q$ interaction function $\Gamma^{(q\bar{q})}_{\y\z}$ given in Eq.~\eqref{Gamma_2P}. As before, for a homogeneous nuclear profile $\mu(u^+)\to \mu$ and the longitudinal integral can be performed trivially, yielding
\begin{align}
\label{d2-hom_GM}
 d_{[a^+,b^+]}(\w',\v') = \exp \Big[ (a^+-b^+)\, {g^2}\, \mu^2\, \Gamma^{(q\bar{q})}_{\w'\v'}\Big].
\end{align}
The NEik interaction contribution to the 3-point function can be computed using Eq.~\eqref{eq:2pts_ev} and the result reads 
\begin{align}
\label{NEik_int_3p}
    \frac{1}{N_c^2-1}\left\langle \mathcal{O}^a(z^+,\z) [\tilde{A}^{a-}(z^+,\w')-\tilde{A}^{a-}(z^+,\v')] \right\rangle = &- 2{g^2} \mu^2(z^+) \left[ G^{j-} - z^+\partial^jG^{--} \right]_{\w'\z} \notag \\
    &+2 {g^2}\mu^2(z^+) \left[ G^{j-} - z^+\partial^jG^{--} \right]_{\v'\z} \ ,
\end{align}
with the relevant components of the function $G^{\mu\nu}$ given in Eqs.~\eqref{def:g--} and~\eqref{def:gi-}. 
Finally, substituting Eq.~\eqref{NEik_int_3p} into Eq.~\eqref{3point_Fact_def}, the NEik 3-point operator can be expressed within the Gaussian model as 
\begin{align}
\label{3_point_non-hom_GM}
\mathcal{O}^{\rm 3- point}_{[\frac{L^+}{2},-\frac{L^+}{2}]}(\w',\v',\z) = 
&\int {dz^+} \, C^{(3)}_{\big[\frac{L^+}{2},z^+\big]}(\w',\v';\z)\;   
d_{\big[z^+,-\frac{L^+}{2}\big]}(\w',\v') 
\nn  \\
&\times \Big\{2 \, C_F \, {g^2} \, \mu^2(z^+)\Big[  \big( G^{j-} - z^+\partial^jG^{--} \big)_{\w'\z} - \big( G^{j-} - z^+\partial^jG^{--} \big)_{\v'\z} \Big]\Big\}
\end{align}
for a target whose color charge density depends on the longitudinal coordinate. Furthermore, the eikonal 3-point function and  the eikonal dipole with  generic longitudinal support are given in Eqs.~\eqref{C3_non-hom} and~\eqref{d2_non-hom}, respectively.
%
In the limit of a target with a homogeneous nuclear profile, $\mu^2(z^+)\to\mu^2$ in the second line of  Eq.~\eqref{3_point_non-hom_GM}, the eikonal 3-point function and the eikonal dipole are given by Eqs.~\eqref{C3-hom_GM} and~\eqref{d2-hom_GM}, respectively. In that case, the longitudinal integral over $z^+$ in Eq.~\eqref{3_point_non-hom_GM} takes the form
\begin{equation}
    \int^{\frac{1}{2}}_{-\frac{1}{2}} dz\ e^{(\frac{1}{2}-z)A} \{1,z \} e^{(z+\frac{1}{2})B},
\end{equation}
which is computed in Appendix~\ref{App:Longi_integrals}.

\section{DIS dijet production cross section at next-to-eikonal accuracy within a Gaussian model}
\label{sec:X_sec_in_Gaussian_model}
%
In this section, we use the eikonal and NEik dipole and quadrupole operators derived within our Gaussian model to express the DIS dijet production cross section in a form suitable for future numerical studies. Although our primary focus is the evaluation of NEik corrections to DIS dijet production, it is useful to briefly comment on the eikonal cross section as well, given that the corresponding eikonal dipole and quadrupole operators were also obtained within the Gaussian model in Section~\ref{sec:DIS_dijet_prod}.

The eikonal-order DIS dijet production cross section is provided in Eq.~\eqref{Eik_dijet_Xsec}, where function $C_\lambda$ is defined in Eqs.~\eqref{CL} and~\eqref{CT} for longitudinally and transversely polarized virtual photons, respectively. The cross section is expressed in terms of eikonal dipole and quadrupole operators, whose Gaussian-model forms are provided in Eqs.~\eqref{eq: def dipole} and~\eqref{eq:quad_T_N}. By performing the change of variable $(\x,\y)\to(\v,\w)$ or $(\w',\v')$ in Eq.~\eqref{eq: def dipole} for the dipole operators, and $(\x_1,\x_2,\x_3,\x_4)\to(\w',\v',\v,\w)$ in Eq.~\eqref{eq:quad_T_N} for the quadrupole operator\footnote{The eikonal quadrupole operator defined in Eq.~\eqref{eq:quad_T_N} is given in terms of tadpole contribution given in Eq.~\eqref{eq: def tadpole quadrupole} and a non-tadpole contribution given in Eq.~\eqref{eq:quad_N_final}. For the non-tadpole contribution, the eigensystem is provided in Eqs.~\eqref{eq: eigen system of F_1},~\eqref{eq: eigen system of F_2} and~\eqref{eq: eigen system of F_3} written in terms of the kinetic factors $W_{ij}$ given in Eq.~\eqref{eq:Wij_def} with the transverse coordinate dependence $\x_i$.}, the eikonal DIS dijet cross section can be cast into a form well suited for numerical implementation. 

Let us now turn to the NEik contributions to DIS dijet production cross section, given in Eq~ \eqref{eq: Neik cross section} together with Eq.~\eqref{NEik_X_Section_T}. We reorganize these expressions and rewrite the cross section as 
\begin{align}
\label{NEik_X_sec_Separated}
\frac{d\sigma^{\gamma^{*}_{\lambda}+ A \rightarrow q \bar q + X}}{d^{2}\k_{1}d^{2}\k_{2}d\eta_{1}d\eta_{2}} \bigg |_{\text{NEik}} 
&= 
\frac{d\sigma^{\gamma^{*}_{\lambda}+ A \rightarrow q \bar q + X}}{d^{2}\k_{1}d^{2}\k_{2}d\eta_{1}d\eta_{2}} \bigg |_{d_j^{(1)}} 
+
\frac{d\sigma^{\gamma^{*}_{\lambda}+ A \rightarrow q \bar q + X}}{d^{2}\k_{1}d^{2}\k_{2}d\eta_{1}d\eta_{2}} \bigg |_{d^{(2)}} 
+
\frac{d\sigma^{\gamma^{*}_{\lambda}+ A \rightarrow q \bar q + X}}{d^{2}\k_{1}d^{2}\k_{2}d\eta_{1}d\eta_{2}} \bigg |_{Q_j^{(1)}}
+
\frac{d\sigma^{\gamma^{*}_{\lambda}+ A \rightarrow q \bar q + X}}{d^{2}\k_{1}d^{2}\k_{2}d\eta_{1}d\eta_{2}} \bigg |_{Q^{(2)}} 
\nn \\
& 
+ \delta^{\lambda T}\bigg\{
\frac{d\sigma^{\gamma^{*}_{\lambda}+ A \rightarrow q \bar q + X}}{d^{2}\k_{1}d^{2}\k_{2}d\eta_{1}d\eta_{2}} \bigg |_{d_{ij}^{(3)}} 
+
\frac{d\sigma^{\gamma^{*}_{\lambda}+ A \rightarrow q \bar q + X}}{d^{2}\k_{1}d^{2}\k_{2}d\eta_{1}d\eta_{2}} \bigg |_{Q_{ij}^{(3)}}
+
\frac{d\sigma^{\gamma^{*}_{\lambda}+ A \rightarrow q \bar q + X}}{d^{2}\k_{1}d^{2}\k_{2}d\eta_{1}d\eta_{2}} \bigg |_{{\cal O}_j^{(\rm 3-point)}}\bigg\} ,
\end{align}
where $\delta^{\lambda T}$ is $1$ when $\lambda =T$ and $0$ when $\lambda=L$. Here, the contributions of the individual operators have been isolated and will be discussed separately in the following. 

The first contribution accounts for the NEik dipole operator of type-1 and it reads 
%

%
\begin{align}
  \frac{d\sigma^{\gamma^{*}_{\lambda}+ A \rightarrow q \bar q + X}}{d^{2}\k_{1}d^{2}\k_{2}d\eta_{1}d\eta_{2}} \bigg |_{d_j^{(1)}} 
  & = 
  \frac{1}{q^{+}} \text{Re} \int_{\v, \v', \w, \w'} e^{i\k_{1} \cdot (\mathbf{v'}-\mathbf{v}) + i \k_{2} \cdot (\w' - \w)} \C_{\lambda}(\w'-\v', \w - \v) \\ \nn
   & \times \Bigg\{ -\frac{1}{z_{1}} \bigg[ \frac{\k_{2}^{j} - \k_{1}^{j}}{2} +\frac{i}{2}\partial_{\w^j} \bigg]  d_{j}^{(1)}(\v_{*} , \w) 
    +\frac{1}{z_{2}} \bigg[ \frac{\k_{2}^{j} - \k_{1}^{j}}{2} -\frac{i}{2}\partial_{\v^j} \bigg]  d_{j}^{(1)}(\w_{*} , \v)^{\dagger} \bigg\}.
\end{align}
Noting that the NEik type-1 decorated dipole obeys 
\begin{align}
    d_{j}^{(1)}(\w_{*}, \v)^{\dagger} = d_{j}^{(1)}(\v_{*}, \w)
\end{align}
and using its expression given within the Gaussian model in Eq.~\eqref{type_1_dip_fin}, the contribution to the DIS dijet production cross section from $d_j^{(1)}$ can be written as 
\begin{align}
\label{Xsec_NEik_dip_1_fin}
  \frac{d\sigma^{\gamma^{*}_{\lambda}+ A \rightarrow q \bar q + X}}{d^{2}\k_{1}d^{2}\k_{2}d\eta_{1}d\eta_{2}} \bigg |_{d_j^{(1)}} 
  & = 
  \frac{1}{q^{+}} \text{Re} \int_{\v, \v', \w, \w'} e^{i\k_{1} \cdot (\mathbf{v'}-\mathbf{v}) + i \k_{2} \cdot (\w' - \w)} \C_{\lambda}(\w'-\v', \w - \v) \,  
  \\ \nn
   & \hspace{-1cm}
   \times 
  \frac{iQ_{s}^{2}}{P_{q}^{-}}
   \bigg[ -\frac{1}{z_{1}} 
   \bigg( \frac{\k_{2}^{j} - \k_{1}^{j}}{2} +\frac{i}{2}\partial_{\w^j} \bigg)
     +
    \frac{1}{z_{2}} 
    \bigg( \frac{\k_{2}^{j} - \k_{1}^{j}}{2} -\frac{i}{2}\partial_{\v^j} \bigg)
     \bigg]
      {\frac{1}{2}}(\v -\w)^{j} \,  \ln\bigg( \frac{1}{|\v - \w| \Lam}\bigg) \, d(\v,\w)\bigg] ,
     %
\end{align}
with $d(\v,\w)$ being the eikonal dipole operator expressed in Eq.~\eqref{eq: def dipole} and function $C_\lambda$ is defined in Eqs.~\eqref{CL} and~\eqref{CT} for longitudinally and transversely polarized virtual photons, respectively. 

The second term in Eq. \eqref{NEik_X_sec_Separated} originates from the NEik dipole operator of type-2 and is given by  
\begin{align}
\label{Xsec_d2_cont}
 \frac{d\sigma^{\gamma^{*}_{\lambda}+ A \rightarrow q \bar q + X}}{d^{2}\k_{1}d^{2}\k_{2}d\eta_{1}d\eta_{2}} \bigg |_{d^{(2)}} 
   &  = 
   \text{Re} \,  \frac{1}{q^{+}} \int_{\v, \v', \w, \w'} e^{i\k_{1} \cdot (\mathbf{v'}-\mathbf{v}) + i \k_{2} \cdot (\w' - \w)} 
   \, \C_{\lambda}(\w'-\v', \w - \v) 
   \nn \\
   & \times\, 
   \bigg[ \frac{i}{z_1} d^{(2)}(\v_{*} , \w)+\frac{i}{z_2}d^{(2)}(\w_{*} , \v)^{\dagger}\bigg],
   \end{align}
with $d^{(2)}(\v_{*}, \w)$ specified in Eq.~\eqref{d2_GM_fin} within the Gaussian model. As is evident from its explicit expression, the NEik type-2 dipole operator is even in the transverse coordinates. Upon substitution into the cross section in  Eq.~\eqref{Xsec_d2_cont}, its Fourier transform -- being real -- leads to the result of the integral in Eq. \eqref{Xsec_d2_cont}  that is purely imaginary because of the explicit factor of $(i)$. Since the cross section requires taking the real part of this result, the contribution of the NEik type-2 dipole operator to DIS dijet production vanishes. 

The third term in Eq.~\eqref{NEik_X_sec_Separated} corresponds to the NEik type-1 quadrupole contribution to the DIS dijet production cross section. It reads 
\begin{align}
\label{Q1_cont_Xsec_fin}
\frac{d\sigma^{\gamma^{*}_{\lambda}+ A \rightarrow q \bar q + X}}{d^{2}\k_{1}d^{2}\k_{2}d\eta_{1}d\eta_{2}} \bigg |_{Q_j^{(1)}} 
& =
\frac{1}{q^{+}} \text{Re} \int_{\v, \v', \w, \w'} e^{i\k_{1} \cdot (\mathbf{v'}-\mathbf{v}) + i \k_{2} \cdot (\w' - \w)} \C_{\lambda}(\w'-\v', \w - \v) \\ \nn
   & \times \bigg[ 
   \frac{1}{z_{1}} 
   \bigg( \frac{\k_{2}^{j} - \k_{1}^{j}}{2} + \frac{i}{2} \partial_{\w^{j}}\bigg) 
   Q_{j}^{(1)}(\w', \v', \v_{*}, \w)  
-\frac{1}{z_{2}}
   \bigg( \frac{\k_{2}^{j} - \k_{1}^{j}}{2} - \frac{i}{2} \partial_{\v^{j}}\bigg) 
   Q_{j}^{(1)}(\v', \w', \w_{*}, \v)^{\dagger}   \Bigg] . \nn
\end{align}
This contribution does not vanish within the Gaussian model and must therefore be included in the numerical studies. Its explicit form is rather lengthy. For this reason, we simply collect  the relevant equations that constitute the final expression suitable for numerical implementation. The NEik type-1 quadrupole operator in the Gaussian model is given in Eq.~\eqref{Op_Q1_fin} with the eikonal  tadpole contribution $T_Q$ in Eq.~\eqref{eq: def tadpole quadrupole}, while the eikonal non-tadpole contributions are implemented through the matrix $S_{[x^+,y^+]}$ defined in Eq.~\eqref{eikS_for_Q1}. The rotation matrix $R$ is introduced in Eq.~\eqref{System_F}, with its eigensystem provided in Eqs.~\eqref{eq: eigen system of F_1},~\eqref{eq: eigen system of F_2} and~\eqref{eq: eigen system of F_3}. The NEik interaction is encoded via the interaction matrix $F^{NEik, j}$, given in Eq.~\eqref{relation_F_V_NEik} together with Eq.~\eqref{eq:app_VNeik}, and its components are provided in Eq.~\eqref{Q1_V_comp}. In order to compute the operator in second term in Eq.~\eqref{Q1_cont_Xsec_fin}, one should replace $\v'\leftrightarrow\w'$, $\v_{*}\rightarrow\w_{*}$ and $\w\rightarrow\v$ in the final result of operator $Q_{j}^{(1)}(\w', \v', \v_{*}, \w)$ and take its dagger.   

The fourth term in Eq.~\eqref{NEik_X_sec_Separated} corresponds to the contribution of the NEik type-2 quadrupole operator to the DIS dijet production cross section which reads 
\begin{align}
\label{Q2_cont_Xsec_fin}
\frac{d\sigma^{\gamma^{*}_{\lambda}+ A \rightarrow q \bar q + X}}{d^{2}\k_{1}d^{2}\k_{2}d\eta_{1}d\eta_{2}} \bigg |_{Q^{(2)}}  
& =
 \text{Re} \,  \frac{1}{q^{+}} \int_{\v, \v', \w, \w'} e^{i\k_{1} \cdot (\mathbf{v'}-\mathbf{v}) + i \k_{2} \cdot (\w' - \w)} 
   \, \C_{\lambda}(\w'-\v', \w - \v) 
   \nn \\
   & \times\, 
   (-1)\bigg[ \frac{i}{z_1} Q^{(2)}(\w', \v', \mathbf{v}_{*}, \w)
   +\frac{i}{z_2}Q^{(2)}(\v', \w', \mathbf{w}_{*}, \v)^{\dagger}\bigg],
   \end{align}
where the NEik quadrupole type-2 operator is given in Eq.~\eqref{eq:quad_type_2_Decomp} together with Eqs.~\eqref{quad_2+3},~\eqref{eq:non-tad_Neik_quad_type_2} and~\eqref{eq:tad_Neik_quad_type_2}. As in the NEik type-2 dipole case,  the NEik type-2 quadrupole operator is real within the Gaussian model and therefore invariant under simultaneous exchange of the coordinate pairs in the amplitude and complex conjugate amplitude. Consequently, when substituted into the cross section in Eq.~\eqref{Q2_cont_Xsec_fin}, the resulting integral is purely imaginary due to the explicit factor of $(i)$. Since only the real part of this result contributes to the cross section, the contribution of the NEik type-2 quadrupole to the DIS dijet cross section vanishes within the Gaussian model.   

The last three terms in Eq.~\eqref{NEik_X_sec_Separated} contribute to the DIS dijet production cross section only for transversely polarized photons. The first two contributions originate  from the NEik type-3 dipole operator and NEik type-3 quadrupole operator, and they read 
\begin{align}
\label{Xsec_d3_fin}
\frac{d\sigma^{\gamma^{*}_{\lambda}+ A \rightarrow q \bar q + X}}{d^{2}\k_{1}d^{2}\k_{2}d\eta_{1}d\eta_{2}} \bigg |_{d_{ij}^{(3)}} 
& =
 \text{Re} \,  \frac{1}{q^{+}}\, 
 \int_{\v, \v', \w, \w'} e^{i\k_{1} \cdot (\mathbf{v'}-\mathbf{v}) + i \k_{2} \cdot (\w' - \w)} 
 \, \mathcal{D}^{ij}_{T,1}(\w'-\v', \w - \v) \nn \\
   & \times (-1) \,  \bigg[ \frac{1}{z_{1}} \, d_{ij}^{(3)}(\v_{*} , \w) + \frac{1}{z_{2}}  d_{ij}^{(3)}(\w_{*} , \v)^{\dagger}\bigg]
\end{align}
and 
\begin{align}
\label{Xsec_Q3_fin}
\frac{d\sigma^{\gamma^{*}_{\lambda}+ A \rightarrow q \bar q + X}}{d^{2}\k_{1}d^{2}\k_{2}d\eta_{1}d\eta_{2}} \bigg |_{Q_{ij}^{(3)}} 
& =
 \text{Re} \,  \frac{1}{q^{+}}\, 
 \int_{\v, \v', \w, \w'} e^{i\k_{1} \cdot (\mathbf{v'}-\mathbf{v}) + i \k_{2} \cdot (\w' - \w)} 
 \, \mathcal{D}^{ij}_{T,1}(\w'-\v', \w - \v) \nn \\
   & \times \bigg[ \frac{1}{z_{1}} Q_{ij}^{(3)}(\w', \v', \v_{*}, \w)  + \frac{1}{z_{2}} Q_{ij}^{(3)}(\v', \w', \mathbf{w}_{*}, \v)^{\dagger} \bigg].
\end{align}
 However, as discussed in Section~\ref{subsec:type_3}, both NEik type-3 dipole and NEik type-3 quadrupole operators vanish within the Gaussian model. Consequently, these operators do not contribute to the DIS dijet production cross section, and the expressions in Eqs.~\eqref{Xsec_d3_fin} and~\eqref{Xsec_Q3_fin} identically vanish.  

Finally, the last term in Eq.~\eqref{NEik_X_sec_Separated} corresponds to the NEik 3-point contribution to the DIS dijet production cross section and reads 
\begin{align}
\label{Neik_Xsec_3point}
\frac{d\sigma^{\gamma^{*}_{\lambda}+ A \rightarrow q \bar q + X}}{d^{2}\k_{1}d^{2}\k_{2}d\eta_{1}d\eta_{2}} \bigg |_{{\cal O}_j^{(\rm 3-point)}}
& =  - \frac{1}{q^{+}} \text{Im} \int_{\z, \v', \w'} e^{i\k_{1} \cdot (\mathbf{v'}-\z) + i \k_{2} \cdot (\w' - \z)} \, \mathcal{D}^{j}_{T,2}(\w' -\v') \,  \mathcal{O}^{\rm 3- point}_{[\frac{L^+}{2},-\frac{L^+}{2}]}(\w',\v',\z),
\end{align}
which yields a non-vanishing contribution. Function $\mathcal{D}^j_{T,2}$ is defined in Eq.~\eqref{DT2} and the NEik 3-point operator introduced in Eq.~\eqref{Op_3Point_def} is given explicitly in Eq.~\eqref{3_point_non-hom_GM} within the Gaussian model. Together, these expressions provide a form well suited for numerical implementation.  

To sum up, NEik corrections to the DIS dijet production cross section within the Gaussian model are given by Eq.~\eqref{NEik_X_sec_Separated}, with the non-vanishing contributions provided in Eqs.~\eqref{Xsec_NEik_dip_1_fin},~\eqref{Q1_cont_Xsec_fin} and~\eqref{Neik_Xsec_3point}.


\section{Summary and outlook}
\label{sec:summary}

In this work, we have developed a Gaussian model for evaluating decorated dipole and quadrupole operators that arise in observables computed beyond the standard eikonal approximation within the CGC framework. Although the method we introduce is general and applicable to arbitrary NEik Wilson-line structures, our analysis focuses on a specific observable -- DIS dijet production at NEik accuracy -- and on the particular NEik operator structures that contribute to this process.

To set the stage, we first summarize the results obtained previously in~\cite{Altinoluk:2022jkk}, highlighting the decorated Wilson-line structures that appear in the NEik dijet cross section. As a preliminary application of the method, we compute the eikonal dipole and quadrupole operators within the Gaussian model and verify that our expressions agree with the known results in the literature~\cite{Gelis:2001da,Blaizot:2004wv}. We then develop the general framework required to evaluate NEik Wilson lines in the Gaussian model and apply it to the explicit NEik operator structures that enter the DIS dijet cross section. Having computed the NEik dipole, quadrupole, and 3-point operators, we substitute these results into the cross section. We find that the NEik type-2 operators do not contribute to this observable, while the NEik type-3 operators vanish identically within the Gaussian model, independent of the observable under consideration. Consequently, the NEik corrections to DIS dijet production arise solely from the NEik type-1 operators and the NEik 3-point correlation functions.

The final expressions are presented in a form suitable for numerical implementation. A full numerical analysis of NEik DIS dijet production, however, remains an extensive task and lies beyond the scope of this manuscript; we leave this for future work. Since our method provides a systematic procedure for computing general NEik Wilson-line structures, it can also be applied to other processes, including NEik single-inclusive and dijet production in proton-nucleus collisions, which we plan to address in the future. 

Finally, we note that the present formulation incorporates only the NEik corrections that stem from a purely gluonic background. An important extension of this work is the inclusion of NEik corrections associated with $t$- channel quark exchange, which we plan to investigate in the future.



\acknowledgements{
PA, NA and FC thank the Physics Department at the University of Connecticut, and SM thanks IGFAE, for warm hospitality during stays when part of this work was done. PA is supported by Conseller\'{\i}a de Cultura, Educaci\'on e Universidade of Xunta de Galicia under grant ED481B-2022-050. TA is supported in part by the National Science Centre (Poland) under the research Grant No. 2023/50/E/ST2/00133 (SONATA BIS 13). PA and NA are supported by European Research Council project ERC-2018-ADG-835105 YoctoLHC, by Xunta de Galicia (CIGUS Network of Research Centres), by European Union ERDF, and by the Spanish Research State Agency under projects PID20231527\-62NB---I00 and CEX2023-001318-M financed by MICIU/AEI/10.13039/501100011033.
 GB and SM are supported in part by the National Science Centre (Poland) under the research Grant No. 2020/38/E/ST2/00122 (SONATA BIS 10). 
{FC acknowledges support from the Polish National Science Center (NCN) grant No. 2022/46/E/ST2/00346.} SM is also supported by the Research Council of Finland, the Centre of Excellence in Quark Matter, and by the European Research Council (ERC, grant agreements No. ERC-2023-101123801 GlueSatLight and ERC-2018-ADG-835105 YoctoLHC).}

\appendix
\section{Color algebra}
\label{appA}
\subsection{Derivation of the identity in the basis $|e_i\rangle$}\label{App:Color_Identity}
We derive the relation~\eqref{eq:int_V} used to recast the identity in the basis spawn by $\{|e_1\rangle,|e_2\rangle \}$.
As a linear map $V\otimes \bar{V} \otimes V \otimes \bar{V} \rightarrow V\otimes \bar{V} \otimes V \otimes \bar{V}$, where $V$ is the complex vector space for the color of a quark ({i.e.} $V = \mathbb{C}^{N_c}$) and $\bar{V}$ the corresponding dual space, the identity $\mathcal{I}$ is represented in birdtracks by
\begin{equation}
\mathcal{I} = 
\vcenter{\hbox{
\begin{tikzpicture}[scale=0.5]
\draw[thick] (-1,1) -- ++(2,0); \draw[thick,->] (-1,1) -- ++(1,0);
\draw[thick] (-1,.5) -- ++(2,0); \draw[thick,-<] (-1,.5) -- ++(1,0);
\draw[thick] (-1,-.5) -- ++(2,0); \draw[thick,->] (-1,-.5) -- ++(1,0);
\draw[thick] (-1,-1) -- ++(2,0); \draw[thick,-<] (-1,-1) -- ++(1,0);
\end{tikzpicture}}}\ .
\end{equation}
Under time evolution, the color state of the quadupole is an overall singlet, thus we only focus on the singlet projection $\mathcal{I}|_{singlet}$. For notation clarity, we will always imply $\mathcal{I}|_{singlet}$ without mention of the subscript in the rest of this Appendix. 
Using the Fierz identity\footnote{The generators in the fundamental irrep are normalized according to $\text{tr}(T^aT^b) = \frac{1}{2}\delta^{ab}$, thus the factor in the Fierz identity is $2$.} on the pair $q_1\otimes \bar{q}_2$ and on the pair $q_3\otimes \bar{q}_4$, we write
\begin{equation}
\mathcal{I} = 
\frac{1}{N_c^2}
\left.\vcenter{\hbox{
\begin{tikzpicture}[scale=0.5]
\draw[thick] (-1,1) --  (-.5,1) to[out=0,in=0] (-.5,.5) -- (-1,.5);\draw[thick,->] (-1,1) -- ++(.3,0);
\draw[thick] (-1,-1) --  (-.5,-1) to[out=0,in=0] (-.5,-.5) -- (-1,-.5);\draw[thick,->] (-1,-.5) -- ++(.3,0);
\draw[thick] (1,1) --  (.5,1) to[out=180,in=180] (.5,.5) -- (1,.5);\draw[thick,->] (.5,1) -- ++(.3,0);
\draw[thick] (1,-1) --  (.5,-1) to[out=180,in=180] (.5,-.5) -- (1,-.5);\draw[thick,->] (.5,-.5) -- ++(.3,0);
\end{tikzpicture}}}\ \right|_{singlet}
+ \frac{2}{N_c}
\left.\vcenter{\hbox{
\begin{tikzpicture}[scale=0.5]
\draw[thick] (-1,1) --  (-.5,1) to[out=0,in=0] (-.5,.5) -- (-1,.5);\draw[thick,->] (-1,1) -- ++(.3,0);
\draw[thick] (-1,-1) --  (-.5,-1) to[out=0,in=0] (-.5,-.5) -- (-1,-.5);\draw[thick,->] (-1,-.5) -- ++(.3,0);
\draw[thick] (1,1) --  (.5,1) to[out=180,in=180] (.5,.5) -- (1,.5);\draw[thick,->] (.5,1) -- ++(.3,0);
\draw[thick] (1,-1) --  (.5,-1) to[out=180,in=180] (.5,-.5) -- (1,-.5);\draw[thick,->] (.5,-.5) -- ++(.3,0);
\draw[thick,gluon] (-.35,.75) -- (.35,.75);
\end{tikzpicture}}}\ \right|_{singlet}
+ \frac{2}{N_c}
\left.\vcenter{\hbox{
\begin{tikzpicture}[scale=0.5]
\draw[thick] (-1,1) --  (-.5,1) to[out=0,in=0] (-.5,.5) -- (-1,.5);\draw[thick,->] (-1,1) -- ++(.3,0);
\draw[thick] (-1,-1) --  (-.5,-1) to[out=0,in=0] (-.5,-.5) -- (-1,-.5);\draw[thick,->] (-1,-.5) -- ++(.3,0);
\draw[thick] (1,1) --  (.5,1) to[out=180,in=180] (.5,.5) -- (1,.5);\draw[thick,->] (.5,1) -- ++(.3,0);
\draw[thick] (1,-1) --  (.5,-1) to[out=180,in=180] (.5,-.5) -- (1,-.5);\draw[thick,->] (.5,-.5) -- ++(.3,0);
\draw[thick,gluon] (-.35,-.75) -- (.35,-.75);
\end{tikzpicture}}}\ \right|_{singlet}
+ 4
\left.\vcenter{\hbox{
\begin{tikzpicture}[scale=0.5]
\draw[thick] (-1,1) --  (-.5,1) to[out=0,in=0] (-.5,.5) -- (-1,.5);\draw[thick,->] (-1,1) -- ++(.3,0);
\draw[thick] (-1,-1) --  (-.5,-1) to[out=0,in=0] (-.5,-.5) -- (-1,-.5);\draw[thick,->] (-1,-.5) -- ++(.3,0);
\draw[thick] (1,1) --  (.5,1) to[out=180,in=180] (.5,.5) -- (1,.5);\draw[thick,->] (.5,1) -- ++(.3,0);
\draw[thick] (1,-1) --  (.5,-1) to[out=180,in=180] (.5,-.5) -- (1,-.5);\draw[thick,->] (.5,-.5) -- ++(.3,0);
\draw[thick,gluon] (-.35,.75) -- (.35,.75);
\draw[thick,gluon] (-.35,-.75) -- (.35,-.75);
\end{tikzpicture}}}\ \right|_{singlet}.
\end{equation}
The second and third tensor lie in the adjoint irrep (there is a gluon in the intermediate state) and, thus, are orthogonal to the singlet projection. The first tensor is already a singlet (there is no color propagating in the intermediate state), and we project the fourth tensor to the singlet piece of the product $8\otimes 8$:
\begin{equation}
\mathcal{I} = 
\frac{1}{N_c^2}
\vcenter{\hbox{
\begin{tikzpicture}[scale=0.5]
\draw[thick] (-1,1) --  (-.5,1) to[out=0,in=0] (-.5,.5) -- (-1,.5);\draw[thick,->] (-1,1) -- ++(.3,0);
\draw[thick] (-1,-1) --  (-.5,-1) to[out=0,in=0] (-.5,-.5) -- (-1,-.5);\draw[thick,->] (-1,-.5) -- ++(.3,0);
\draw[thick] (1,1) --  (.5,1) to[out=180,in=180] (.5,.5) -- (1,.5);\draw[thick,->] (.5,1) -- ++(.3,0);
\draw[thick] (1,-1) --  (.5,-1) to[out=180,in=180] (.5,-.5) -- (1,-.5);\draw[thick,->] (.5,-.5) -- ++(.3,0);
\end{tikzpicture}}}
\ + 4\times \frac{1}{N_c^2-1}
\vcenter{\hbox{
\begin{tikzpicture}[scale=0.5]
\draw[thick] (-1.5,1) --  (-1,1) to[out=0,in=0] (-1,.5) -- (-1.5,.5);\draw[thick,->] (-1.5,1) -- ++(.3,0);
\draw[thick] (-1.5,-1) --  (-1,-1) to[out=0,in=0] (-1,-.5) -- (-1.5,-.5);\draw[thick,->] (-1.5,-.5) -- ++(.3,0);
\draw[thick] (1.5,1) --  (1,1) to[out=180,in=180] (1,.5) -- (1.5,.5);\draw[thick,->] (1,1) -- ++(.3,0);
\draw[thick] (1.5,-1) --  (1,-1) to[out=180,in=180] (1,-.5) -- (1.5,-.5);\draw[thick,->] (1,-.5) -- ++(.3,0);
\draw[thick,gluon] (-.85,.75) to[out=0,in=0] (-.85,-.75);
\draw[thick,gluon] (.85,.75) to[out=180,in=180] (.85,-.75);
\end{tikzpicture}}}\ .
\end{equation}
The latter diagram can be expressed in terms of $\left|e_i\rangle \langle e_j \right|$ by recasting the two gluons with the Fierz identity one last time. One finally recovers
\begin{align}
\mathcal{I} &= 
\frac{1}{N_c^2-1} \left[
\vcenter{\hbox{
\begin{tikzpicture}[scale=0.5]
\draw[thick] (-1,1) --  (-.5,1) to[out=0,in=0] (-.5,.5) -- (-1,.5);\draw[thick,->] (-1,1) -- ++(.3,0);
\draw[thick] (-1,-1) --  (-.5,-1) to[out=0,in=0] (-.5,-.5) -- (-1,-.5);\draw[thick,->] (-1,-.5) -- ++(.3,0);
\draw[thick] (1,1) --  (.5,1) to[out=180,in=180] (.5,.5) -- (1,.5);\draw[thick,->] (.5,1) -- ++(.3,0);
\draw[thick] (1,-1) --  (.5,-1) to[out=180,in=180] (.5,-.5) -- (1,-.5);\draw[thick,->] (.5,-.5) -- ++(.3,0);
\end{tikzpicture}}}
\ +
\vcenter{\hbox{
\begin{tikzpicture}[scale=0.5]
\draw[thick] (-1.5,1) --  (-1,1) to[out=0,in=0] (-1,-1) -- (-1.5,-1);\draw[thick,->] (-1.5,1) -- ++(.3,0);
\draw[thick] (-1.5,.5) --  (-1,.5) to[out=0,in=0] (-1,-.5) -- (-1.5,-.5);\draw[thick,->] (-1.5,-.5) -- ++(.3,0);
\draw[thick] (1.5,1) --  (1,1) to[out=180,in=180] (1,-1) -- (1.5,-1);\draw[thick,->] (1,1) -- ++(.3,0);
\draw[thick] (1.5,.5) --  (1,.5) to[out=180,in=180] (1,-.5) -- (1.5,-.5);\draw[thick,->] (1,-.5) -- ++(.3,0);
\end{tikzpicture}}}
\ -\frac{1}{N_c}
\vcenter{\hbox{
\begin{tikzpicture}[scale=0.5]
\draw[thick] (-1.5,1) --  (-1,1) to[out=0,in=0] (-1,-1) -- (-1.5,-1);\draw[thick,->] (-1.5,1) -- ++(.3,0);
\draw[thick] (-1.5,.5) --  (-1,.5) to[out=0,in=0] (-1,-.5) -- (-1.5,-.5);\draw[thick,->] (-1.5,-.5) -- ++(.3,0);
\draw[thick] (1,1) --  (.5,1) to[out=180,in=180] (.5,.5) -- (1,.5);\draw[thick,->] (.5,1) -- ++(.3,0);
\draw[thick] (1,-1) --  (.5,-1) to[out=180,in=180] (.5,-.5) -- (1,-.5);\draw[thick,->] (.5,-.5) -- ++(.3,0);
\end{tikzpicture}}}
\ -\frac{1}{N_c}
\vcenter{\hbox{
\begin{tikzpicture}[scale=0.5]
\draw[thick] (-1,1) --  (-.5,1) to[out=0,in=0] (-.5,.5) -- (-1,.5);\draw[thick,->] (-1,1) -- ++(.3,0);
\draw[thick] (-1,-1) --  (-.5,-1) to[out=0,in=0] (-.5,-.5) -- (-1,-.5);\draw[thick,->] (-1,-.5) -- ++(.3,0);
\draw[thick] (1.5,1) --  (1,1) to[out=180,in=180] (1,-1) -- (1.5,-1);\draw[thick,->] (1,1) -- ++(.3,0);
\draw[thick] (1.5,.5) --  (1,.5) to[out=180,in=180] (1,-.5) -- (1.5,-.5);\draw[thick,->] (1,-.5) -- ++(.3,0);
\end{tikzpicture}}}
\right] \notag \\
&=\frac{1}{N_c^2-1}\left[ 
\left|e_1\right\rangle \left\langle e_1 \right| 
+
\left|e_2\right\rangle \left\langle e_2 \right|
-\frac{1}{N_c}
\left|e_2\right\rangle \left\langle e_1 \right|
-\frac{1}{N_c}
\left|e_1\right\rangle \left\langle e_2 \right|
\right] = c_{ij} \left|e_i\right\rangle \left\langle e_j \right|, \label{eq:id_quadr}
\end{align}
which we use in the main text.
Notice that, in matrix form, $c_{ij}$ is simply related to the inverse of the overlaps between elements of the basis:
\begin{equation}
    c_{ij} = \left[ \langle e_i | e_j \rangle\right]^{-1} =
    \left[
    \begin{pmatrix}
        N_c^2 & N_c \\
        N_c & N_c^2
    \end{pmatrix}
    \right]^{-1}
    = \frac{1}{N_c^2-1}
    \begin{pmatrix}
        1 & -1/N_c \\
        -1/N_c & 1
    \end{pmatrix}.
\end{equation}

\subsection{Derivation of $\mathcal{V}$ in the basis $|e_i\rangle$}\label{App:Color_Interaction}
In order to derive the non-tadpole interactions, that are resumed from the time evolution of the  $q_1\otimes \bar{q}_2 \otimes q_3 \otimes \bar{q}_4$ system into the quadrupole, one only needs to compute two distinct interactions between either $q\otimes\bar{q}$ or $q\otimes q$ out of the six possible pairings from a system of four partons.
Interaction are mediated by
\begin{equation}
    \mathcal{V} = \sum_{ij} \mathcal{C}_{ij} w_{ij}\ .
\end{equation}
\paragraph{\underline{12 and 34 exchanges}:} We have
\begin{equation}
\mathcal{C}_{12} = -T_1^a \otimes T_2^a \otimes \mathbb{1}_3 \otimes \mathbb{1}_4 
= -
\vcenter{\hbox{
\begin{tikzpicture}[scale=0.5]
\draw[thick] (-1,1) -- ++(2,0); \draw[thick,->] (-1,1) -- ++(1,0);
\draw[thick] (-1,.5) -- ++(2,0); \draw[thick,-<] (-1,.5) -- ++(1,0);
\draw[thick] (-1,-.5) -- ++(2,0); \draw[thick,->] (-1,-.5) -- ++(1,0);
\draw[thick] (-1,-1) -- ++(2,0); \draw[thick,-<] (-1,-1) -- ++(1,0);
\draw[Red,gluon] (-.5,1) to[out=-90,in=120] (0,-3);
\draw[Red,gluon] (.5,.5) to[out=-90,in=60] (0,-3);
\end{tikzpicture}}}.
\end{equation}
We first replace the exchanged red gluon using the Fierz identity. It reads 
\begin{equation}
\mathcal{C}_{12} =
\frac{1}{2} \left[ 
\vcenter{\hbox{
\begin{tikzpicture}[scale=0.5]
\draw[thick] (-1,1) --  (-.5,1) to[out=0,in=0] (-.5,.5) -- (-1,.5);\draw[thick,->] (-1,1) -- ++(.3,0);
\draw[thick] (1,1) --  (.5,1) to[out=180,in=180] (.5,.5) -- (1,.5);\draw[thick,->] (.5,1) -- ++(.3,0);
\draw (-1,-.5) -- (1,-.5);\draw[thick,->] (-1,-.5) -- ++(1,0);
\draw (-1,-1) -- (1,-1);\draw[thick,-<] (-1,-1) -- ++(1,0);
\end{tikzpicture}}}
\ - \frac{1}{N_c}
\vcenter{\hbox{
\begin{tikzpicture}[scale=0.5]
\draw[thick] (-1,1) -- ++(2,0); \draw[thick,->] (-1,1) -- ++(1,0);
\draw[thick] (-1,.5) -- ++(2,0); \draw[thick,-<] (-1,.5) -- ++(1,0);
\draw[thick] (-1,-.5) -- ++(2,0); \draw[thick,->] (-1,-.5) -- ++(1,0);
\draw[thick] (-1,-1) -- ++(2,0); \draw[thick,-<] (-1,-1) -- ++(1,0);
\end{tikzpicture}}}
\right].
\end{equation}
Using the the result for $\mathcal{I}$, we find
\begin{align}
\mathcal{C}_{12} &= 
\frac{1}{N_c^2-1} \left[ \frac{N_c^2-2}{2N_c}
\vcenter{\hbox{
\begin{tikzpicture}[scale=0.5]
\draw[thick] (-1,1) --  (-.5,1) to[out=0,in=0] (-.5,.5) -- (-1,.5);\draw[thick,->] (-1,1) -- ++(.3,0);
\draw[thick] (-1,-1) --  (-.5,-1) to[out=0,in=0] (-.5,-.5) -- (-1,-.5);\draw[thick,->] (-1,-.5) -- ++(.3,0);
\draw[thick] (1,1) --  (.5,1) to[out=180,in=180] (.5,.5) -- (1,.5);\draw[thick,->] (.5,1) -- ++(.3,0);
\draw[thick] (1,-1) --  (.5,-1) to[out=180,in=180] (.5,-.5) -- (1,-.5);\draw[thick,->] (.5,-.5) -- ++(.3,0);
\end{tikzpicture}}}
\ - \frac{1}{2N_c}
\vcenter{\hbox{
\begin{tikzpicture}[scale=0.5]
\draw[thick] (-1.5,1) --  (-1,1) to[out=0,in=0] (-1,-1) -- (-1.5,-1);\draw[thick,->] (-1.5,1) -- ++(.3,0);
\draw[thick] (-1.5,.5) --  (-1,.5) to[out=0,in=0] (-1,-.5) -- (-1.5,-.5);\draw[thick,->] (-1.5,-.5) -- ++(.3,0);
\draw[thick] (1.5,1) --  (1,1) to[out=180,in=180] (1,-1) -- (1.5,-1);\draw[thick,->] (1,1) -- ++(.3,0);
\draw[thick] (1.5,.5) --  (1,.5) to[out=180,in=180] (1,-.5) -- (1.5,-.5);\draw[thick,->] (1,-.5) -- ++(.3,0);
\end{tikzpicture}}}
\ +\frac{1}{2N_c^2}
\vcenter{\hbox{
\begin{tikzpicture}[scale=0.5]
\draw[thick] (-1.5,1) --  (-1,1) to[out=0,in=0] (-1,-1) -- (-1.5,-1);\draw[thick,->] (-1.5,1) -- ++(.3,0);
\draw[thick] (-1.5,.5) --  (-1,.5) to[out=0,in=0] (-1,-.5) -- (-1.5,-.5);\draw[thick,->] (-1.5,-.5) -- ++(.3,0);
\draw[thick] (1,1) --  (.5,1) to[out=180,in=180] (.5,.5) -- (1,.5);\draw[thick,->] (.5,1) -- ++(.3,0);
\draw[thick] (1,-1) --  (.5,-1) to[out=180,in=180] (.5,-.5) -- (1,-.5);\draw[thick,->] (.5,-.5) -- ++(.3,0);
\end{tikzpicture}}}
\ +\frac{1}{2N_c^2}
\vcenter{\hbox{
\begin{tikzpicture}[scale=0.5]
\draw[thick] (-1,1) --  (-.5,1) to[out=0,in=0] (-.5,.5) -- (-1,.5);\draw[thick,->] (-1,1) -- ++(.3,0);
\draw[thick] (-1,-1) --  (-.5,-1) to[out=0,in=0] (-.5,-.5) -- (-1,-.5);\draw[thick,->] (-1,-.5) -- ++(.3,0);
\draw[thick] (1.5,1) --  (1,1) to[out=180,in=180] (1,-1) -- (1.5,-1);\draw[thick,->] (1,1) -- ++(.3,0);
\draw[thick] (1.5,.5) --  (1,.5) to[out=180,in=180] (1,-.5) -- (1.5,-.5);\draw[thick,->] (1,-.5) -- ++(.3,0);
\end{tikzpicture}}}
\right] \notag \\
&=\frac{1}{N_c^2-1}\left[ 
\frac{N_c^2-2}{2N_c}
\left|e_1\right\rangle \left\langle e_1 \right| 
- \frac{1}{2N_c}
\left|e_2\right\rangle \left\langle e_2 \right|
+\frac{1}{2N_c^2}
\left|e_2\right\rangle \left\langle e_1 \right|
+\frac{1}{2N_c^2}
\left|e_1\right\rangle \left\langle e_2 \right|
\right].
\end{align}
By symmetry under reflection, we readily see that $\mathcal{C}_{34} = \mathcal{C}_{12}$.

\paragraph{\underline{13 and 24 exchanges}:} We have
\begin{equation}
\mathcal{C}_{13} = -T_1^a \otimes \mathbb{1}_2 \otimes T^a_3 \otimes \mathbb{1}_4
= -
\vcenter{\hbox{
\begin{tikzpicture}[scale=0.5]
\draw[thick] (-1,1) -- ++(2,0); \draw[thick,->] (-1,1) -- ++(1,0);
\draw[thick] (-1,.5) -- ++(2,0); \draw[thick,-<] (-1,.5) -- ++(1,0);
\draw[thick] (-1,-.5) -- ++(2,0); \draw[thick,->] (-1,-.5) -- ++(1,0);
\draw[thick] (-1,-1) -- ++(2,0); \draw[thick,-<] (-1,-1) -- ++(1,0);
\draw[Red,gluon] (-.5,1) to[out=-90,in=120] (0,-3);
\draw[Red,gluon] (.5,-.5) to[out=-90,in=60] (0,-3);
\end{tikzpicture}}}.
\end{equation}
We replace the exchanged gluon using the Fierz identity (the sign is due to the convention that $T_{\overline{F}} \equiv -(T_F)^T$):
\begin{align}
\mathcal{C}_{13} = 
\frac{-1}{2} \left[ 
\vcenter{\hbox{
\begin{tikzpicture}[scale=0.5]
\draw[thick] (-1,1) -- (-.5,1) to[out=0,in=180] (.5,-.5) -- (1,-.5);\draw[thick,->] (-1,1) -- ++(.3,0);
\draw[thick] (-1,-.5) -- (-.5,-.5) to[out=0,in=180] (.5,1) -- (1,1);
\draw[thick,->] (-1,-.5) -- ++(.3,0);
\draw[thick] (-1,.5) -- (1,.5); \draw[thick,-<] (-1,.5) -- ++(.3,0);
\draw[thick] (-1,-1) -- (1,-1);\draw[thick,-<] (-1,-1) -- ++(1,0);
\end{tikzpicture}}}
\ - \frac{1}{N_c}
\vcenter{\hbox{
\begin{tikzpicture}[scale=0.5]
\draw[thick] (-1,1) -- ++(2,0); \draw[thick,->] (-1,1) -- ++(1,0);
\draw[thick] (-1,.5) -- ++(2,0); \draw[thick,-<] (-1,.5) -- ++(1,0);
\draw[thick] (-1,-.5) -- ++(2,0); \draw[thick,->] (-1,-.5) -- ++(1,0);
\draw[thick] (-1,-1) -- ++(2,0); \draw[thick,-<] (-1,-1) -- ++(1,0);
\end{tikzpicture}}}
\right] .
\end{align}
Let us call the first tensor $\hat{\sigma}_{13}$. It permutes the color flow at location $\x_1$ and $\x_3$.
Acting on the basis $\{|e_1\rangle,|e_2\rangle \}$, it has the following property:
\begin{equation}
\hat{\sigma}_{13} |e_1\rangle = |e_2\rangle, \qquad \hat{\sigma}_{13} \cdot \hat{\sigma}_{13} = 1.
\end{equation}
Using the previous result for $\mathcal{I}$, we get
\begin{align}
\mathcal{C}_{13} &= -\frac{1}{2} \left(\hat{\sigma}_{13} - \frac{1}{N_c} \mathcal{I} \right) \\
&= 
\frac{1}{N_c^2-1} \left[ 
\frac{1}{N_c}
\vcenter{\hbox{
\begin{tikzpicture}[scale=0.5]
\draw[thick] (-1,1) --  (-.5,1) to[out=0,in=0] (-.5,.5) -- (-1,.5);\draw[thick,->] (-1,1) -- ++(.3,0);
\draw[thick] (-1,-1) --  (-.5,-1) to[out=0,in=0] (-.5,-.5) -- (-1,-.5);\draw[thick,->] (-1,-.5) -- ++(.3,0);
\draw[thick] (1,1) --  (.5,1) to[out=180,in=180] (.5,.5) -- (1,.5);\draw[thick,->] (.5,1) -- ++(.3,0);
\draw[thick] (1,-1) --  (.5,-1) to[out=180,in=180] (.5,-.5) -- (1,-.5);\draw[thick,->] (.5,-.5) -- ++(.3,0);
\end{tikzpicture}}}
\ + \frac{1}{N_c}
\vcenter{\hbox{
\begin{tikzpicture}[scale=0.5]
\draw[thick] (-1.5,1) --  (-1,1) to[out=0,in=0] (-1,-1) -- (-1.5,-1);\draw[thick,->] (-1.5,1) -- ++(.3,0);
\draw[thick] (-1.5,.5) --  (-1,.5) to[out=0,in=0] (-1,-.5) -- (-1.5,-.5);\draw[thick,->] (-1.5,-.5) -- ++(.3,0);
\draw[thick] (1.5,1) --  (1,1) to[out=180,in=180] (1,-1) -- (1.5,-1);\draw[thick,->] (1,1) -- ++(.3,0);
\draw[thick] (1.5,.5) --  (1,.5) to[out=180,in=180] (1,-.5) -- (1.5,-.5);\draw[thick,->] (1,-.5) -- ++(.3,0);
\end{tikzpicture}}}
\ -\frac{N_c^2+1}{2N_c^2}
\vcenter{\hbox{
\begin{tikzpicture}[scale=0.5]
\draw[thick] (-1.5,1) --  (-1,1) to[out=0,in=0] (-1,-1) -- (-1.5,-1);\draw[thick,->] (-1.5,1) -- ++(.3,0);
\draw[thick] (-1.5,.5) --  (-1,.5) to[out=0,in=0] (-1,-.5) -- (-1.5,-.5);\draw[thick,->] (-1.5,-.5) -- ++(.3,0);
\draw[thick] (1,1) --  (.5,1) to[out=180,in=180] (.5,.5) -- (1,.5);\draw[thick,->] (.5,1) -- ++(.3,0);
\draw[thick] (1,-1) --  (.5,-1) to[out=180,in=180] (.5,-.5) -- (1,-.5);\draw[thick,->] (.5,-.5) -- ++(.3,0);
\end{tikzpicture}}}
\ -\frac{N_c^2+1}{2N_c^2}
\vcenter{\hbox{
\begin{tikzpicture}[scale=0.5]
\draw[thick] (-1,1) --  (-.5,1) to[out=0,in=0] (-.5,.5) -- (-1,.5);\draw[thick,->] (-1,1) -- ++(.3,0);
\draw[thick] (-1,-1) --  (-.5,-1) to[out=0,in=0] (-.5,-.5) -- (-1,-.5);\draw[thick,->] (-1,-.5) -- ++(.3,0);
\draw[thick] (1.5,1) --  (1,1) to[out=180,in=180] (1,-1) -- (1.5,-1);\draw[thick,->] (1,1) -- ++(.3,0);
\draw[thick] (1.5,.5) --  (1,.5) to[out=180,in=180] (1,-.5) -- (1.5,-.5);\draw[thick,->] (1,-.5) -- ++(.3,0);
\end{tikzpicture}}}
\right] \notag \\
&=\frac{1}{N_c^2-1}\left[ 
\frac{1}{N_c}
\left|e_1\right\rangle \left\langle e_1 \right| 
+ \frac{1}{N_c}
\left|e_2\right\rangle \left\langle e_2 \right|
-\frac{N_c^2+1}{2N_c^2}
\left|e_2\right\rangle \left\langle e_1 \right|
-\frac{N_c^2+1}{2N_c^2}
\left|e_1\right\rangle \left\langle e_2 \right|
\right].
\end{align}
By reflection symmetry, we see that $\mathcal{C}_{24} = \mathcal{C}_{13}$.
\\
\paragraph{\underline{14 and 23 exchanges}:} We finally have
\begin{equation}
\mathcal{C}_{23} = - \mathbb{1}_1 \otimes T_2^a  \otimes T^a_3 \otimes \mathbb{1}_4
= -
\vcenter{\hbox{
\begin{tikzpicture}[scale=0.5]
\draw[thick] (-1,1) -- ++(2,0); \draw[thick,->] (-1,1) -- ++(1,0);
\draw[thick] (-1,.5) -- ++(2,0); \draw[thick,-<] (-1,.5) -- ++(1,0);
\draw[thick] (-1,-.5) -- ++(2,0); \draw[thick,->] (-1,-.5) -- ++(1,0);
\draw[thick] (-1,-1) -- ++(2,0); \draw[thick,-<] (-1,-1) -- ++(1,0);
\draw[Red,gluon] (-.5,.5) to[out=-90,in=120] (0,-3);
\draw[Red,gluon] (.5,-.5) to[out=-90,in=60] (0,-3);
\end{tikzpicture}}}.
\end{equation}
This factor can easily be related to $\mathcal{C}_{12}$ using $\hat{\sigma}_{13}$:
\begin{equation}
\mathcal{C}_{23} = \hat{\sigma}_{13} \cdot \mathcal{C}_{12} \cdot \hat{\sigma}_{13} 
=
\vcenter{\hbox{
\begin{tikzpicture}[scale=0.5]
\draw[thick] (-1,1) -- ++(2,0); \draw[thick,->] (-1,1) -- ++(1,0);
\draw[thick] (-1,.5) -- ++(2,0); \draw[thick,-<] (-1,.5) -- ++(1,0);
\draw[thick] (-1,-.5) -- ++(2,0); \draw[thick,->] (-1,-.5) -- ++(1,0);
\draw[thick] (-1,-1) -- ++(2,0); \draw[thick,-<] (-1,-1) -- ++(1,0);
\draw[Red,gluon] (-.5,1) to[out=-90,in=120] (0,-3);
\draw[Red,gluon] (.5,.5) to[out=-90,in=60] (0,-3);

\draw[thick] (-1.2,1) to[out=180,in=0] (-2.2,-.5);
\draw[thick] (-1.2,-.5) to[out=180,in=0] (-2.2,1);
\draw[thick] (-2.2,.5) -- ++(1,0);
\draw[thick] (-2.2,-1) -- ++(1,0);

\draw[thick] (1.2,1) to[out=0,in=180] (2.2,-.5);
\draw[thick] (1.2,-.5) to[out=0,in=180] (2.2,1);
\draw[thick] (1.2,.5) -- ++(1,0);
\draw[thick] (1.2,-1) -- ++(1,0);
\end{tikzpicture}}}.
\end{equation}
From the action of $\hat{\sigma}_{13}$ on the basis $\{|e_1\rangle,|e_2\rangle \}$ and the result for $\mathcal{C}_{12}$, we automatically conclude that
\begin{align}
\mathcal{C}_{12} &= 
\frac{1}{N_c^2-1} \left[ \frac{N_c^2-2}{2N_c}
\vcenter{\hbox{
\begin{tikzpicture}[scale=0.5]
\draw[thick] (-1.5,1) --  (-1,1) to[out=0,in=0] (-1,-1) -- (-1.5,-1);\draw[thick,->] (-1.5,1) -- ++(.3,0);
\draw[thick] (-1.5,.5) --  (-1,.5) to[out=0,in=0] (-1,-.5) -- (-1.5,-.5);\draw[thick,->] (-1.5,-.5) -- ++(.3,0);
\draw[thick] (1.5,1) --  (1,1) to[out=180,in=180] (1,-1) -- (1.5,-1);\draw[thick,->] (1,1) -- ++(.3,0);
\draw[thick] (1.5,.5) --  (1,.5) to[out=180,in=180] (1,-.5) -- (1.5,-.5);\draw[thick,->] (1,-.5) -- ++(.3,0);
\end{tikzpicture}}}
\ - \frac{1}{2N_c}
\vcenter{\hbox{
\begin{tikzpicture}[scale=0.5]
\draw[thick] (-1,1) --  (-.5,1) to[out=0,in=0] (-.5,.5) -- (-1,.5);\draw[thick,->] (-1,1) -- ++(.3,0);
\draw[thick] (-1,-1) --  (-.5,-1) to[out=0,in=0] (-.5,-.5) -- (-1,-.5);\draw[thick,->] (-1,-.5) -- ++(.3,0);
\draw[thick] (1,1) --  (.5,1) to[out=180,in=180] (.5,.5) -- (1,.5);\draw[thick,->] (.5,1) -- ++(.3,0);
\draw[thick] (1,-1) --  (.5,-1) to[out=180,in=180] (.5,-.5) -- (1,-.5);\draw[thick,->] (.5,-.5) -- ++(.3,0);
\end{tikzpicture}}}
\ +\frac{1}{2N_c^2}
\vcenter{\hbox{
\begin{tikzpicture}[scale=0.5]
\draw[thick] (-1.5,1) --  (-1,1) to[out=0,in=0] (-1,-1) -- (-1.5,-1);\draw[thick,->] (-1.5,1) -- ++(.3,0);
\draw[thick] (-1.5,.5) --  (-1,.5) to[out=0,in=0] (-1,-.5) -- (-1.5,-.5);\draw[thick,->] (-1.5,-.5) -- ++(.3,0);
\draw[thick] (1,1) --  (.5,1) to[out=180,in=180] (.5,.5) -- (1,.5);\draw[thick,->] (.5,1) -- ++(.3,0);
\draw[thick] (1,-1) --  (.5,-1) to[out=180,in=180] (.5,-.5) -- (1,-.5);\draw[thick,->] (.5,-.5) -- ++(.3,0);
\end{tikzpicture}}}
\ +\frac{1}{2N_c^2}
\vcenter{\hbox{
\begin{tikzpicture}[scale=0.5]
\draw[thick] (-1,1) --  (-.5,1) to[out=0,in=0] (-.5,.5) -- (-1,.5);\draw[thick,->] (-1,1) -- ++(.3,0);
\draw[thick] (-1,-1) --  (-.5,-1) to[out=0,in=0] (-.5,-.5) -- (-1,-.5);\draw[thick,->] (-1,-.5) -- ++(.3,0);
\draw[thick] (1.5,1) --  (1,1) to[out=180,in=180] (1,-1) -- (1.5,-1);\draw[thick,->] (1,1) -- ++(.3,0);
\draw[thick] (1.5,.5) --  (1,.5) to[out=180,in=180] (1,-.5) -- (1.5,-.5);\draw[thick,->] (1,-.5) -- ++(.3,0);
\end{tikzpicture}}}
\right] \notag \\
&=\frac{1}{N_c^2-1}\left[ 
\frac{N_c^2-2}{2N_c}
\left|e_2\right\rangle \left\langle e_2 \right| 
- \frac{1}{2N_c}
\left|e_1\right\rangle \left\langle e_1 \right|
+\frac{1}{2N_c^2}
\left|e_2\right\rangle \left\langle e_1 \right|
+\frac{1}{2N_c^2}
\left|e_1\right\rangle \left\langle e_2 \right|
\right].
\end{align}
\\
\paragraph{\underline{Combining eikonal and next-to-eikonal results}:}
The interaction at eikonal accuracy can be combined into the matrix (in the $\{|e_1\rangle,|e_2\rangle\}$-basis)
\begin{equation}
V^{eik} = \frac{1}{N_c^2-1} 
\begin{pmatrix}
\frac{2\beta - \gamma  + \alpha(N_c^2-2)}{2N_c} & \frac{\alpha+\gamma - \beta(N_c^2+1)}{2N_c^2} \\
\frac{\alpha+\gamma - \beta(N_c^2+1)}{2N_c^2} & \frac{2\beta - \alpha  + \gamma(N_c^2-2)}{2N_c}
\end{pmatrix},
\label{eq:veik}
\end{equation}
where $\alpha  = w_{12} + w_{34}$, $\beta = w_{13}+w_{24}$ and $\gamma = w_{14} + w_{23}$, with $w_{ij}$ the kinematic part of the interaction.
The interaction at Neik accuracy can be combined into a matrix formally identical to that in~\eqref{eq:veik} but with different $\alpha,\beta,\gamma$:
\begin{equation}\label{eq:app_VNeik}
V^{Neik}(\x_1) =V^{eik} [\alpha  = w_{12}, \beta = w_{13},  \gamma = w_{14}]
\end{equation}
when the decoration lies at transverse location $\x_1$.

\section{Integrals over the longitudinal coordinates}\label{App:Longi_integrals}
Under the assumption that the nuclear profile of the target is homogeneous, $\mu(v^+) = \mu$, the longitudinal integrals for the evaluation of the NEik operators of type 1 and type 2 can be performed by using the following ones:

\paragraph{\underline{One point}.} We assume non-degenerate eigenvalues $\lambda_i$ when the labels are distinct. One has to consider the following two cases:
\begin{align}
    \int^{\tfrac{1}{2}}_{-\tfrac{1}{2}} dv\ e^{(\frac{1}{2}-v) \lambda_i} \left\{ 1, v \right\} e^{(v+\frac{1}{2}) \lambda_i} 
    &= \left\{ e^{\lambda_i}, 0 \right\}, \\
    \int^{\tfrac{1}{2}}_{-\tfrac{1}{2}} dv\ e^{(\frac{1}{2}-v) \lambda_i} \left\{ 1, v \right\} e^{(v+\frac{1}{2}) \lambda_j}
    &= \left\{\frac{e^{\lambda_i}-e^{\lambda_j}}{\lambda_i - \lambda_j}, \frac{e^{\lambda_i}[2+\lambda_j-\lambda_i] - e^{\lambda_j}[2+\lambda_i-\lambda_j]}{2(\lambda_i-\lambda_j)^2} \right\}.
\end{align}

\paragraph{\underline{Two points}.} Again, we assume non-degenerate eigenvalues $\lambda_i$ when the labels are distinct. One has to consider the following cases:
\begin{equation}
    (iii):\quad \int^{\tfrac{1}{2}}_{-\tfrac{1}{2}}dw \int^{w}_{-\tfrac{1}{2}} dv\  
    \begin{pmatrix}
        1 & w \\
        v & wv
    \end{pmatrix}
    e^{\lambda_i(\frac{1}{2}-w)} e^{\lambda_i(w-v)}  e^{\lambda_i(v+\frac{1}{2})}
    =
    \begin{pmatrix}
    1/2 & 1/12\\
    -1/12 & 0
    \end{pmatrix}
    e^{\lambda_i};
\end{equation}

\begin{align}
    (iij)_{\{1,w\}}:\quad &\int^{\tfrac{1}{2}}_{-\tfrac{1}{2}}dw \int^{w}_{-\tfrac{1}{2}} dv\  e^{\lambda_i(\frac{1}{2}-w)} \left\{ 1, w\right\} e^{\lambda_i(w-v)} e^{\lambda_j(v+\frac{1}{2})} \notag \\
    &= 
    \int^{\tfrac{1}{2}}_{-\tfrac{1}{2}}dw \int^{w}_{-\tfrac{1}{2}} dv\  \left\{ 1, w\right\} e^{\lambda_i(\frac{1}{2}-v)} e^{\lambda_j(v+\frac{1}{2})} \notag \\
    &= 
    \int^{\tfrac{1}{2}}_{-\tfrac{1}{2}}dv \ 
    \left\{\left(\frac{1}{2}-v\right), \frac{1}{2}\left( \frac{1}{2}-v\right) \left( \frac{1}{2}+v\right)\right\}
    e^{\lambda_i(\frac{1}{2}-v)} e^{\lambda_j(v+\frac{1}{2})} \notag \\
    &=
    \left\{ \partial_{\lambda_i}, \frac{1}{2} \partial_{\lambda_i}\partial_{\lambda_j}\right\} \frac{e^{\lambda_i} - e^{\lambda_j}}{\lambda_i - \lambda_j}\ .
\end{align}
Linear terms in $v$ are obtained by acting
$ (-\frac{1}{2} + \partial_{\lambda_j})$ on $(iij)_{\{1,w\}}$.

\begin{align}
    (iji):\quad &\int^{\tfrac{1}{2}}_{-\tfrac{1}{2}}dw \int^{w}_{-\tfrac{1}{2}} dv\  e^{\lambda_i(\frac{1}{2}-w)} \left\{ 1, w\right\} e^{\lambda_j(w-v)} \left\{ 1, v\right\} e^{\lambda_i(v+\frac{1}{2})} \notag \\
    &=e^{\lambda_i} \int dwdv\ \theta(w-v) e^{(\lambda_j-\lambda_i)(w-v)} \left\{ 1,w\right\}\times\left\{ 1,v\right\} \notag \\
    &= e^{\lambda_i} \int dwdv\ \theta(w-v) e^{(\lambda_j-\lambda_i)(w-v)}
    \begin{pmatrix}
    1 & \frac{1}{2}(w+v) +\frac{w-v}{2} \\
    \frac{1}{2}(w+v) - \frac{w-v}{2} & w v
    \end{pmatrix} \notag \\
    &= e^{\lambda_i} \int dwdv\ \theta(w-v) e^{(\lambda_j-\lambda_i)(w-v)}
    \begin{pmatrix}
    1 & 0 +\frac{1}{2}\partial_{(\lambda_j-\lambda_i)} \\
    0 - \frac{1}{2}\partial_{(\lambda_j-\lambda_i)} & w v
    \end{pmatrix} \notag \\
    &=
    e^{\lambda_i}\begin{pmatrix}
    \frac{e^{\mathcal{K}}-1-\mathcal{K}}{\mathcal{K}^2} & \frac{e^{\mathcal{K}}(\mathcal{K}-2)+2+\mathcal{K}}{2\mathcal{K}^3} \\
    -\frac{e^{\mathcal{K}}(\mathcal{K}-2)+2+\mathcal{K}}{2\mathcal{K}^3} & \frac{3e^{\mathcal{K}}(\mathcal{K}-2)^2-12+\mathcal{K}^3+3 \mathcal{K}^2}{12\mathcal{K}^4} 
    \end{pmatrix} ,
\end{align}
with $\mathcal{K}=\lambda_j-\lambda_i$.

\begin{align}
    (ijj)_{\{1,v\}}:\quad &\int^{\tfrac{1}{2}}_{-\tfrac{1}{2}}dw \int^{w}_{-\tfrac{1}{2}} dv\  e^{\lambda_i(\frac{1}{2}-w)} e^{\lambda_j(w-v)} \left\{ 1, v\right\} e^{\lambda_j(v+\frac{1}{2})} \notag \\
    &= 
    \int^{\tfrac{1}{2}}_{-\tfrac{1}{2}}dw \int^{w}_{-\tfrac{1}{2}} dv\  e^{\lambda_i(\frac{1}{2}-w)} e^{\lambda_j(w+\frac{1}{2})} \left\{ 1, v\right\} \notag \\
    &= 
    \int^{\tfrac{1}{2}}_{-\tfrac{1}{2}}dw \  e^{\lambda_i(\frac{1}{2}-w)} e^{\lambda_j(w+\frac{1}{2})} \left\{ \left(\frac{1}{2}+w\right) , -\frac{1}{2}\left(\frac{1}{2}-w\right)\left(w +\frac{1}{2}\right) \right\} \notag \\
    &=  \left\{ \partial_{\lambda_j},\frac{-1}{2}\partial_{\lambda_i}\partial_{\lambda_j}\right\} \frac{e^{\lambda_i} - e^{\lambda_j}}{\lambda_i - \lambda_j}\ .
\end{align}
Linear terms in $w$ are obtained by acting $\left(\frac{1}{2} - \partial_{\lambda_i}\right)$ on $(ijj)_{\{1,v\}}$.

\begin{equation}
    (ijk):\quad \int^{\tfrac{1}{2}}_{-\tfrac{1}{2}}dw \int^{w}_{-\tfrac{1}{2}} dv\  e^{\lambda_i(\frac{1}{2}-w)} e^{\lambda_j(w-v)} e^{\lambda_k(v+\frac{1}{2})}
    =  \frac{\lambda_i(e^{\lambda_j}-e^{\lambda_k})
        + \lambda_j(e^{\lambda_k}-e^{\lambda_i})
        + \lambda_k(e^{\lambda_i}-e^{\lambda_j})
        }{(\lambda_i-\lambda_j)(\lambda_j-\lambda_k)(\lambda_k-\lambda_i)} \ .
\end{equation}
Linear terms in  $w$ are obtained by acting with $\left(\frac{1}{2} - \partial_{\lambda_i}\right)$ on $(ijk)$;
those in $v$, by acting with $\left(-\frac{1}{2} + \partial_{\lambda_k}\right)$;
those in $w v$, by acting with $\left(\frac{1}{2} - \partial_{\lambda_i}\right)\left(-\frac{1}{2} + \partial_{\lambda_k}\right)$.

\section{Comparison of the eikonal quadrupole with the results in the literature}
\label{app:Compare_to_literature}

In this section, we compare the result of the eikonal quadrupole to the known results in the literature, in particular to~\cite{Dominguez:2008aa}. 
We start with the operator $\mathcal{Q}_{[x^+,y^+]}$, Eq.~\eqref{eq:calQ_def}, composed of four fundamental Wilson lines with open color indices. All possible color contractions for this operator in the two-dimensional color basis ${|e_i\rangle}$ defined in Eq.~\eqref{def:basis_ei} can be encoded in the matrix $\mathcal{H}$ which reads 
%
\begin{align}\label{eq: def H}
    {\mathcal{H}_{[x^+, y^+]}}_{ij} = \bra{e_i}  \mathcal{Q}_{[x^{+}, y^{+}]}\ket{e_j}  
    =
    \begin{pmatrix}
        N_c^2 \, DD_{[x^{+},y^+]} & N_c \, \widetilde{Q}_{[x^{+},y^+]} \\
        N_c \, Q_{[x^{+},y^+]} & N_c^2 \, \widetilde{DD}_{[x^{+},y^+]}
    \end{pmatrix}_{ij}\ ,
\end{align}
%
where we define
\begin{align}
    DD_{[x^{+},y^+]} = \frac{1}{N_{c}^{2}} \Braket{\text{Tr}[\U(\x_1) \U^{\dagger}(\x_2)] \text{Tr}[\U(\x_3) \U^{\dagger}(\x_4)]}_{[x^+, y^+]} ,\\
    Q_{[x^+, y^+]} = \frac{1}{N_{c}} \Braket{\text{Tr}[\U(\x_1) \U^{\dagger}(\x_2)] \U(\x_3) \U^{\dagger}(\x_4)]}_{[x^+, y^+]}, \\
    \widetilde{ DD}_{[x^{+},y^+]} = \frac{1}{N_{c}^{2}} \Braket{\text{Tr}[\U(\x_1) \U^{\dagger}(\x_4)] \text{Tr}[\U(\x_3) \U^{\dagger}(\x_2)]}_{[x^+, y^+]}, \\
    \widetilde{Q}_{[x^+, y^+]} = \frac{1}{N_{c}} \Braket{\text{Tr}[\U(\x_1) \U^{\dagger}(\x_4)] \U(\x_3) \U^{\dagger}(\x_2)]}_{[x^+, y^+]}.
\end{align}


In order to write the operator $\mathcal{Q}$ in terms of matrix $\mathcal{H}$, we can multiply it by the identity, $\mathbb{1}$, defined in Eq.~\eqref{eq: def c and identity}, on both sides of $\mathcal{Q}$, to get
\begin{align}
    \mathcal{Q} &= \ket{e_i} c_{ij} \bra{e_j}\mathcal{Q} \ket{e_m} c_{mn} \bra{e_n} \\ \nn
    &= \ket{e_i} \big[ c \ \mathcal{H} \ c \big]_{ij}\bra{e_j}\ .
\end{align}
Indeed, we obtain the non-tadpole part of the quadrupole by projecting it onto $\bra{e_2}...\ket{e_1}$ and using Eq.~\eqref{eq: def c and identity}:
\begin{align}
    \N = \bra{e_2} \mathcal{Q} \ket{e_1} = c_{2i}^{-1} \big[c \mathcal{H} c\big]_{ij} c_{j1}^{-1} = \mathcal{H}_{21}\ .
\end{align}
Now, comparing with Eq.~\eqref{NQ}, we can write
\begin{align}
    \mathcal{H} = c^{-1} R \begin{pmatrix}
        e^{\mu^{2}\lambda_{+}} & 0 \\
        0 & e^{\mu^{2}\lambda_{-}} \\
    \end{pmatrix} R^{-1},
\end{align}
with $R$ being the rotation matrix defined below and $\lambda_{\pm}$ the eigenvalues defined in Eq.~\eqref{eq: eigen system of F_1}. 
By including the tadpole contribution given in Eq.~\eqref{eq: def tadpole quadrupole}, the complete quadrupole can be written as 
%
\begin{align}
    Q = \T \mathcal{H} =  c^{-1} R \begin{pmatrix}
        e^{\mu^{2}\lambda_{+}} & 0 \\
        0 & e^{\mu^{2}\lambda_{-}} \\
    \end{pmatrix} R^{-1} \ e^{-4 \pi Q_{s}^{2} G^{--}(0)}\ ,
\end{align}
where
\begin{align}
    c^{-1} = \begin{pmatrix}
        N_{c}^2 & N_c \\
        N_{c }  & N_{c}^{2}
    \end{pmatrix} ,
    \hspace{1 cm}
    R = \begin{pmatrix}
        a_{+} & a_{-} \\
        1 & 1
    \end{pmatrix},
\end{align}
and $a_{\pm}$ defined in Eq.~\eqref{eq: eigen system of F_2}.

In order to solve the quadrupole numerically, it is convenient to define it in terms of dipoles.
Therefore, analogous to~\cite{Dominguez:2008aa}, let us define
\begin{align}
    \Gamma_{xy} &= \mu^{2} C_{F} g^{2} \bigg[G^{--}(0) - G^{--}(\x - \y)\bigg] 
   = \frac{Q^{2}_{s}}{4} (\x - \y)^{2} \ln\frac{1}{|\x -\y| \Lam}
\end{align}
and
\begin{align}
    F_{\x_1 \x_2 \x_3 \x_4} = \big(\Gamma_{\x_1 \x_3} - \Gamma_{\x_1 \x_4} + \Gamma_{\x_2 \x_4} - \Gamma_{\x_2 \x_3} \big) \left( \frac{-1}{\mu^{2} C_{F}} \right).
\end{align}
On the other hand, following the definition of $\alpha, \beta$ and $\gamma$ in Eq.~\eqref{def:abg}, we can write
\begin{align}
    &\alpha = g^{2} \big[G^{--}(\x_1 - \x_2) + G^{--}(\x_3 - \x_4)\big] ,\\
    &\beta =  g^{2} \big[G^{--}(\x_1 - \x_3) + G^{--}(\x_2 - \x_4)\big] ,\\
    &\gamma =  g^{2} \big[G^{--}(\x_1 - \x_4) + G^{--}(\x_2 - \x_3)\big] ,  
\end{align}
so
\begin{align}
    \beta - \gamma &=  g^{2} \big[G^{--}_{\x_1 \x_3} + G^{--}_{\x_2 \x_4} - G^{--}_{\x_1 \x_4} - G^{--}_{\x_2 \x_3}\big] \\ \nn
    &= - \frac{1}{\mu^{2}C_{F}g^{2}} g^{2} \big[ \Gamma_{\x_1 \x_3} + \Gamma_{\x_2 \x_4} - \Gamma_{\x_1 \x_4} - \Gamma_{\x_2 \x_3}\big]
    = F_{\x_1 \x_2 \x_3 \x_4}\ , \nn \\
    \alpha - \gamma &=  - \frac{1}{\mu^{2}C_{F}}  \big[ \Gamma_{\x_1 \x_2} + \Gamma_{\x_3 \x_4} - \Gamma_{\x_1 \x_4} - \Gamma_{\x_2 \x_3}\big]
    = F_{\x_1 \x_3 \x_2 \x_4}\ , \\
    \alpha - \beta &=  - \frac{1}{\mu^{2}C_{F}}  \big[ \Gamma_{\x_1 \x_2} + \Gamma_{\x_3 \x_4} - \Gamma_{\x_1 \x_3} - \Gamma_{\x_2 \x_4}\big]
    = - F_{\x_1 \x_4 \x_3 \x_2} \ .
\end{align}
Thus, comparing to Eqs.~\eqref{eq: eigen system of F_1},~\eqref{eq: eigen system of F_2} and~\eqref{eq: eigen system of F_3}, we can write
\begin{align}
    \Delta = (F_{\x_1 \x_3 \x_2 \x_4})^{2} + \frac{4}{N_{C}^{2}} F_{\x_1 \x_2 \x_3 \x_4} F_{\x_1 \x_4 \x_3 \x_2} 
\end{align}
and 
\begin{align}
    a_{\pm} = \frac{(-F_{\x_1 \x_3 \x_2 \x_4} \pm \sqrt{\Delta}) N_{c}}{2 F_{\x_1 \x_2 \x_3 \x_4}}\ .
\end{align}
Now we can evaluate $\begin{pmatrix}
    e^{\mu^{2}\lambda_{+}} & 0 \\
        0 & e^{\mu^{2}\lambda_{-}} \\
    \end{pmatrix} R^{-1} \ e^{-4 \pi Q_{s}^{2} G^{--}(0)}
$, where $4 \pi Q_{s}^{2} G^{--}(0) = 2g^{2}C_{F} \mu^{2} G^{--}_{0}$,  in terms of $\alpha, \beta$ and $\gamma$.
We obtain
\begin{align}
    \mu^{2} \lambda_{\pm} &= \frac{\mu^{2}}{2 N_{c}} (\beta - \alpha - \gamma) + \mu^{2} \frac{N_{c}}{4} (\alpha + \gamma \pm \sqrt{\Delta}) \\ \nn
    &= \frac{\mu^{2}}{2 N_{c}} (-\alpha) + \frac{\mu^{2}}{2 N_{c}}(\beta - \gamma) + \frac{\mu^{2} N_{c}}{4} [2 \alpha + (\beta - \gamma) + (\gamma - \beta) \pm \sqrt{\Delta}] \\ \nn
    &= \mu^{2} C_{F} \alpha + \frac{\mu^{2}}{2 N_{c}} F_{\x_1 \x_2 \x_3 \x_4} + \frac{\mu^{2} N_{c}}{4} (- F_{\x_1 \x_3 \x_2 \x_4} \pm \sqrt{\Delta}),
\end{align}
which yields
\begin{align}
\label{eq: eikonal quadrupole in terms of Fx1x2x3x4}
    \mu^{2} \lambda_{\pm} - 2g^{2}C_{F} \mu^{2} G_{0}^{--} &= - \Gamma_{\x_1 \x_2} - \Gamma_{\x_3 \x_4} + \frac{\mu^{2}}{2N_{c}} F_{\x_1 \x_2 \x_3 \x_4} + \frac{\mu^{2} N_{c}}{4} (- F_{\x_1 \x_3 \x_2 \x_4} \pm \sqrt{\Delta}) \\ \nn
    &= A \pm B,
\end{align}
with 
\begin{align}
A &= - \Gamma_{\x_1 \x_2} - \Gamma_{\x_3 \x_4} + \frac{\mu^{2}}{2N_{c}} F_{\x_1 \x_2 \x_3 \x_4} + \frac{\mu^{2} N_{c}}{4} (- F_{\x_1 \x_3 \x_2 \x_4})\ , \\
B &= \frac{\mu^{2} N_{c}}{4} \sqrt{\Delta} \ ,
\end{align}
in agreement with the results obtained in~\cite{Dominguez:2008aa}.  

Analogously, we can compute the double dipole function in the same manner by projecting onto $\braket{e_1 | ...| e_1}$. The
result reads
\begin{align}
    \N_{DD} = \bigg[ \frac{\sqrt{\Delta} + \alpha - \gamma }{2 \sqrt{\Delta}} - \frac{\beta-\gamma}{N_{c}^{2}\sqrt{\Delta}}\bigg] e^{\tilde{\mu}^{2}\lambda_{-}} - \bigg[ \frac{-\sqrt{\Delta} + \alpha - \gamma }{2 \sqrt{\Delta}} - \frac{\beta-\gamma}{N_{c}^{2}\sqrt{\Delta}}\bigg] e^{\tilde{\mu}^{2}\lambda_{+}} ,
\end{align}
which agrees with the one obtained in~\cite{Dominguez:2008aa}.

\bibliography{mybib_New}
\end{document}